\documentclass[a4paper,12pt,english]{article}

\usepackage[T1]{fontenc} 
\usepackage[pdftex]{graphicx} 
\usepackage{caption}
\usepackage{hyperref} 
\hypersetup{colorlinks=true, linkcolor=black, urlcolor=blue}
\usepackage{amsfonts} 
\usepackage{lmodern}
\usepackage{amsmath}
\usepackage{epsfig}
\usepackage{palatino}
\usepackage[rftl]{floatflt}
\usepackage{colordvi}
\usepackage{fancybox}
\usepackage{bbold}
\usepackage{amssymb}
\usepackage{ae}
\usepackage{aecompl}
\usepackage{cancel}
\usepackage{verbatim} 
\usepackage{color}

\newcommand{\beq}{\begin{equation}}
\newcommand{\eeq}{\end{equation}}
\newcommand{\bea}{\begin{eqnarray}}
\newcommand{\eea}{\end{eqnarray}}
\newcommand{\bem}{\begin{pmatrix}}
\newcommand{\eem}{\end{pmatrix}}
\definecolor{purpleheart}{rgb}{0.41, 0.21, 0.61}

\newcommand{\non}{\nonumber}

\newcommand{\TeV}{\, \mathrm{TeV}}
\newcommand{\GeV}{\, \mathrm{GeV}}
\newcommand{\MeV}{\, \mathrm{MeV}}

\newcommand\lsim{\mathrel{\rlap{\lower4pt\hbox{\hskip1pt$\sim$}}
    \raise1pt\hbox{$<$}}}
\newcommand\gsim{\mathrel{\rlap{\lower4pt\hbox{\hskip1pt$\sim$}}
    \raise1pt\hbox{$>$}}}

\DeclareMathOperator{\Tr}{Tr}

\begin{document}

\numberwithin{equation}{section}

\title{{\bf  Resurrecting the minimal renormalizable\\ supersymmetric SU(5) model}
\vskip .5cm}

\author{
{\bf
Borut Bajc$^{a}$\footnote{borut.bajc@ijs.si} ,
St\'ephane Lavignac$^{b}$\footnote{stephane.lavignac@cea.fr}$\; $ and 
Timon Mede$^{c}$\footnote{mede@ipnp.troja.mff.cuni.cz}
}
\\
\\
$^a$ {\sl Jo\v zef Stefan Institute, 1000 Ljubljana, Slovenia}\\[0.3em]
$^b$ {\sl Institut de Physique Th\'eorique, CEA-Saclay,}\\
{\sl F-91191 Gif-sur-Yvette, France}\footnote{Unit\'e Mixte de Recherche
du CEA/DSM et du CNRS (UMR 3681).}\\[0.3em]
$^c$ {\sl Institute of Particle and Nuclear Physics,}\\
{\sl Faculty of Mathematics and Physics, Charles University in Prague,}\\
{\sl V Hole\v{s}ovi\v{c}k\'{a}ch 2, 180 00 Praha 8, Czech Republic}
}

\date{}

\maketitle


\vskip .3cm

\centerline{\bf Abstract}
\vskip .3cm

\noindent
It is a well-known fact that the minimal renormalizable supersymmetric SU(5)
model is ruled out assuming superpartner masses of the order of a few TeV.
Giving up this constraint and assuming only SU(5) boundary conditions 
for the soft terms, we find that the model is still alive. The viable region
of the parameter space typically features superpartner masses
of order $10^2$ to $10^4 \TeV$, with $\tan \beta$ values between $2$ and $5$,
but lighter spectra with single states around $10$ TeV are also possible.
The main constraints come from proton decay, the Higgs mass, the requirement 
of the SU(5) spectrum being reasonably below the Planck scale, and the lifetime of 
the universe. A generic feature of the model is metastability of the electroweak vacuum. 
In the absence of a suitable dark matter particle in the neutralino sector,
a light (order GeV or smaller) gravitino is a natural candidate.

\vfill
\eject

\tableofcontents

\hypersetup{colorlinks=true, linkcolor=red, urlcolor=blue}


\section{Introduction}

The minimal renormalizable supersymmetric SU(5) model~\cite{Dimopoulos:1981zb},
with just 3 pairs of $\bf 10 \oplus \bar 5$ fermion representations and an adjoint $\bf 24$ 
plus a $\bf 5 \oplus \bar 5$ pair in the Higgs sector,
is the simplest supersymmetric 
Grand Unified extension of the Standard Model.
It is therefore particularly important
to test this model in detail, and possibly to rule it out. 
Although the choice of the gauge group, supersymmetry and minimality do not need a special
motivation, it is more difficult to justify the absence of non-renormalizable terms in the superpotential.
Experimental evidence tells us that some of these terms must be strongly suppressed.
For instance, the superpotential operators $Q_1 Q_1 Q_2 L_k / M_{Planck}$ ($k=1,2,3$)
induce proton decay at an unacceptable rate unless they come with coefficients smaller
than about $10^{-7}$.
Their smallness will lead us to assume in this paper that for some (to us unknown) reason
all non-renormalizable operators can be neglected, and to adopt the minimal renormalizable
supersymmetric SU(5) model as a benchmark\footnote{Another option is to give up renormalizability
and to assume that the non-renormalizable operators giving rise to fast proton decay are
suppressed, while harmless higher-dimensional couplings can be sizable. Under this assumption,
it is possible to avoid fast proton decay from heavy colour triplet 
exchange~\cite{Chkareuli:1998wi,Bajc:2002bv,Bajc:2002pg,Bachas:1995yt} 
and to correct the (phenomenologically inaccurate) SU(5) relations between
charged lepton and down-type quark masses~\cite{EmmanuelCosta:2003pu}
(for a review on these issues, see Ref.~\cite{Bajc:2002pg}).}.

It has been shown long ago~\cite{Murayama:2001ur} that the region of the parameter
space of the minimal renormalizable supersymmetric SU(5) model corresponding
to TeV-scale soft terms is excluded\footnote{In the case of decoupling 
scenario of heavy first two generations of sfermions considered in \cite{Murayama:2001ur} 
proton decay can still be consistent with the experimental bounds providing the flavor sfermion 
sector gets a very specific form \cite{Bajc:2002bv}.}. The reason put forward was the incompatibility
between the colour triplet mass constraints associated with gauge coupling unification
on the one hand, and with proton decay on the other hand.
This conclusion relies however on the assumption of a relatively light superpartner spectrum
(although masses as large as $10 \TeV$ for the first two generations of sfermions
were considered in Ref.~\cite{Murayama:2001ur}).
The purpose of this paper is to ascertain whether it can be extended to the region of
larger superpartner masses.

In order to answer this question, several constraints have to be imposed on the model:
precise gauge coupling unification, correct predictions for charged fermion masses,
the Higgs mass constraint, the experimental bound on the proton lifetime, and finally
the experimental bounds from flavour physics and from direct searches for supersymmetric particles. 
We also require perturbativity of the model.
The observed down-type quark and charged lepton masses are accounted for by
generation-dependent supersymmetric threshold corrections with large
A-terms~\cite{Hall:1985dx,Hall:1993gn,Hempfling:1993kv,DiazCruz:2000mn,
Enkhbat:2009jt,Iskrzynski:2014zla,Anandakrishnan:2014qxa},
which in turn are bounded by considerations related to the stability of the electroweak
vacuum~\cite{Frere:1983ag,Claudson:1983et,Kusenko:1996jn}.
The soft supersymmetry breaking terms are required to respect SU(5) invariance
but are not assumed to be flavour universal as in the so-called mSUGRA model
(as a matter of fact, generation-dependent A-terms are needed to correct the SU(5)
fermion mass relations). To avoid potentially dangerous flavour effects,
we will therefore take the GUT-scale sfermion soft terms to be
aligned with fermion masses\footnote{Flavour violating soft terms
are going to be generated from renormalization group running (due to the CKM matrix,
and possibly to right-handed neutrino couplings and R-parity violating couplings),
but this will lead only to small effects in flavour-changing processes, especially in view
of the large superpartner masses. We shall therefore neglect these small RG effects.}~\cite{Nir:1993mx}.
Neutrino masses can be generated either 
through bilinear R-parity violation\footnote{The SU(5)-invariant
bilinear R-parity violating operators $\mu_i \bar 5_i 5_H$ also contain baryon number
violating terms, which however are harmless by virtue of the doublet-triplet splitting
if $\mu_i \ll M_{GUT}$~\cite{Hall:1983id,Smirnov:1995ey}. 
Contrary to Ref.~\cite{Bajc:2015zja}, we assume negligible trilinear R-parity violating
couplings.}~\cite{Hall:1983id,Hempfling:1995wj,Nilles:1996ij,Diaz:2003as} or 
by adding right-handed
neutrinos to the model in order to implement the seesaw mechanism~\cite{seesaw}.
Finally, if the neutralino sector does not contain a suitable candidate for dark matter 
(either because the lightest neutralino is not the lightest supersymmetric particle,
or because it is too heavy), this role may be played by the gravitino.

Not surprisingly, the main constraint on the superpartner mass scale 
comes from proton decay, which pushes the spectrum above the $10 \TeV$ scale
(the conflict between the proton decay and unification constraints pointed out
in Ref.~\cite{Murayama:2001ur} is resolved by relaxing the upper bound on
the superpartner masses). Perturbativity imposes an upper bound on the coloured
triplet mass, which translates into an upper limit on superpartner masses
once gauge coupling unification is imposed.
The observed Higgs mass, together with vacuum
metastability constraints associated with the stop A-term, also excludes large portions
of the parameter space. A priori, there is no guarantee that the minimal
renormalizable supersymmetric SU(5) model will survive all these constraints,
even if very large values of the soft terms are allowed.

\begin{figure}[h!]
\centering
\includegraphics[width=0.5\textwidth]{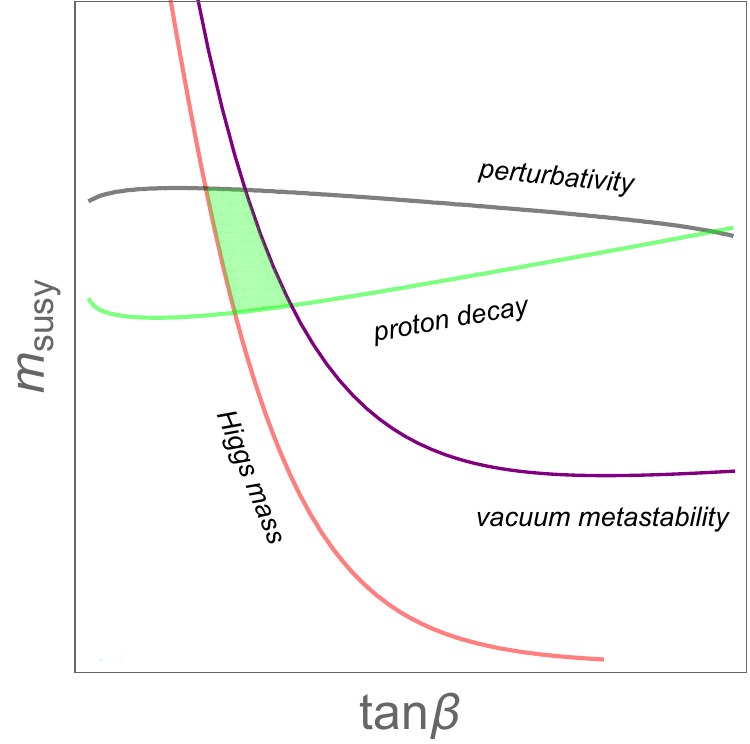}
\caption{Approximate constraints on the minimal renormalizable supersymmetric SU(5) model
in the $(\tan \beta, m_{susy})$ plane. The region to the left of the \textit{red curve} is excluded
by the measured Higgs boson mass, while the area to the right of the 
\textit{purple curve} is excluded
by vacuum stability considerations (namely, the value of the stop A-term needed
to reproduce the observed Higgs boson mass would render the electroweak vacuum
unstable with a lifetime shorter than the age of the universe).
The experimental lower bound on the proton lifetime is satisfied above the \textit{green line},
and the perturbativity constraint $m_T < M_{Planck} / 10$, where $m_T$ is the colour triplet mass,
is satisfied below the \textit{black line}. The \textit{green area} is consistent with all four constraints.
In this plot, the sfermion masses, the $\mu$ parameter and $m_{H_u}$ are all equal to $m_{susy}$,
as well as the common gaugino mass at the GUT scale. Since these values are not consistent
with SU(5) boundary conditions, the figure should be considered as illustrative only.}
\label{fig:Parameter space}
\end{figure}

To give an idea of how the parameter space is restricted by the various
phenomenological requirements, we show in Fig.~\ref{fig:Parameter space}
the approximate constraints in the ($\tan \beta$, $m_{susy}$) plane obtained
by making several simplifying assumptions.
Namely, all sfermion masses are taken to be equal to $m_{susy}$ at the
low scale, as well as the $\mu$ parameter and $m_{H_u}$ ($m^2_{H_u} > 0$
is assumed, and $m^2_{H_d}$ and the $B$ parameter are computed
from the electroweak symmetry breaking conditions),
while the gaugino masses assume a common value $M_{1/2} = m_{susy}$ at the GUT
scale and are split by  renormalization group running.
Obviously, these inputs are not consistent with SU(5) symmetry of the soft terms
at the GUT scale, but they make it possible to show several constraints in a single plot.
Imposing SU(5) boundary conditions will significantly affect quantities such as
the heavy colour triplet mass and more crucially the proton lifetime, making the
investigation of the parameter space of the minimal renormalizable supersymmetric
SU(5) model more involved than suggested by Fig.~\ref{fig:Parameter space}.
As we are going to see, phenomenologically viable points
typically feature superpartners in the ${\cal O} (10^2-10^4) \TeV$ range, with
values of $\tan \beta$ between $2$ and $5$, but lighter spectra with supersymmetric
particles as light as a few $10 \TeV$ can also be found.

In the process we have generalized the procedure of Refs.~\cite{Drees:1991ab,Rattazzi:1995gk} 
for deriving approximate semi-analytic solutions to the one-loop renormalization
group equations (RGEs) for the MSSM soft terms. 
In this way we are able to write the low-energy soft terms as linear or quadratic functions of
the initial (GUT-scale) parameters, making it possible to explore the parameter space 
without having to solve the RGEs for each point. In practice, one just needs to solve
numerically the RGEs for gauge and Yukawa couplings for each
choice of $\tan{\beta}$ and $m_{susy}$, the matching scale between the SM and the MSSM.
The low-energy soft terms and their dependence on the other model parameters are
then simply given by linear and quadratic algebraic equations.

The paper is organized as follows. Section~\ref{sec:model} introduces the minimal
renormalizable supersymmetric SU(5) model. In Section~\ref{sec:RGEs}, the running
of the model parameters is discussed, and the semi-analytical procedure used to solve
the renormalization group equations (RGEs) is presented in Section~\ref{sec:solutions}
(more details can be found in Appendix~\ref{app:RGEs}). Proton decay and the constraints
associated with the metastability of the electroweak vacuum are addressed in
Sections~\ref{sec:pdk} and~\ref{sec:vacuum}, respectively, with technical details
relegated to Appendices~\ref{app:pdk}, \ref{app:UFB} and~\ref{app:CCB}.
Finally, we present our results in Section~\ref{sec:results}.


\section{The minimal renormalizable supersymmetric SU(5) model}
\label{sec:model}

In this section, we briefly describe the minimal renormalizable supersymmetric SU(5)
model~\cite{Dimopoulos:1981zb} and present our notations. The Higgs sector includes
the adjoint $24_H$, which spontaneously breaks the SU(5) gauge group to
SU(3)$\times$SU(2)$\times$U(1), and a fundamental and anti-fundamental
representations $5_H$ and $\bar{5}_H$ containing the two light Higgs doublets
responsible for electroweak symmetry breaking. The $5_H \oplus \bar{5}_H$
Higgs fields also includes a heavy pair of colour triplet and antitriplet that
mediate proton decay through $d = 5$ operators. All matter fields belong
to $10_i$ and $\bar{5}_i$ representations (leaving aside right-handed neutrinos
in the singlet representation that may also be present), where $i=1,2,3$ is the generation index. 

In order to connect the minimal renormalizable supersymmetric SU(5)
model with experimental data,
one has to deal with three different theories: SU(5) above the unification scale
$M_{GUT}$, the Minimal Supersymmetric Standard Model (MSSM) between
$M_{GUT}$ and the supersymmetry scale $m_{susy}$, and the Standard Model (SM)
between $m_{susy}$ and the weak scale $m_Z$.
Since the heavy GUT states (resp. the superpartners) are not degenerate in mass,
the matching between the SU(5) theory and the MSSM at $M_{GUT}$ (resp.
between the MSSM and the SM at $m_{susy}$) will involve threshold corrections.

In the next subsections we give the relevant parts of the corresponding Lagrangians
(i.e. the Higgs and Yukawa sectors and the soft supersymmetry breaking terms,
which determine the superpartner spectrum) and we specify our notations and assumptions.

\subsection{The SU(5) model}
\label{subsec:SU5}

The superpotential of the minimal renormalizable supersymmetric SU(5) model
is determined by its field content, gauge invariance and renormalizability.
It can be divided into two parts describing the Higgs and Yukawa sectors, respectively:
\begin{align}
  W_H & \ = \ \frac{\mu_5}{2} \Tr 24_H^2 + \, \frac{\lambda_5}{3} \Tr 24_H^3
    + \bar{5}_H \left (\mu_H+\eta_H \,24_H \right ) 5_H\, ,
\label{eq:W_H}  \\
  W_Y & \ = \ \frac{1}{4}\, \Lambda^{10}_{ij}\, 10_i  10_j  5_H
    - \sqrt{2}\, \Lambda^{\bar 5}_{ij}\, \bar{5}_i 10_j \bar{5}_H\, ,
\label{eq:W_Y}
\end{align}
in which we have omitted terms involving right-handed neutrinos as well as R-parity
violating couplings that may be present, depending on how neutrino masses are generated. 
After having solved the equations of motion for the $24_H$ in the SM singlet direction:
\beq
  \langle 24_H \rangle\, =\, \sigma_0\, \mbox{Diag} (2,2,2,-3,-3)\; , \qquad \sigma_0\, =\, \mu_5/\lambda_5\, ,
\eeq
and performed the fine-tuning needed to achieve doublet-triplet splitting in the Higgs sector,
one can write down the masses of the heavy states
in terms of the SU(5) superpotential parameters:
\begin{align}
  m_V = 5 g_{\scriptscriptstyle GUT} \sigma_0\, , \ \  \quad
  m_T = 5 \eta_H \sigma_0\, , \ \  \quad
  m_8 = - m_3 = -5 m_1 = 5 \lambda_5 \sigma_0\,  ,
\end{align}
where $m_V$ is the mass of the SU(5) gauge bosons in the representations
$(3,2)_{-5/3} \oplus (\bar 3, \bar 2)_{+5/3}$ of the SM gauge group; $m_T$ is the mass
of the colour triplet and antitriplet pair $(T, \bar T)$ contained in $5_H \oplus \bar{5}_H$;
and $m_1$, $m_3$ and $m_8$ are the masses of the SM singlet, SU(2) triplet and
SU(3) octet components of $24_H$, respectively.
Demanding that the superpotential couplings $\lambda_5$ and $\eta_H$
be in the perturbative regime and taking into account the fact that the unified gauge coupling
$g_{\scriptscriptstyle GUT}$ is of order $1$, one obtains the following constraint:
\begin{align}
  m_T, \, m_8, \, m_3 & \ \lsim \ m_V\, .
\end{align}
We also assume that supersymmetry breaking is coming from above the GUT scale,
as for example in supergravity. In practice this means that the soft terms should be SU(5)
symmetric at the GUT scale:
\bea
  -{\cal L}_{soft} & = & (m^2_{10})_{ij}\, \tilde{10}^\dagger_i \tilde{10}_j
    + (m^2_{\bar 5})_{ij}\, \tilde{\bar 5}^\dagger_i \tilde{\bar 5}_j
    + \left( \frac{1}{2}\, A^{10}_{ij}\,\tilde{10}_i \tilde{10}_j 5_H
    - A^{\bar 5}_{ij}\,\tilde{\bar 5}_i \tilde{10}_j \bar 5_H + \mbox{h.c.} \right)  \non \\
  && +\ m_{5_H}^2\, 5^\dagger_H 5_H + m_{\bar 5_H}^2\, \bar 5^\dagger_H \bar 5_H
    + \left( B_5\, 5_H \bar 5_H + \mbox{h.c.} \right) + \frac{1}{2}\, M_{1/2}\, \bar \lambda^a \lambda^a .
\eea
We will consider the possibility of generation-dependent soft terms (as explained
in the introduction, generation-dependent A-terms are needed to correct the SU(5)
fermion mass relations), but in order to comply with the strong constraints coming
from flavour physics we must ensure that they do not induce large flavour-violating
effects. To this end, we assume that the soft sfermion mass matrices
are diagonal in the basis in which the Yukawa couplings $\Lambda^{\bar 5}$ are diagonal,
and that the A-term matrices are diagonal in the corresponding fermion mass
eigenstate basis:
\bea
  (m^2_{10})_{ij} = m^2_{10_i} \delta_{ij}\, ,\ \  (m^2_{\bar 5})_{ij} = m^2_{\bar 5_i} \delta_{ij}\, ,\ \
    A^{\bar 5}_{ij} = A^{\bar 5}_i \delta_{ij} & & \mbox{in the basis}\
    \Lambda^{\bar 5}_{ij} = \Lambda^{\bar 5}_i \delta_{ij}\, ,  \hskip 1cm
  \label{eq:soft_mass_BC} \\
  A^{10}_{ij} = A^{10}_i \delta_{ij} & & \mbox{in the basis}\
    \Lambda^{10}_{ij} = \Lambda^{10}_i \delta_{ij}\, ,
  \label{eq:A_10_BC}
\eea
so that all flavour violation at the GUT scale is concentrated in the up squark sector
and controlled by the CKM angles, yielding an effective alignment of sfermion
soft terms with fermion masses~\cite{Nir:1993mx}. In addition, we assume that the soft 
masses of the first two generations of sfermions are degenerate:
\beq
  m^2_{10_1}\, =\, m^2_{10_2}\, ,  \quad   m^2_{\bar 5_1}\, =\, m^2_{\bar 5_2}\, .
\eeq
Finally, we will take $M_{1/2}$ and the A-terms (as well as the $\mu$ parameter
$\mu \equiv \mu_H - 3 \eta_H \sigma_0$) to be real.
This may be more than what we need to evade flavour and CP constraints from
low-energy experiments, especially in view of the fact that the superpartner spectrum
is heavy, but this choice also helps reducing the number of parameters. 
In our subsequent exploration of the parameter space of the minimal renormalizable
supersymmetric SU(5) model we shall completely neglect flavour violation
in the sfermion sector, including the small amount of flavour violation that is generated
from the running of the soft terms.

\subsection{MSSM}

Below the scale $M_{GUT}$, the relevant theory is the MSSM, with superpotential
\begin{align}
  W_{MSSM} & \ = \ \Lambda_{ij}^U\, U_i^c Q_j H_u\, -\, \Lambda_{ij}^D\, D_i^c Q_j H_d\,
  -\, \Lambda_{ij}^E\, E_i^c L_j H_d\, +\, \mu\, H_u H_d\, ,
\end{align}
and soft supersymmetry breaking terms
\bea   
  -\, {\cal L}_{soft} & = & (m^2_{\tilde Q})_{ij}\, {\tilde q}^{\dagger}_i {\tilde q}_j
  + (m^2_{\tilde u^c})_{ij}\, {\tilde u}^c_i {\tilde u}^{c \dagger}_j
  + (m^2_{\tilde e^c})_{ij}\, {\tilde e}^c_i {\tilde e}^{c \dagger}_j
  + (m^2_{\tilde L})_{ij}\, {\tilde l}^{\dagger}_i {\tilde l}_j
  + (m^2_{\tilde d^c})_{ij}\, {\tilde d}^c_i {\tilde d}^{c \dagger}_j  \nonumber \\
  && + \left( A^U_{ij}\, {\tilde u^c}_i {\tilde q}_j h_u - A^D_{ij}\, {\tilde d^c}_i {\tilde q}_j h_d 
    - A^E_{ij}\, {\tilde e}^c_i {\tilde l}_j h_d + \mbox{h.c.} \right)  \nonumber \\
  && +\ m_{H_u}^2 h_u^{\dagger} h_u + m_{H_d}^2 h_d^{\dagger} h_d
    + \left(  B \, h_u h_d + \mbox{h.c.} \right)  \nonumber \\ 
  && +\ \frac{1}{2} M_1\, \bar{\tilde B} {\tilde B} + \frac{1}{2} M_2\, \bar{\tilde W}^i {\tilde W}^i
    + \frac{1}{2} M_3\, \bar{\tilde g}^a {\tilde g}^a ,
\eea
where the contraction of $SU(2)_L$ indices is understood (for instance,
$H_u H_d$ stands for $\epsilon_{ij} H^i_u H^j_d = H^+_u H^-_d - H^0_u H^0_d$,
where $i,j = 1,2$ are $SU(2)_L$ indices and $\epsilon_{ij}$ is the totally
antisymmetric tensor with $\epsilon_{12} = +1$).
Due to the boundary conditions~(\ref{eq:soft_mass_BC})
and~(\ref{eq:A_10_BC}), and to the fact that we are neglecting the effects of
the CKM matrix in the running, the sfermion soft terms keep a diagonal form
all the way down to low energies.

\subsection{Standard Model}

Below the matching scale
\begin{align}
\label{msusy}
  m_{susy} & \, \equiv \, \sqrt{m_{\tilde t_1}(m_{susy})\, m_{\tilde t_2}(m_{susy})}\,
    \simeq\, \sqrt{m_{\tilde Q_3} (m_{susy})\, m_{\tilde u^c_3} (m_{susy})}
\end{align}
(where the last approximation is valid as long as the mixing in the stop sector is small,
i.e. $m_t X_t \ll m^2_{\tilde Q_3}, m^2_{\tilde u^c_3}$), the relevant theory is the Standard
Model. The Higgs potential is given by (with the SM Higgs doublet given by
$h = \sqrt{2} \left( \cos \beta\, h_d + \sin \beta\,  i \sigma^2 h^*_u \right)$
in the decoupling limit):
\begin{align}
  V_{SM} & = - \, m_{h}^2 \, h^{\dagger} h + \frac{\lambda}{2} \left ( h^{\dagger} h \right )^2 ,
\end{align}
while the Yukawa Lagrangian is
\beq
  {\cal L}_{Yukawa} = - H_{ij}^U\, \bar u_{Ri} \tilde h^\dagger q_j - H_{ij}^D\, \bar d_{Ri} h^\dagger q_j
    -H_{ij}^E\, \bar e_{Ri} h^\dagger l_j + \mbox{h.c.}\, ,
\eeq
where $\tilde h \equiv i \sigma^2 h^*$.


\section{Renormalization group equations}
\label{sec:RGEs}

In this section, we collect the SM and MSSM renormalization group equations (RGEs)
and various expressions used in our analysis (from boundary to matching conditions). 
We use the SM RGEs~\cite{Luo:2002ey} between $m_Z$ and $m_{susy}$,
and the MSSM RGEs~\cite{Martin:1993zk} between $m_{susy}$ and $M_{GUT}$.
The gauge and Yukawa couplings, as well as the Higgs quartic coupling are evolved
with the 2-loop RGEs, with the 1-loop threshold corrections accounting for the splitting
of superpartner masses added at the scale $m_{susy}$. 
All soft parameters (A-terms, gaugino masses and soft scalar masses) are run at 1 loop.

\subsection{Gauge couplings}

The 2-loop RGEs for the gauge couplings $g_i$ ($i = 1,2,3$ for the gauge groups
$U(1)_Y$, $SU(2)_L$ and $SU(3)_C$, respectively, with $g_1 = \sqrt{5/3}\, g'$) read:
\begin{align}
\frac{d}{dt}g_i & \ = \ \frac{g_i^3}{(4 \pi)^2} b_i \, +\, \frac{g_i^3}{(4 \pi)^4} \left(\, \sum_{j=1}^3 B_{ij} \, g_j^2
- \! \sum_{\alpha=u,d,e} \!\! C_{i\alpha} \Tr(\Lambda_{\alpha}^{\dagger}\Lambda_{\alpha}) \right) ,
\label{eq:RGEgauge}
\end{align}
where $t \equiv \ln (m / m_0)$, $m$ being the renormalization group scale, and the $\beta$-function
coefficients below and above $m_{susy}$ are given by:
\begin{align}
b_i^{SM} & =  (41/10,-19/6,-7)\, , & & \,\, b_i^{MSSM}  =  (33/5,1,-3)\, , \\
B_{ij}^{SM} & =  \left[ \begin{array}{ccc} 199/50 & 27/10 & 44/5 \\ 9/10 & 35/6 & 12 \\ 11/10 & 9/2 & -26 \end{array} \right]\, ,
& & B_{ij}^{MSSM}  =  \left[ \begin{array}{ccc} 199/25 & 27/5 & 88/5 \\ 9/5 & 25 & 24 \\ 11/5 & 9 & 14 \end{array} \right]\, , \\
C_{i\alpha}^{SM} & =  \left[ \begin{array}{ccc} 17/10 & 1/2 & 3/2 \\ 3/2 & 3/2 & 1/2 \\ 2 & 2 & 0 \end{array} \right]\, , & &
C_{i\alpha}^{MSSM} = \left[ \begin{array}{ccc} 26/5 & 14/5 & 18/5 \\ 6 & 6 & 2 \\ 4 & 4 & 0 \end{array} \right] .
\end{align}
In the last term of Eq.~(\ref{eq:RGEgauge}), the MSSM Yukawa couplings $\Lambda_{\alpha}$
should be replaced with the SM ones ($H_{\alpha}$) below $m_{susy}$.

At $m_{susy}$, the running gauge couplings should be converted from the $\overline{\textrm{MS}}$
scheme to the $\overline{\textrm{DR}}$ scheme, in which the MSSM RGEs are written:
\begin{align}
\alpha_1^{-1}(m_{susy}+\epsilon) &\, =\, \alpha_1^{-1}(m_{susy}-\epsilon)\, ,  \label{eq:g1_conversion}  \\
\alpha_2^{-1}(m_{susy}+\epsilon) &\, =\, \alpha_2^{-1}(m_{susy}-\epsilon) -\frac{1}{6\pi}\, ,  \\
\alpha_3^{-1}(m_{susy}+\epsilon) &\, =\, \alpha_3^{-1}(m_{susy}-\epsilon) -\frac{1}{4\pi}\, ,  \label{eq:g3_conversion}
\end{align}
where  $\alpha_i \equiv g_i^2 / (4\pi)$ and $\epsilon\to0^+$.

\subsubsection{Threshold corrections to gauge couplings}

Imposing gauge coupling unification at the GUT scale:
\begin{align}
\alpha_1(M_{GUT}) \ = \ \alpha_2(M_{GUT}) \ = \ \alpha_3(M_{GUT}) & \ \equiv \ \alpha_{GUT}
\end{align}
implies certain relations among the masses of the various thresholds (supersymmetric partners
of the SM fields and heavy GUT fields). Adding 1-loop threshold
corrections~\cite{Weinberg:1980wa,Hall:1980kf,Yamada:1992kv,HMY93,Hisano:2013cqa}
to the running gauge couplings evolved with the 2-loop MSSM RGEs between the scales
$m_{susy}$ and $M_{GUT}$ yields the following relations ($i = 1, 2, 3$):
\begin{align}
\alpha_{GUT}^{-1} &\ = \overbrace{{}\bar \alpha_i^{-1}(M_{GUT})}^{\textrm{2-loop single-scale MSSM}}
  - \overbrace{\frac{1}{2\pi}\ \sum_n \Delta b_i^{(n)}
  \ln\frac{m_{susy}}{m_n}}^{\textrm{1-loop low-energy threshold corrections}} \nonumber \\
& \qquad - \underbrace{\frac{1}{2\pi}\left [\Delta b_i^{V}\ln\frac{M_{GUT}}{m_V} + \Delta b_i^{8}\ln\frac{M_{GUT}}{m_8}
  + \Delta b_i^{3}\ln\frac{M_{GUT}}{m_3}+ \Delta b_i^{T}\ln\frac{M_{GUT}}{m_T} \right  ]}_{\textrm{1-loop high-energy
  threshold corrections}}\ ,
\label{eq:unification}
\end{align}
where the ${}\bar \alpha_i$ denote the values of the gauge couplings obtained by solving numerically
the 2-loop MSSM RGEs with all superpartner masses at the scale $m_{susy}$, and $n$ runs over
the superpartners. Their contributions $\Delta b_i^{(n)}$ to the $\beta$-function coefficients are given by:
\beq
\begin{array}{rclrcl}
\Delta b_i^{\tilde{Q}_i} & = & (1/30,1/2,1/3)\, , \quad &  \Delta b_i^{\tilde{h}} & = & (2/5,2/3,0)\, , \\
\Delta b_i^{\tilde{u}_i} & = & (4/15,0,1/6)\, , &  \Delta b_i^{A} & = & (1/10,1/6,0)\, , \\
\Delta b_i^{\tilde{d}_i} & = & (1/15,0,1/6)\, , & \Delta b_i^{\tilde{b}} & = & (0,0,0)\, , \\
\Delta b_i^{\tilde{L}_i} & = & (1/10,1/6,0)\, , & \Delta b_i^{\tilde{w}} & = & (0,4/3,0)\, , \\
\Delta b_i^{\tilde{e}_i} & = & (1/5,0,0)\, , & \Delta b_i^{\tilde{g}} & = & (0,0,2)\, ,
\end{array}
\eeq
while the contributions of the heavy GUT fields are:
\beq
\begin{array}{rclrcl}
\Delta b_i^{8} & = & (0,0,3)\, , \quad & \Delta b_i^{3} & = & (0,2,0)\, , \\
\Delta b_i^{1} & = & (0,0,0)\, , & \Delta b_i^{T} & = & (2/5,0,1)\, , \\
\Delta b_i^{V} & = & (-10,-6,-4)\, . & & &
\end{array}
\eeq
Taking appropriate combinations of the three equations~(\ref{eq:unification}), one obtains:
\begin{align}
\label{mT}
\frac{m_T}{M_{GUT}} & \ = \ \exp{\left[\frac{5\pi}{6}\left(-\,{}\bar \alpha_1^{-1}+3\,{}\bar \alpha_2^{-1}-2\,{}\bar \alpha_3^{-1}\right) (M_{GUT})\right]}
\left(\frac{m_3}{m_8}\right)^{5/2}\nonumber\\
& \ \times \ \left(\frac{m_{\tilde w}}{m_{\tilde g}}\right)^{5/3}
\prod_{i=1}^3\left(\frac{m_{\tilde Q_i}^4}{m_{\tilde u^c_i}^3m_{\tilde e^c_i}}\frac{m_{\tilde L_i}^2}{m_{\tilde d^c_i}^2}\right)^{1/12}
\left(\frac{m_{\tilde h}^4m_A}{m_{susy}^5}\right)^{1/6} ,  \\
\label{mV}
\frac{\left[m_V^2(m_3m_8)^{1/2}\right]^{1/3}}{M_{GUT}} & \ = \
\exp{\left[\frac{\pi}{18}\left(5\,{}\bar \alpha_1^{-1}-3\,{}\bar \alpha_2^{-1}-2\,{}\bar \alpha_3^{-1}\right) (M_{GUT})\right]}\nonumber\\
& \ \times \ \left(\frac{m_{susy}^2}{m_{\tilde w}m_{\tilde g}}\right)^{1/9}
\prod_{i=1}^3\left(\frac{m_{\tilde u^c_i}m_{\tilde e^c_i}}{m_{\tilde Q_i}^2}\right)^{1/36} ,  \\
\label{aGUT}
\alpha_{GUT}^{-1} & \ = \ \left [ \left (-\frac{5}{9}\,{}\bar \alpha_1^{-1}+\frac{12}{9}\, {}\bar \alpha_2^{-1}+\frac{2}{9}\, {}\bar \alpha_3^{-1}\right ) (M_{GUT}) \right ]  \nonumber \\
+ \ \frac{1}{2\pi} \ln & \!\left [ \frac{m_8^{1/5} m_3^{4/5}}{m_V} \right ]^{10/3}\!
+\ \frac{1}{2\pi} \ln\! \left [\frac{m_{\tilde{h}}^{12} m_A^3 m_{\tilde{w}}^{32} m_{\tilde{g}}^{8}}{m_{susy}^{91}} \, \prod_{i=1}^3 \left (\frac{m_{\tilde{Q}_i}^{13} m_{\tilde{L}_i}^{3}}{m_{\tilde{u}_i^c}^{2} m_{\tilde{e}_i^c}^{2}}\right )  \right ]^{1/18} .
\end{align}
%
At the 1-loop level, the matching scales $M_{GUT}$ and $m_{susy}$ drop out from
Eqs.~(\ref{mT}) and~(\ref{mV}), while only $m_{susy}$ drops out from Eq.~(\ref{aGUT}). 
Since $m_{susy} \gg m_Z$, one can neglect the mixing between the higgsinos and the electroweak
gauginos and identify:
\begin{align}
m_{\tilde h} & \ = \ \mu\, , \\
m_A & \ = \ \sqrt{(|\mu|^2+m_{H_d}^2 (m_{susy}))(1+1/\tan^2{\beta})} \nonumber \\
& \ \simeq \ \sqrt{(m_{H_d}^2 (m_{susy}) - m_{H_u}^2 (m_{susy})) (\tan^2\beta + 1) \, / \, (\tan^2\beta - 1)}\ ,
\label{eq:mA} \\
m_{\tilde w} / m_{\tilde g} & \ \simeq \ \alpha_2(m_{\tilde g}) / \alpha_3(m_{\tilde g})\, ,
\end{align}
where $\tan \beta \equiv \langle h^0_u \rangle / \langle h^0_d \rangle$ is the ratio
of the two MSSM Higgs doublet vevs, and $\mu$ satisfies the electroweak symmetry
breaking (EWSB) condition (again neglecting $m^2_Z$):
\begin{align}
|\mu|^2 & \ \simeq \ \frac{m_{H_d}^2(m_{susy})-m_{H_u}^2(m_{susy})\tan^2{\beta}}{\tan^2{\beta}-1}\ .
\label{eq:mu}
\end{align}

In spite of the initial historical success~\cite{Dimopoulos:1981yj,Ibanez:1981yh,Einhorn:1981sx,Marciano:1981un},
it is well known that gauge couplings do not unify accurately at the 2-loop level
in the MSSM with TeV-scale superpartners. High-energy threshold corrections
thus play a crucial role in achieving precise unification~\cite{HMY93}.
Using Eqs.~(\ref{mT}) and (\ref{mV}),
one can express the combinations of GUT state masses needed for
exact 2-loop unification in terms of the superpartner masses~\cite{HMY93}.
Assuming that all superpartners have masses equal to $m_{susy}$, one obtains
for the colour triplet mass and for the combination of heavy gauge boson
and adjoint Higgs masses $m_V^{2/3} m_3^{1/3}$:
\begin{align}
  m_T & \ \sim\, 2 \times 10^{15} \GeV\, \left (\frac{m_{susy}}{1\TeV} \right )^{5/6}\, ,
  \label{eq:mT_msusy}  \\
  m_V^{2/3}m_3^{1/3} & \ \sim\, 2 \times 10^{16} \GeV\, \left (\frac{m_{susy}}{1\TeV} \right )^{-2/9}\, ,
\end{align}
where we have used the fact that $m_3 = m_8$ in the minimal renormalizable supersymmetric
SU(5) model.  For superpartner masses in the TeV range, the colour triplet is far too light
and makes the proton decay too fast, which led Ref.~\cite{Murayama:2001ur} to conclude that
the minimal renormalizable supersymmetric SU(5) model is excluded (this conclusion has 
been found to be mitigated at the three-loop level~\cite{Martens:2010nm} though, and can 
be avoided for a specific choice of the soft terms~\cite{Bajc:2002bv}).

\subsection{Yukawa couplings}
\label{subsec:Yukawa}

For the Yukawa couplings, we use the 2-loop Standard Model RGEs~\cite{Luo:2002ey}
below the scale $m_{susy}$, and the 2-loop MSSM RGEs~\cite{Martin:1993zk} above it.
We neglect all CKM contributions and work with diagonal Yukawa matrices.
In the absence of threshold corrections, the matching conditions at $m_{susy}$ are:
\begin{align}
  \lambda_{t,c,u}(m_{susy})\, & =\, h_{t,c,u}(m_{susy})/\sin\beta\, ,  \label{eq:Yu_conversion} \\
  \lambda_{b,s,d}(m_{susy})\, & =\, h_{b,s,d}(m_{susy})/\cos\beta\, , \\
  \lambda_{\tau,\mu,e}(m_{susy})\, & =\, h_{\tau,\mu,e}(m_{susy})/\cos\beta\, .  \label{eq:Ye_conversion}
\end{align}
where the $h$ couplings (resp. the $\lambda$ couplings) are the diagonal entries of
the SM Yukawa matrices $H^{U,D,E}$ (resp. of the MSSM Yukawa matrices
$\Lambda^{U,D,E}$).

\subsubsection{Threshold corrections to Yukawa couplings}

At the GUT scale, the SU(5)-invariant boundary conditions apply:
\begin{align}
  \Lambda_U(M_{GUT}) & \ = \ \Lambda_U^T(M_{GUT})\, ,  \\
  \Lambda_D(M_{GUT}) & \ = \ \Lambda_E^T(M_{GUT})\, .
  \label{eq:Md_Me}
\end{align}
Threshold corrections due to the splitting of the heavy GUT state masses slightly
modify these relations (see later).
After running down to low energy, Eqs.~(\ref{eq:Md_Me}) lead to predictions for down-type quark
and charged lepton masses that are in gross contradiction with the data. Supersymmetric
threshold corrections at the scale $m_{susy}$ may cure this problem.

\paragraph{Supersymmetric threshold corrections to light fermion masses\\}

In the following, we will neglect supersymmetric threshold corrections to the leptonic
Yukawa couplings, and consider only the corrections to the down-type quark
Yukawa couplings\footnote{Supersymmetric threshold corrections to up-type quark
masses remain under control, as no large A-terms are needed to correct the SU(5)
prediction. As for the top quark mass, even the large stop mixing that may be needed 
to reproduce the measured Higgs boson mass does not induce sizable threshold corrections.},
whose dominant contributions are proportional to $\alpha_3$ and $\lambda^2_t$ 
(see however Ref.~\cite{Antusch:2008tf}). 
This will allow us to derive the SU(5) Yukawa couplings $\Lambda^{\bar 5}$ by simply running
the charged lepton couplings up to the GUT scale. 
The leading supersymmetric threshold corrections to down-type quark masses are given
by the gluino and higgsino contributions
\cite{Hall:1985dx,Hall:1993gn,Hempfling:1993kv,Enkhbat:2009jt,Iskrzynski:2014zla,Anandakrishnan:2014qxa} 
(the latter can safely be neglected for the first two generations):
\begin{align}
  \frac{\Delta m_{d_i}}{m_{d_i}} & \, = \, - \, \frac{2 \alpha_3}{3 \pi}\, m_{\tilde g} \, X_{d_i} \,
    I_3 (m^2_{\tilde g}, m_{\tilde d_{L_i}}^2, m_{\tilde d_{R_i}}^2) \,
    + \frac{\lambda_t^2}{16\pi^2} \, \mu \, X_t \, \tan \beta \, I_3 (|\mu|^2,m_{\tilde{t}_1}^2,m_{\tilde{t}_2}^2)\, \delta_{i3}\, ,
\label{eq:m_di_thresholds}
\end{align}
where
\begin{align}
X_t & \, \equiv\, A_t / \lambda_t - \mu \cot \beta\, , \\
X_{d_i} & \, \equiv\, A_{d_i} / \lambda_{d_i} - \mu \tan \beta\, , \\
X_{e_i} & \, \equiv\, A_{e_i} / \lambda_{e_i} - \mu \tan \beta\, ,
\end{align}
and the loop function $I_3$ is defined by:
\begin{align}
  I_3(x,y,z) & \, = \, -\frac{xy\ln{(x/y)}+yz\ln{(y/z)}+zx\ln{(z/x)}}{(x-y)(y-z)(z-x)}\ ,
\label{eq:I3}
\end{align}
with the limits
\begin{align}
 I_3(x,x,z) & \, = \, \frac{1-(z/x)+(z/x)\ln{(z/x)}}{x(1-(z/x))^2}\ ,  \\
 I_3(x,x,x) & \, = \, \frac{1}{2x}\ .
\end{align}
The matching is done at the scale $m_{susy}$:
\begin{align}
  m^{\rm SM}_{d_i} & \, = \, m^{\rm MSSM}_{d_i} \left( 1 + \frac{\Delta m_{d_i}}{m_{d_i}} \right) ,
\label{eq:Susy_thresholds_Md}
\end{align}
where
\beq
  m^{\rm SM}_{d_i}\, =\, h_{d_i} (m_{susy}) \, v\, ,  \qquad
  m^{\rm MSSM}_{d_i}\, =\, \lambda_{d_i} (m_{susy}) \, v \cos \beta\, ,
\eeq
in which $v = \langle h^0 \rangle$.
As a first approximation, $b\, $--$\, \tau$ Yukawa unification is a relatively successful
prediction of SU(5), while the discrepancy between the prediction and the data
is much more important for the first two generations. 
As can be seen from Eq.~(\ref{eq:m_di_thresholds}), the non-holomorphic
($\propto\mu\tan{\beta}$) contributions to $\Delta m_d / m_d$ and $\Delta m_s / m_s$
are the same for equal first two generation squark masses, while the ratios $m_s/m_\mu$
and $m_d/m_e$ are widely different. This implies that large A-terms $A_d$ and $A_s$
are needed to bring these ratios into agreement with experimental data\footnote{Incidentally,
it turns out that the correct $m_b / m_\tau$ ratio cannot be obtained
from the corrections proportional to $\mu\tan{\beta}$ and to $X_t$ alone,
and that a large $A_b$ is also needed.}, which in turn
makes the electroweak vacuum metastable~\cite{Enkhbat:2009jt,Iskrzynski:2014zla}.
This issue will be discussed in Section~\ref{sec:vacuum}.

\paragraph{High-scale thresholds corrections to $\lambda_b$ and $\lambda_\tau$\\}

In addition to supersymmetric corrections at the superpartner mass scale,
Yukawa couplings are also subject to high-scale threshold corrections due
to the heavy GUT states. These may affect in particular bottom-tau Yukawa
unification, which as explained before is an important constraint on the model,
hence one must take them into account.
In practice, all one needs is the difference between $\lambda_b (M_{GUT})$ and
$\lambda_\tau (M_{GUT})$ induced by the GUT-scale threshold corrections.
One can check that it is given by:
\begin{align}
  \lambda_b (M_{GUT}) - \lambda_{\tau} (M_{GUT}) & \ = \
    \frac{\lambda_\tau (M_{GUT})}{(4\pi)^2} \left[ \lambda^2_t (M_{GUT}) \ln{\frac{M_{GUT}}{m_T}}
    - 4 \, g_{\scriptscriptstyle GUT}^2 \, \ln{\frac{M_{GUT}}{m_V}} \right]\, .
\label{eq:GUT_thresholds_Yuk}
\end{align}
GUT threshold corrections also affect strange quark-mu and down quark-electron
Yukawa unification, but the numerical effect is negligible compared with
the size of the low-scale (supersymmetric) threshold corrections that are needed
to account for the observed masses.

\subsection{Higgs quartic coupling}

For heavy stop masses, the proper way to compute the lightest
supersymmetric Higgs boson mass (for the standard computation, see
Refs.~\cite{Ellis:1990nz,Ellis:1991zd}) is to consider the effective theory
below the scale $m_{susy}$, in which all superpartners and heavy Higgs
bosons have been integrated out and the Higgs boson mass is determined
from the Higgs quartic coupling $\lambda$, with the value of $\lambda (m_{susy})$
determined by the supersymmetric theory valid above $m_{susy}$.
At tree level, the matching condition is
$\lambda = \left( 3 g_1^2 / 5 + g_2^2 \right) \cos^2 2\beta / 4$,
while at the 1-loop level it is given by:
\beq
  \lambda(m_{susy})\ =\ \left(\frac{3}{5} g_1^2(m_{susy})+g_2^2(m_{susy})\right)\frac{\cos^2{(2\beta)}}{4}\
    +\, \Delta\lambda^{reg}\, +\, \Delta\lambda^{\phi}\, +\, \Delta\lambda^{\chi}\, ,
\label{eq:lambda_matching}
\eeq
where the first term on the right-hand side of Eq.~(\ref{eq:lambda_matching}) is the tree-level
contribution, $\Delta\lambda^{reg}$ accounts for the conversion of the gauge couplings
from the $\overline{\textrm{DR}}$ scheme to the $\overline{\textrm{MS}}$ scheme,
and $\Delta\lambda^{\phi}$ and $\Delta\lambda^\chi$ are the one-loop threshold corrections
due to scalars and electroweak gauginos/higgsinos, respectively, whose expressions can
be found in Ref.~\cite{Bagnaschi:2014rsa}. The dominant contributions are the ones
proportional to $(m_t/v)^4$ in $\Delta\lambda^{\phi}$, which in the case where
all sparticle masses lie close to $m_{susy}$ (and in particular
$m_{\tilde Q_3} \simeq m_{\tilde u_3} \simeq m_{susy}$)
reduce to the leading stop mixing term:
\beq
  \Delta\lambda^{\phi}\, \simeq\, \frac{6h_t^4}{(4 \pi)^2}\frac{X_t^2}{m_{\tilde t_1} m_{\tilde t_2}}
    \left[1-\frac{1}{12}\frac{ X_t^2}{m_{\tilde t_1} m_{\tilde t_2}}\right]\, .
\label{eq:Xt}
\eeq
For large values of $\tan \beta$ and/or large values of $X_{b,\tau}$,
one should also include on the RHS of Eq.~(\ref{eq:Xt})
the leading sbottom and stau contributions, which in the limit
$m_{\tilde Q_3} \simeq m_{\tilde d_3} \simeq m_{\tilde L_3} \simeq m_{\tilde e_3} \simeq m_{susy}$
read:
\beq
\label{eq:Xb}
  \frac{6 h_b^4}{(4 \pi)^2} \frac{X_b^2}{m_{\tilde b_1} m_{\tilde b_2}}
    \left [ 1-\frac{1}{12} \frac{X_b^2}{m_{\tilde b_1} m_{\tilde b_2}} \right ]\,
    +\, \frac{2 h_{\tau}^4}{(4 \pi)^2} \frac{X_{\tau}^2}{m_{\tilde \tau_1} m_{\tilde \tau_2}}
    \left [ 1-\frac{1}{12} \frac{X_{\tau}^2}{m_{\tilde \tau_1} m_{\tilde \tau_2}} \right ] \, .
\eeq

Considering only the leading term~(\ref{eq:Xt}), one can see that for each value of
$m_{susy}$ and $\tan \beta$ there exist either $4$, $2$ or $0$ different values of $X_t$
satisfying the matching condition~(\ref{eq:lambda_matching}).
There also exists a lower bound on $m_{susy}$ for each value of 
$\tan\beta$, reached when $X_t $ becomes
\beq
  X_t^{max}=\pm \sqrt{6} \, m_{susy}\, ,
\label{eq:Xtmax}
\eeq
where the stop mixing contribution reaches its maximum~\cite{Delgado:2013gza}
(when higher order corrections are included the Higgs mass also depends
on the sign of $X_t$).
\begin{figure}[h!]
\centering
\includegraphics[width=0.7\textwidth]{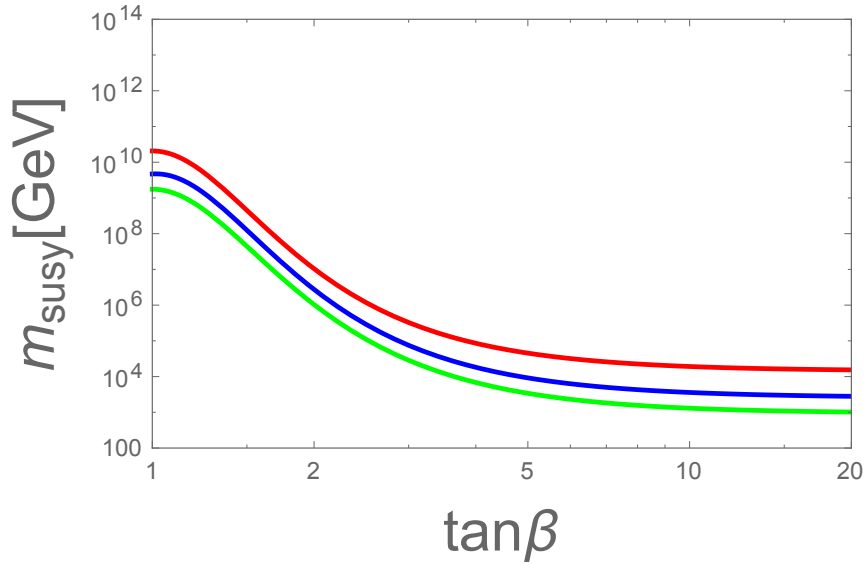} 
\caption{Higgs mass constraint in the ($\tan \beta$, $m_{susy}$) plane for different
matching conditions, with all SM parameters set to their central values. The red
curve corresponds to the tree-level matching condition and the blue curve
to the full 1-loop matching condition~(\ref{eq:lambda_matching}), while
for the green curve only the leading 1-loop threshold correction~(\ref{eq:Xt})
was used. Both the blue and green curves assume maximal positive stop mixing
($X_t=\sqrt{6} \, m_{susy}$). The superpartner spectrum is chosen
as in Fig.~\ref{fig:Parameter space}.
}
\label{fig:Higgs}
\end{figure}
This is illustrated in Fig.~\ref{fig:Higgs}, in which the lower bound on $m_{susy}$
is represented by the green curve, assuming a superpartner spectrum as
in Fig.~\ref{fig:Parameter space}. The comparison with the red curve shows
the importance of the 1-loop threshold corrections to the Higgs quartic coupling.
Note that the SM parameters were set to their central values in this figure;
taking into account the uncertainty on the top quark mass would spread the curves
into bands which become very broad at small $\tan\beta$ values. 
For a discussion on this point, see Ref.~\cite{Delgado:2013gza}.

\subsection{A-terms}

As discussed in the introduction and in Subsection~\ref{subsec:Yukawa},
we assume the A-term matrices to be diagonal\footnote{Due to the CKM mixing
in the quark sector, this assumption is not renormalization group-invariant. However,
the off-diagonal entries generated by the running from the GUT scale to low energy
are suppressed by the small CKM angles (the same statement holds for
the soft sfermion mass matrices). We shall therefore neglect this effect in the following.}
in the corresponding fermion mass eigenstate basis, with generation-dependent entries:
\begin{align}
A_U\, & =\, \text{Diag}\,(A_u,A_c,A_t)\, ,  \\
A_D\, & =\, \text{Diag}\,(A_d,A_s,A_b)\, ,  \\
A_E\, & =\, \text{Diag}\,(A_e,A_{\mu},A_{\tau})\, . 
\end{align}
These matrices are run with the 1-loop MSSM RGEs, with 
the SU(5) boundary conditions imposed at the GUT scale:
\begin{align}
  A_U(M_{GUT})\, &=\, A_U^T(M_{GUT})\, ,  \\
  A_D(M_{GUT})\, &=\, A_E^T(M_{GUT})\, .
  \label{eq:Ad_Ae}
\end{align}
%

\subsection{Gaugino masses}

For the running of gaugino masses we use the 1-loop RGEs:
\begin{align}
  \frac{d}{dt}M_i\, & =\, \frac{b_i}{2 \pi} \alpha_i \, M_i\, ,
\end{align}
and impose the SU(5) boundary condition at the GUT scale:
\begin{align}  
  M_1(M_{GUT}) \, = \, M_2(M_{GUT}) \, = \, M_3(M_{GUT}) & \, \equiv \, M_{1/2}\, .
\label{eq:gaugino_unification}
\end{align}
%

\subsection{Soft scalar masses}

We also assume the soft sfermion mass matrices $m^2_{\tilde X}$ $(X = Q, u^c, e^c, L, d^c)$
to be diagonal with generation-dependent entries:
\begin{align}
  m^2_{\tilde X}\, & =\, \text{Diag}\,(m_{\tilde X_1}^2,m_{\tilde X_2}^2,m_{\tilde X_3}^2)\, ,
\end{align}
however with $m_{\tilde X_1}^2 = m_{\tilde X_2}^2$ imposed at the GUT scale.
Furthermore, SU(5) invariance requires the following relations to hold at $M_{GUT}$ ($i = 1,2,3$):
\begin{align}
  m_{\tilde Q_i}(M_{GUT})\, =\, m_{\tilde u^c_i}(M_{GUT})\, =\, m_{\tilde e^c_i}(M_{GUT})\,
    & \equiv\, m_{10_i}\, ,  \label{eq:GUT_10} \\
  m_{\tilde L_i}(M_{GUT})\, =\, m_{\tilde d^c_i}(M_{GUT})\, & \equiv\, m_{\bar 5_i}\, . \label{eq:GUT_5}
\end{align}
Hence the splitting of the soft sfermion masses within SU(5) representations and between
the first two generations is only due to the running, performed at the 1-loop level
as for the other soft terms.
As for the soft Higgs masses, we allow the possibility of different boundary conditions
for the two Higgs doublets of the MSSM, namely
$m^2_{H_u} (M_{GUT}) \neq	 m^2_{ H_d} (M_{GUT})$.

Note that our boundary conditions are less restrictive than the so-called minimal supergravity
ansatz (mSUGRA), which assumes universal scalar and gaugino masses as well as A-terms
proportional to the Yukawa couplings. By contrast we allow for some generation dependence
in the sfermion soft terms, but we require them to be aligned with fermion masses
(exactly at the GUT scale, approximately at low energy due to the running)
in order to minimize supersymmetric contributions to flavour-violating processes.

\subsection{$\mu$ and $B \mu$ terms}

Contrary to the other MSSM parameters, the $\mu$ and $B\mu$ terms are not fixed
at the GUT scale and renormalized down to low energy, but rather
determined from the tree-level electroweak symmetry breaking conditions
at the scale $m_{susy}$, as is usually done (by inserting the running soft terms in the
tree-level Higgs potential one ensures that the most relevant 1-loop radiative
corrections are taken into account). This procedure yields:
\begin{align} \label{eq:EWSB}
|\mu|^2 & \ = \ \frac{m_{H_d}^2-m_{H_u}^2\tan^2{\beta}}{\tan^2{\beta}-1} - \frac{m_Z^2}{2}\, ,  \\
B & \ = \ (m^2_{H_u} + m^2_{H_d}+ 2 |\mu|^2) \, \frac{\sin{2 \beta}}{2}  \nonumber \\
& \ = \ \frac{(m_{H_d}^2-m_{H_u}^2)\tan{\beta}}{\tan^2{\beta}-1} - \frac{m_Z^2  \tan\beta}{\tan^2\beta+1}\ ,
\end{align}
while the sign of $\mu$ remains undetermined.

\section{Solutions of the RGEs for soft terms}
\label{sec:solutions}

Here we present (approximate) semi-analytic solutions to the 1-loop renormalization
group equations for the MSSM soft terms. More details about the procedure used
to derive them can be found in Appendix~\ref{app:RGEs}.

\subsection{Gauge and Yukawa couplings}
\label{subsec:gauge_Yukawa}

As a first step, for each choice of $m_{susy}$ and $\tan{\beta}$, one runs the gauge
couplings $g_i(t)$ ($i=1,2,3$) and the Yukawa couplings  $\lambda_n(t)$
($n=t,b,\tau,c,s,\mu,u,d,e$) from $t_Z \equiv \ln m_Z$ to $t_S \equiv \ln{m_{susy}}$
with the 2-loop SM RGEs. Then, after applying the matching
conditions~(\ref{eq:g1_conversion})--(\ref{eq:g3_conversion}),~(\ref{eq:Yu_conversion})
and~(\ref{eq:Ye_conversion}) at the scale $m_{susy}$, the gauge couplings
and the up-type quark and charged lepton Yukawa couplings are further evolved
up to the GUT scale with the 2-loop MSSM RGEs. 
At $M_{GUT}$ the SU(5) boundary conditions~(\ref{eq:Md_Me}) and
the heavy threshold corrections~(\ref{eq:GUT_thresholds_Yuk}) are imposed,
then the down-type quark Yukawa couplings are run back from $t_0 \equiv \ln{M_{GUT}}$
to $t_S \equiv \ln{m_{susy}}$ (in this procedure one needs to fix the heavy GUT state
masses $m_V$ and $m_T$, in addition to the GUT scale itself).
This provides us with numerical solutions for the running
gauge couplings $g_i(t)$ ($i=1,2,3$) and Yukawa couplings  $\lambda_n(t)$
($n=t,b,\tau,c,s,\mu,u,d,e$) in the range $t_S \leq t \leq t_0$, accurate at the 2-loop level.

\subsection{Gaugino masses}

The solutions to the 1-loop RGEs for gaugino masses read:
\beq
M_i (m)\, =\, M_{1/2} \, e^{-\int_{\ln{m}}^{\ln{M_{GUT}}} dt \, \alpha_i(t) \, b_i^{MSSM} / (2\pi)}\, .
\eeq
These 3 equations connect linearly 4 variables: $M_{1,2,3}(m_{susy})$ and $M_{1/2}$.

\subsection{A-terms}

As explained in Appendix~\ref{app:RGEs}, the hierarchy of Yukawa couplings makes it
possible to solve the 1-loop RGEs for A-terms in a sequential way, from $A_t$ to $A_e$.
With the convention that the index $n$ runs over the ordered values $\{ t, b, \tau, c, s, \mu, u, d, e \}$
(such that e.g. $p < c$ means $p = t, b$ or $\tau$), one can write the (approximate) solutions as:
\beq
  A_n(t)\ =\ A_n(t_0)\, e^{\, \int_{t_0}^t dt' \beta^A_n(t')}\, +\ e^{\, \int_{t_0}^t dt' \beta^A_n(t')}
    \int_{t_0}^t dt'\, \left ( \gamma^A_n(t')\, e^{- \int_{t_0}^{t'} dt'' \beta^A_n(t'')} \right )\, ,
\label{eq:Aterms_solution}
\eeq
where $\beta^A_n$ is a function of the gauge and Yukawa couplings,
and $\gamma^A_n$ is the product of $\lambda_n$ times a linear combination
of $g_i^2 M_i$ and $\lambda_m A_m$ with $m<n$, all of which are
already known quantities, since the RGEs for the $A_m$'s are solved in order
of increasing $m$.
One can therefore rewrite Eq.~(\ref{eq:Aterms_solution}) in the form:
\beq
A_n(t)\, =\, \sum_{m=1}^n a_{nm}(t)A_m(t_0) + b_n(t)M_{1/2}\, ,
\label{eq:An}
\eeq
in which the coefficients $a_{nm}(t)$ and $b_n(t)$ are integrals that can be evaluated
numerically after having solved the MSSM RGEs for the gauge and Yukawa couplings.
Imposing the SU(5) boundary conditions $A_{d_i}(t_0) = A_{e_i}(t_0)$,
one obtains the running A-terms at an arbitrary scale as linear combinations of
the SU(5) soft parameters $M_{1/2}$, $A^{10}_i$, $A^{\bar 5}_i$ ($i=1,2,3$), with
numerical coefficients depending on the choice of $m_{susy}$ and $\tan\beta$
(and very mildly on $M_{GUT}$, $m_T$ and $m_V$).

In practice it may be more convenient to express the $A_n(t)$ in terms of other input parameters,
for example $M_3(m_{susy})$ and the $A_{d_i}(m_{susy})$ ($i=1,2,3$), since these quantities
enter the supersymmetric threshold corrections to the down-type quark masses needed
to fit the experimental values. One may also want to trade $A^{10}_3$ for $A_t(m_{susy})$,
which is directly constrained by the measured value of the Higgs mass. 
To this end, it suffices to invert the linear relations~(\ref{eq:An}) so that the $A_n(t)$
can be rewritten as a function of the desired input parameters.
In the following, we shall choose:
\beq
  M_3(m_{susy}), \ A_{d_i}(m_{susy}), \ A_{u_i}(m_{susy}) \qquad (i=1,2,3)\, .
\eeq

\subsection{Soft scalar masses}

Following Appendix~\ref{app:RGEs}, we first introduce the combinations of masses
(which appear in the RGEs for soft scalar masses):
\beq
  S\ =\ m^2_{H_u}-m^2_{H_d} + \sum_{i=1}^{3} \left[ m_{\tilde Q_i}^2 - m_{\tilde L_i}^2
    - 2 m_{\tilde u_i^c}^2 + m_{\tilde d_i^c}^2 + m_{\tilde e_i^c}^2 \right] ,
\eeq
and
\begin{align}
  \Sigma_{u_i} & \ \equiv \ m^2_{\tilde Q_i} + m^2_{\tilde u^c_i} + m^2_{H_u}\, ,  \\
  \Sigma_{d_i} & \ \equiv \ m^2_{\tilde Q_i} + m^2_{\tilde d^c_i} + m^2_{H_d}\, ,  \\
  \Sigma_{e_i} & \ \equiv \ m^2_{\tilde L_i} + m^2_{\tilde e^c_i} + m^2_{H_d}\, .
\end{align}
The 1-loop RGE for $S$ is easily integrated to give:
\beq
  S(t)\, =\, S(t_0) \, e^{\int_{t_0}^{t} dt'  \, \alpha_1(t') \, b_1^{MSSM} / (2\pi)}\, ,
\eeq
where $S(t_0) = m^2_{H_u}(t_0) - m^2_{H_d}(t_0)$ due to the SU(5)
boundary conditions on soft scalar masses,
while the RGEs for the $\Sigma_n$'s can be solved sequentially in a similar way
to the A-term RGEs, taking advantage of the hierarchy of Yukawa couplings to
neglect subdominant terms. 
We then define:
\begin{align}
I_{A_n} (t) & \, \equiv \, \frac{1}{(4\pi)^2} \int_{t_0}^{t} dt' \, A_n^2(t')\, ,  \\
I_{M_i} (t) & \, \equiv \, \frac{1}{(4\pi)^2} \int_{t_0}^{t} dt' \, g_i^2 (t') M_i^2(t')\, ,  \\
I_{S} (t)\, & \equiv\, \frac{1}{(4\pi)^2} \int_{t_0}^{t} dt' \, g_1^2 (t') S(t')\, .
\end{align}

With these ingredients one can express the solutions to the 1-loop RGEs for soft scalar masses as:
\bea
  m_\alpha^2(t) = m_\alpha^2(t_0) + \sum_n T^\alpha_n \left(\Sigma_n(t)-\Sigma_n(t_0)\right)
    +\sum_n U^\alpha_n I_{A_n}(t) + \sum_i V^\alpha_i I_{M_i}(t) + R^\alpha I_S(t)\, ,  \hskip .5cm
\label{eq:m2}
\eea
where the index $\alpha$ runs over
$\{H_u,H_d,\tilde Q_{3,2,1},\tilde u^c_{3,2,1},\tilde e^c_{3,2,1}, \tilde L_{3,2,1},\tilde d^c_{3,2,1}\}$,
and the numerical coefficients $T^\alpha_{n}$, $U^\alpha_{n}$ and $V^\alpha_{i}$ (resp. $R^\alpha$)
are given in Table~\ref{Table:softmasses2} (resp. Table~\ref{Table:softmasses1})
of Appendix \ref{app:RGEs}.
Plugging the previously derived semi-analytic expressions for $A_n(t)$, Eq.~(\ref{eq:An}),
and for $\Sigma_n(t)$ into Eq.~(\ref{eq:m2}), one then arrives at
\begin{align}
  m_\alpha^2(t)\, =\, \sum_\beta c_{\alpha\beta}(t)m_\beta^2(t_0)
    & + \sum_{n,m}d_{\alpha nm}(t)A_n(t_0)A_m(t_0)  \nonumber \\
  & + \sum_{n}e_{\alpha n}(t)A_n(t_0)M_{1/2} + f_{\alpha}(t)M_{1/2}^2\, ,
\label{eq:m2_alpha}
\end{align}
in which the coefficients $c_{\alpha\beta}(t)$, $d_{\alpha nm}(t)$, $e_{\alpha n}(t)$
and $f_{\alpha}(t)$ are integrals that can be evaluated numerically just from the knowledge
of the solutions to the 2-loop MSSM RGEs for gauge and Yukawa couplings.
Finally, imposing the SU(5) boundary conditions on soft terms,
one obtains the running soft scalar masses $m^2_\alpha(t)$ at an arbitrary scale
as quadratic functions of the SU(5) soft parameters $m_{10_i}$, $m_{\bar 5_i}$,
$m_{H_u}(M_{GUT})$, $m_{H_d}(M_{GUT})$, $M_{1/2}$, $A^{10}_i$ and $A^{\bar 5}_i$
($i=1,2,3$), with numerical coefficients depending on the choice of $m_{susy}$
and $\tan\beta$ (and very mildly on $M_{GUT}$, $m_T$ and $m_V$).

As was done for A-terms, one can easily rewrite the running soft scalar masses
in terms of other input variables by inverting the system of equations~(\ref{eq:An})
and/or~(\ref{eq:m2_alpha}). As explained before, a convenient choice for exploring
the parameter space of the minimal renormalizable supersymmetric SU(5) model
is to trade the 7 SU(5) parameters $M_{1/2}$, $A^{10}_i$ and $A^{\bar 5}_i$
for $M_3(m_{susy})$, $A_{d_i}(m_{susy})$ and $A_{u_i}(m_{susy})$.
One may also invert 8 of the 17 equations~(\ref{eq:m2_alpha}) in order to replace
the GUT-scale masses $m_{10_i}$, $m_{\bar 5_i}$, $m_{H_u}(M_{GUT})$
and $m_{H_d}(M_{GUT})$ by 8 low-energy soft scalar masses, so that
(for given values of $m_{susy}$ and $\tan\beta$)
the whole supersymmetric spectrum is parametrized by 15 low-energy input variables.

\section{Proton decay}
\label{sec:pdk}

Proton decay is one of the main prediction of Grand Unified Theories, and since
it has not been observed yet, it sets strong constraints on the parameter space
of the minimal renormalizable supersymmetric SU(5) model. Qualitatively,
the proton lifetime behaves as:
\beq
  \tau_p\, \simeq\, \tau (p \to K^+ \bar \nu)\, \propto\, m_T^2\, m_{susy}^2
    \tan^2\beta / (1+\tan^2\beta)^2\, ,
\eeq
implying a $\tan \beta$-dependent lower bound on the superpartner mass scale.

To compute precisely the proton lifetime we will need the following input parameters:
\bea
  m_p=0.9383 \GeV\, , & & m_{K^+}=0.4937 \GeV\, ,  \\
  \alpha_3 (2\GeV)=0.31\, , & & \alpha_3 (m_b)=0.22\, , \\
  W_0^{112} (2 \GeV)=(0.111 \pm 0.027) \GeV^2\, , & &
    W_0^{121} (2 \GeV)=(0.036 \pm 0.014) \GeV^2\, , \hskip .8cm
\eea
where $W_0^{112}$ and $W_0^{121}$ appear in the hadronic matrix elements
for proton decay, as well as the entries of the CKM matrix:
\begin{align}
V_{CKM} (m_Z) & \ = \ \left[ \begin{array}{ccc}
1-\lambda^2/2 & \lambda & A \lambda^3 (\rho - {i\mkern1mu} \eta) \\
-\lambda & 1-\lambda^2/2 & A \lambda^2 \\
A \lambda^3 (1-\rho-{i\mkern1mu} \eta) & -A \lambda^2 & 1
 \end{array} \right] ,
\end{align}
here written in the Wolfenstein's parametrization, with~\cite{Agashe:2014kda}:
\beq
  \lambda = 0.22537\, , \ \ A = 0.814\, , \ \ \rho = 0.117\, , \ \ \eta = 0.353\, .
\eeq
Details about the computation of the proton lifetime can be found in Appendix~\ref{app:pdk}. 
The predicted proton lifetime is compared with the experimental constraint
$\tau (p \rightarrow K^+ \bar \nu) > 2.3 \times 10^{33}$~yrs ($90 \%$~C.L.)~\cite{p_K+nubar}.


\section{Vacuum (meta)stability}
\label{sec:vacuum}

In the general MSSM, some regions of the parameter space lead to instabilities
of the electroweak vacuum. One may encounter two kinds of dangerous situations.

The first one is the possible existence 
of directions in field space along which the potential is unbounded from below (UFB). To remain on 
the safe side, we will allow only the points in parameter space that do not possess any such 
direction. The associated constraints on the  model parameters will be summarized
in Subsection~\ref{avoidUFB}.

The second one is the metastability of the electroweak vacuum, 
which due to the large trilinear soft terms 
needed to correct the SU(5) predictions for fermion masses cannot be avoided.
This means that there are minima in field space, lower than the electroweak vacuum,
into which it will eventually decay. These new 
minima are the so-called charge and colour breaking (CCB) vacua. Although one cannot
forbid the decay of the electroweak vacuum into a lower minimum,
one can check whether its lifetime is long enough on cosmological time scales.
The procedure followed in our analysis is summarized in Subsection~\ref{checkCCB}.

All computations in this section are done at the tree level. The generalization to higher orders
is conceptually straightforward, although technically much more involved, so we will leave it
for future work. Below we summarize the discussion presented in Appendices~\ref{app:UFB}
and~\ref{app:CCB}, where  technical details and relevant references can be found.

\subsection{\label{avoidUFB}Unbounded from below directions}

As explained in Appendix~\ref{app:UFB}, the tree-level constraints associated with the
absence of UFB directions can be written as (neglecting $m_Z$):
\begin{align}
m_{H_u}^2 +m_{\tilde L_i}^2& \, > \, 0\, ,
\end{align}
which must be satisfied for any of the three slepton generations.

\subsection{\label{checkCCB}Charge and colour breaking vacua}

We shall discuss the constraint applying to $A_t$ separately from the ones applying
to all other A-terms. The reason to treat them differently is that in the second case
the D-terms can be considered to be vanishing to a good approximation, thus providing
constraints on the fields, while in the first case this assumption is not justified due to the
large top Yukawa coupling\footnote{Strictly speaking, the same comment applies
to $A_b$ in the large $\tan \beta$ regime. In the minimal renormalizable supersymmetric
SU(5) model, however, large values of $\tan \beta$ are excluded by a combination
of constraints (see Section~\ref{sec:results}), so only the case of a large $A_t$ needs
to be discussed separately.}. We will nevertheless set the colour D-terms to zero
for simplicity, but allow for non-vanishing hypercharge and SU(2) D-terms.

\subsubsection{Constraints on $A_n$ ($n \neq t$)}

Let us first define:
\begin{align}
M_2^2(\psi,\bar\psi,H,\bar H)\, &\equiv\, M_\psi^2\psi^2+M_{\bar\psi}^2\bar\psi^2
+\left(m_{H}^2+|\mu|^2\right)H^2+\left(m_{\bar H}^2+|\mu|^2\right)\bar H^2-2BH \bar H ,  \\
M_3(\psi,H,\bar H)\, &\equiv\, -2\psi^2(A_\psi H-\lambda_\psi \mu \bar H)\, ,  \\
z(\psi,\bar\psi,H,\bar H, c_z)\, &\equiv\, 2\left(2\psi^2+ c_z\bar\psi^2+H^2+\bar H^2\right) ,
\end {align}
and write the SU(2) D-term constraint:
\begin{align}
\label{Dterm1}
0\, &=\, -\psi^2+(3-2c)\bar \psi^2+ H^2-\bar H^2 ,
\end{align}
where $\psi$, $\bar \psi$ are sfermion fields and $H$, $\bar H$ Higgs fields that parametrize
a specific direction in field space along which a CCB minimum is present. 
The different possibilities for the constant fields $\psi$, $\bar\psi$, $H$, $\bar H$, their mass 
parameters and the coefficients $c$ and $c_z$ are summarized in Table~\ref{cases}
of Appendix~\ref{app:CCB}.

To evaluate the lifetime of the electroweak vacuum, we consider the (normalized) bounce action, 
which for $A_n \neq A_t$ can be approximated by:
\beq
\label{Sapprox1}
{\cal S}\, =\, \frac{z^2M_2^2}{M_3^2}(\psi,\bar\psi,H,\bar H)\, .
\eeq
This action is then minimized by varying the direction in field space $(\psi, \bar\psi, H, \bar H)$, 
subject to the constraint~(\ref{Dterm1}). In order for the lifetime of the electroweak
vacuum to be larger than the age of the universe, the minimum of ${\cal S}$ must satisfy
\beq
{\cal S}^{min}\, \gsim\, 9\, .
\label{eq:Smin}
\eeq
Notice that one needs to minimize $\cal S$ with respect to two variables only. Indeed, one of the variables
$\psi,\bar\psi,H,\bar H$ is fixed by Eq.~(\ref{Dterm1}), and the quantity~(\ref{Sapprox1})
only depends on ratios of fields. 
The minimization is performed numerically. A point of the parameter space is admitted
if the bounce action satisfies Eq.~(\ref{eq:Smin}).

\subsubsection{Constraint on $A_t$}

We now define:
\bea
M_2^2&\equiv&\left(m_{\tilde Q}^2+ m_{\tilde t}^2\right)t^2
+\left(m_{H_u}^2+|\mu|^2\right)H_u^2+\left(m_{H_d}^2+|\mu|^2\right)H_d^2-2BH_u H_d\, ,  \\
M_3&\equiv&-2t^2(A_tH_u-\lambda_t\mu H_d)\, ,  \\
z&\equiv&2\left(2t^2+H_u^2+H_d^2\right) ,  \\
\lambda&\equiv&\lambda_t^2t^2(t^2+2H_u^2)+\frac{g'^2 + g_2^2}{8}\left(H_u^2-H_d^2-t^2\right)^2 .
\eea
Due to the large value of the top quark Yukawa coupling, the bounce action
can no longer be approximated by Eq.~(\ref{Sapprox1}). One should instead minimize
\beq
{\cal S}\, =\, \left(1+\frac{f(\kappa)-f(0)}{\hat S(0)}\right)\frac{z^2M_2^2}{M_3^2}\, ,
\eeq
with
\beq
f(\kappa)\, =\, \frac{\pi^2/6}{(1-4\kappa)^3}+\frac{16.5}{(1-4\kappa)^2}+\frac{28}{1-4\kappa}\ , \
\quad \kappa = \lambda\frac{M_2^2}{M_3^2}\ ,\ \quad \hat S(0) = 45.4\ .
\eeq
The minimization goes again over two variables (for example $H_u/t$ and $H_d/t$),
and is done numerically.


\section{Results and discussion}
\label{sec:results}

We are now ready to address the question we asked in the introduction,
namely whether the minimal renormalizable supersymmetric SU(5) model
can be considered as a viable extension of the Standard Model.
To answer this question, we must scan over the parameter space of the model
and search for points passing all phenomenological and theoretical
constraints\footnote{As discussed in the introduction, dark matter and neutrino
masses can be accounted for by separate sectors (in particular when the lightest
neutralino is not a suitable dark matter candidate), so we do not include them
in the list of constraints to be imposed on the model.}
(precise gauge coupling unification, correct predictions
for the Higgs boson and charged fermion masses, proton lifetime, experimental
lower bounds on superpartner masses, metastability of the vacuum
and perturbativity of the model).
This is not a straightforward task, as the model involves a large number
of parameters with boundary conditions defined at the GUT scale, while
most constraints apply at low energy. In order to ease the whole analysis,
we will use the semi-analytic solutions to the soft term RGEs obtained
in Section~\ref{sec:solutions}. This will enable us to perform a much
more efficient scan -- even though it still involves a large number of parameters.

Before going into this programme, let us first try to identify the region of the
parameter space in which viable points are likely to be found. For a qualitative
discussion of how each phenomenological or theoretical requirement
constrains the model, we will consider only two parameters, $\tan \beta$
and the overall superpartner mass scale $m_{susy}$:
\begin{itemize}
\item
a powerful constraint on the parameter space comes from accommodating the 
measured Higgs mass, which provides a $\tan \beta$-dependent lower bound
on $m_{susy}$ (the lower the value of $\tan \beta$, the higher the value of $m_{susy}$).
Sizable mass splittings among superpartners can change this bound,
but not very drastically.
\item
another important constraint comes from the non-observation of proton decay,
which provides another lower bound on $m_{susy}$, with a different dependence on $\tan \beta$.
The actual value of the bound depends on the details of the superpartner spectrum.
\item
requiring the masses of the heavy GUT states derived from gauge coupling unification
to remain below the cut-off scale $M_{Planck}$ bounds $m_{susy}$ from above.
The main constraint comes from the Higgs colour triplet mediating proton decay,
whose mass $m_T$ strongly depends (through the gauge coupling unification condition)
on the higgsino mass $\mu$.
\item
the parameter space is also bounded by vacuum metastability constraints associated
with large values of the A-terms. These are unavoidably present due to the sizable threshold 
corrections needed in the down-type quark sector. Furthermore, a large $A_t$ is necessary
to accommodate the observed Higgs mass in some regions of the parameter space.
Although the required values of the A-terms do not necessarily threaten the metastability
of the electroweak vacuum, significant regions of the parameter space where the lifetime
of the universe would be too short are excluded.
\item
the requirement that the third generation Yukawa couplings should remain perturbative
up to the GUT scale excludes some portions of the parameter space in the small
$\tan\beta$ region (top quark) and potentially also in the large $\tan\beta$ region
(bottom quark and tau Yukawa couplings).
\item
the perturbativity of the parameters of the SU(5) superpotential, reflected in the condition 
$m_T, m_3, m_8 \lsim m_V$, can easily be satisfied by using the freedom
allowed by gauge coupling unification. Indeed, for fixed $M_{GUT}$ and superpartner
masses, $m_T$ and $m^2_V m_3$ are constant (using the fact that the colour octet
and weak triplet components of the adjoint Higgs field have equal masses, $m_3 = m_8$),
see Eqs.~(\ref{mT}) and~(\ref{mV}).
One can therefore increase $m_V$ and decrease $m_3 = m_8$
by diminishing $\lambda_5$ and $\mu_5$ in the superpotential~(\ref{eq:W_H}) in such a way
that $\sigma_0 = \mu_5 / \lambda_5$ increases. Simultaneously, $\eta_H$ should be
made smaller so that $m_T = 5 \eta_H \sigma_0$ stays constant.
\end{itemize}

One can illustrate how these constraints restrict the parameter space of the model
by plotting them in the ($\tan \beta$, $m_{susy}$) plane,
in the spirit of Fig.~\ref{fig:Parameter space} in the introduction.
To be able to do this, we assume a simplified superpartner spectrum
with a common scale $m_{susy}$ for all sfermion masses,
the $\mu$ parameter, $m_{H_u}$ and the SU(5) gaugino mass parameter $M_{1/2}$.
The down-type squark A-terms and $A_t$ are chosen to fit the fermion masses and
the Higgs mass, while the slepton A-terms are taken to be 2/3 of their down-type quark
counterparts\footnote{This empirical factor roughly mimics the effect of the running
from $M_{GUT}$, where the SU(5) relations $A_{d_i} = A_{ei}$ hold, to $m_{susy}$.},
and $A_u$ and $A_c$ are assumed to vanish.
With $m^2_{H_d}$ and the $B$ parameter determined by the electroweak
symmetry breaking conditions, the whole MSSM spectrum,
including the heavy Higgs masses, can be computed.
While these inputs are not fully consistent with SU(5)-invariant boundary conditions
at the GUT scale, they make it possible to display all the constraints discussed above
in a two-dimensional plot. The result can be seen in
Fig.~\ref{fig:parameter_space_with_point}, a simplified version of which was shown
in the introduction. Let us now comment it:
\begin{figure}[t]
\centering
\includegraphics[width=0.5\textwidth]{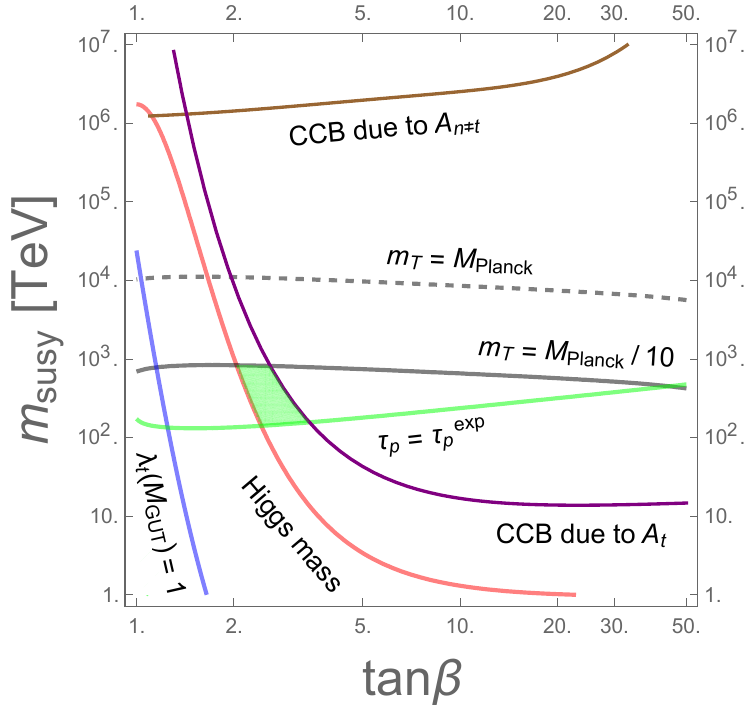}        
\caption{A naive estimate of the allowed parameter space of the minimal renormalizable
supersymmetric SU(5) model (\textit{pale green region}) in the approximation described
in the text, where all sfermions are assumed to have the same mass $m_{susy}$.}
\label{fig:parameter_space_with_point}
\end{figure}
\begin{itemize}
\item
the measured Higgs boson mass excludes the region below the \textit{red curve}
(which corresponds to maximal stop mixing);
\item
the region below the \textit{green line} is ruled out by the experimental lower bound
on the proton lifetime (assuming the high-energy phases appearing in the proton decay
amplitude, see Eq.~(\ref{eq:kappa_tilde}), all vanish). The shape of this line can
be understood by noting that the proton lifetime approximately scales as
\beq
  \tau (p \to K^+ \bar \nu)\, \propto\, m_T^2\, m_{susy}^2 \tan^2\beta / (1+\tan^2\beta)^2\, ,
\eeq
while given the assumptions made on the superpartner spectrum, the colour triplet mass
is proportional to\footnote{Indeed, given the choice $\mu^2 = m^2_{H_u} = m^2_{susy}$,
one has from the electroweak symmetry breaking conditions
$m^2_{H_d} = m^2_{susy} (2 \tan^2 \beta - 1)$ and $m^2_A = 2 m^2_{susy} (\tan^2 \beta + 1)$.
Plugging this, together with the assumption of equal sfermion masses, into Eq.~(\ref{mT}),
one arrives at $m_T \propto  m_{susy}^{5/6} (\tan^2 \beta + 1)^{1/12}$.}
\beq
  m_T\, \propto\,  m_{susy}^{5/6} \, (\tan^2 \beta + 1)^{1/12}\, ,
\eeq
giving 
\beq
  \tau (p \to K^+ \bar \nu)\, \propto\, m_{susy}^{11/3} \, \tan^2 \beta\, (\tan^2 \beta + 1)^{-11/6}\ ;
\eeq
\item
the region above the \textit{purple curve} is excluded by the vacuum metastability constraint
on $A_t$, while the region above the \textit{brown curve} (which is almost entirely due to $A_b$)
is ruled out by the constraints on all other $A$-terms.
In drawing these curves we used the metastability conditions derived in Appendix~\ref{app:CCB},
but they can be approximated to a very good degree by:
\beq
  |A_t|\, <\, \sqrt{m_{H_u}^2+m_{\tilde Q_3}^2+m_{\tilde u_3}^2}\ , \qquad
      |A_b|\, <\, \sqrt{m_{H_d}^2+m_{\tilde Q_3}^2+m_{\tilde d_3}^2}\ ,
\eeq
for $A_t$ and $A_b$, respectively. Note that the constraint on $|A_b|$ is more easily
satisfied for large $\tan \beta$ values due to $m^2_{H_d} = m^2_{susy} (2 \tan^2 \beta - 1)$;
\item
perturbativity of the top quark Yukawa coupling excludes the region to the left of
the \textit{blue curve} (which corresponds to $\lambda_t (M_{GUT}) = 1$).
There is no similar constraint for $\lambda_b$ and $\lambda_{\tau}$,
which are always in the perturbative regime in the region of the parameter space
shown in Fig.~\ref{fig:parameter_space_with_point};
\item
the constraint $m_T < M_{Planck}/10$ (resp. $m_T < M_{Planck}$) rules out $m_{susy}$
values above the \textit{solid black line} (resp. the \textit{dashed black line}), while there are
no additional constraints from $m_3, m_8, m_T < m_V < M_{Planck} / 10$;
\item
the \textit{pale green area} is the region of the parameter space that is allowed
by all the above constraints.
\end{itemize}

The tentative conclusion one can draw from Fig.~\ref{fig:parameter_space_with_point},
even though the assumptions made on the soft terms are not consistent with SU(5)-invariant
boundary conditions at the GUT scale, is that viable points satisfying all phenomenological
and theoretical constraints are likely to be found in the region bounded by
$2 \lsim \tan{\beta} \lsim 5$ and $100 \TeV \lsim m_{susy} \lsim 1000 \TeV$.
This definitely needs confirmation from a more careful investigation of the parameter space.
Namely, one should scan over the parameters of the model,
which besides $\tan \beta$, $\mbox{sign}(\mu)$ and the heavy
state masses include 15 soft terms:
8 sfermion ($m_{10_{1,2,3}}$, $m_{\bar 5_{1,2,3}}$) and Higgs ($m_5$, $m_{\bar 5}$)
soft masses, 1 gaugino mass parameter ($M_{1/2}$) and 6 A-terms
($A^{10}_{1,2,3}$, $A^{\bar 5}_{1,2,3}$). Since these soft parameters are defined
at the GUT scale, one needs to run them down to the scale $m_{susy}$,
where most constraints apply. In order to simplify the problem, we shall use
the semi-analytic approximate solutions to the soft term RGEs derived
in Section~\ref{sec:solutions}. This will allow us to trade the GUT-scale
soft parameters for low-energy ones (namely the $m_{susy}$ values of the
running parameters $M_3$, $A_{u,c,t}$, $A_{d,s,b}$ and of 8 suitably chosen
soft scalar masses) and to use the constraints at $m_{susy}$
to effectively reduce the number of free parameters.

Let us describe more precisely the procedure that we are going to employ:
\begin{enumerate}
\item
first  choose a random point in the $(\tan\beta, m_{susy})$ plane (where $m_{susy}$
is now the \mbox{matching} scale between the SM and the MSSM),
together with $\mbox{sign}(\mu)$ and some sensible values of $M_{GUT}$, $m_T$ and $m_V$;
\item
solve numerically the 2-loop MSSM RGEs for the gauge and Yukawa couplings
(taking into account the GUT-scale relations~(\ref{eq:Md_Me}) and the heavy
threshold corrections to Yukawa couplings~(\ref{eq:GUT_thresholds_Yuk}),
as explained in Subsection~\ref{subsec:gauge_Yukawa}). Then, following the procedure
of Section~\ref{sec:solutions}, express the running soft terms at an arbitrary scale
as algebraic combinations of the $m_{susy}$ values of $M_3$, $A_{u,c,t}$, $A_{d,s,b}$,
$m^2_{\tilde Q_{1,2,3}}$, $m^2_{\tilde d^c_{1,2,3}}$, $m^2_{\tilde u^c_3}$
and $|\mu|^2$, which through the EWSB condition~(\ref{eq:EWSB}) can be
written as a linear combination of $m^2_{H_u}(m_{susy})$ and $m^2_{H_d}(m_{susy})$;
\item
impose the following 8 constraints on the $m_{susy}$ values of the soft terms:
{\it (i)} the definition of the matching scale
$m_{susy} \equiv \sqrt{m_{\tilde Q_3}(m_{susy})m_{\tilde u^c_3}(m_{susy})}$;
{\it (ii)}-{\it (iii)} the equality of the first and second generation soft
sfermion masses, which for simplicity we impose at the scale $m_{susy}$
rather than $M_{GUT}$: $m^2_{\tilde Q_{1}} (m_{susy}) = m^2_{\tilde Q_{2}} (m_{susy})$
and $m^2_{\tilde d^c_{1}} (m_{susy}) = m^2_{\tilde d^c_{2}} (m_{susy})$;
{\it (iv)} the gauge coupling unification condition~(\ref{mT}),
in which $m_T$ and $M_{GUT}$ are fixed and $\mu$ and $m_A$ are functions
of $m^2_{H_u}$ and $m^2_{H_d}$ through the EWSB conditions;
{\it (v)} the 1-loop matching condition for the Higgs quartic coupling~(\ref{eq:lambda_matching}),
keeping as a first approximation only the leading term~(\ref{eq:Xt});
{\it (vi)}-{\it (viii)} the supersymmetric threshold corrections~(\ref{eq:m_di_thresholds})
needed for the down-type quark masses to match their measured values.
Whenever these constraints admit several solutions, we explore all of them.
Imposing them allows us to express 8 of the 15 input parameters at the scale $m_{susy}$
in terms of the remaining 7 ones and of already known quantities; in practice
we choose the remaining 7 free parameters (the ones to which we assign
random values in order to explore the model parameter space) 
to be $m_{\tilde Q_{1,3}} (m_{susy})$,
$m_{\tilde d^c_{1,3}} (m_{susy})$, $M_3(m_{susy})$ and $A_{u,c} (m_{susy})$.
\item
for each point of the parameter space, defined by the chosen values of $m_{\tilde Q_{1,3}} (m_{susy})$,
$m_{\tilde d^c_{1,3}} (m_{susy})$, $M_3(m_{susy})$ and $A_{u,c} (m_{susy})$
(in addition to the values of $\tan\beta$, $m_{susy}$, $M_{GUT}$, $m_T$ and $m_V$,
already fixed in the first step), one can improve the analysis by running the soft terms again,
this time with the full RGEs, and use the complete expression for the Higgs mass. 
Once this is done, the phenomenological viability of this point must be further checked:
proton lifetime, consistency of the superpartner mass spectrum with the experimental limits,
absence of UFB directions and metastability of the electroweak vacuum
(which is checked following the procedure described in Appendix~\ref{app:CCB}), 
perturbativity constraints.
\end{enumerate}
\begin{figure}[t]
\centering
\includegraphics[width=0.45\textwidth]{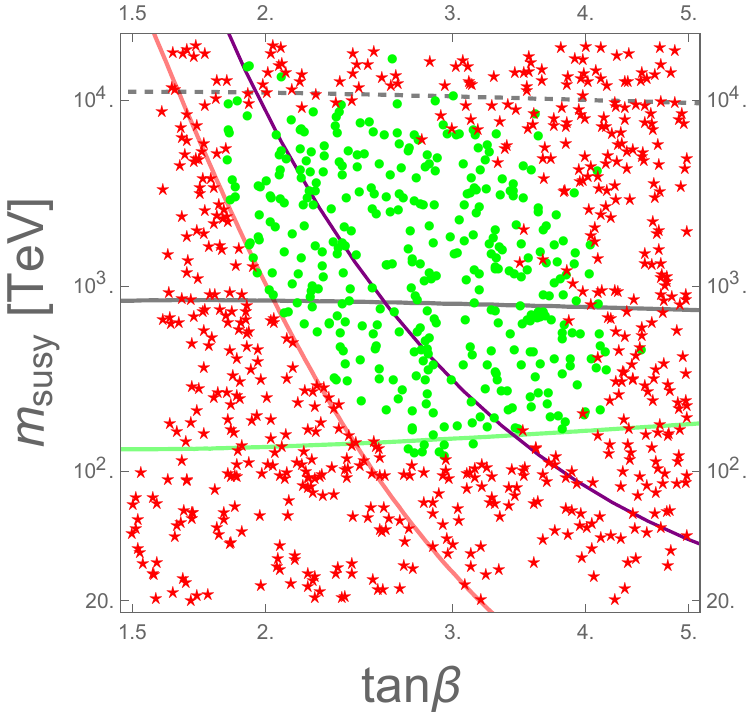}
\qquad
\includegraphics[width=0.45\textwidth]{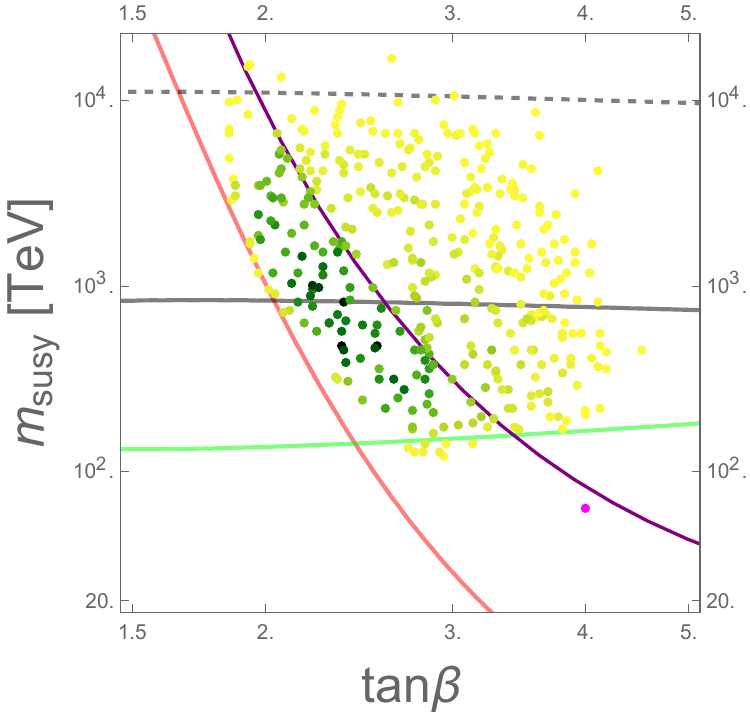}        
\caption{Left panel: successful (\textit{green dots}) and unsuccessful (\textit{red stars})
points among 1000 randomly selected points in the $(\tan\beta, m_{susy})$ plane. 
Right panel: successful points only, where the colour of a given dot represents the number
of superpartner spectrum configurations
respecting SU(5) symmetry at $M_{GUT}$ that passed all phenomenological constraints
(the number of solutions found increases from \textit{light yellow} to \textit{dark green}).
Also shown is a particular solution (\textit{magenta dot}) characterized by a low value
of the matching scale $m_{susy} \equiv \sqrt{m_{\tilde Q_3} (m_{susy})\, m_{\tilde u^c_3} (m_{susy})}$
and described in greater detail in the text below. 
In both panels, the \textit{red}, \textit{green}, \textit{purple} and \textit{black lines}
represent the naive constraints from Fig.~\ref{fig:parameter_space_with_point}
associated with the measured Higgs mass, the proton lifetime,
the metastability of the electroweak vacuum and the condition $m_T < M_{Planck} / 10$,
respectively.}
\label{fig:parameter_space_full}
\end{figure}
The procedure described above was applied by testing each randomly chosen point
in the $(\tan\beta, m_{susy})$ plane for 1000 different random configurations
of the superpartner spectrum, which were obtained by scanning over $m_{\tilde Q_{1,3}} (m_{susy})$,
$m_{\tilde d^c_{1,3}} (m_{susy})$ and $M_3(M_{GUT})$ in the range $[0.5 \, m_{susy},2 \, m_{susy}]$,
and over  $A_{u,c} (m_{susy})$ in the range $[-2 \, m_{susy},2 \, m_{susy}]$.
The results of this exploration of the parameter space of the minimal renormalizable
supersymmetric SU(5) model can be seen in Fig.~\ref{fig:parameter_space_full}.
The plot in the left panel displays all the points of the $(\tan\beta, m_{susy})$ plane
that have been checked; the ones for which at least one configuration of the
superpartner spectrum passed all constraints are in green, while the ones
that failed this test are in red. The plot in the right panel shows how frequently a solution
was found in the model parameter space for each green point. 
The more vivid the colour of the point, the more spectrum configurations
survived all the constraints. 
It turns out that the naive estimate of the allowed parameter space (the green region
in Fig.~\ref{fig:parameter_space_with_point}), despite not respecting the SU(5) symmetry, 
tells us something about how likely it is for a randomly chosen point in the parameter space
with given values of $\tan{\beta}$ and  $m_{susy}$ to be compatible with all experimental
and theoretical constraints discussed at the beginning of this section. In the part of the green 
region of Fig.~\ref{fig:parameter_space_with_point} around several hundreds of TeV
(close to the upper limit given by the requirement $m_T\leq M_{Planck}/10$),
a randomly chosen point in the parameter space has a few percent 
probability of success, while the probability decreases when one exits the green region.
One can see that the naive constraints associated with the Higgs mass (\textit{red line})
and with the proton lifetime (\textit{green line}) 
are very robust, which in the case of the Higgs mass can be traced back to the fact
that the leading term in the 1-loop matching condition for the Higgs quartic coupling,
Eq.~(\ref{eq:Xt}), only depends on $m_{susy}$ in the maximal stop mixing case. 
The naive vacuum metastability bound (\textit{purple line}) is less robust
but remains a reasonable approximation to the exact condition, contrary to the
naive $m_T < M_{Planck} / 10$ constraint (\textit{black line}). 
The main reason for this is that the black line assumed $\mu(m_{susy}) = m_{susy}$,
while a sizable portion of the viable points of the parameter space
feature smaller values of $\mu$, 
thus effectively decreasing the value of $m_T$ by virtue of the gauge coupling
unification condition~(\ref{mT}).

While these results confirm the naive expectation from Fig.~\ref{fig:parameter_space_with_point}
that the allowed parameter space of the minimal renormalizable supersymmetric SU(5) model
is restricted to the region of heavy superpartners, isolated points with lighter supersymmetric
particles are likely to be missed in this random search. In fact we were able to find
a viable point in the parameter space with $m_{susy} = 63 \TeV$ and some squarks as light
as $22 \TeV$, depicted by a \textit{magenta dot} in Fig.~\ref{fig:parameter_space_full}. 
We give below the input parameters at the GUT scale as well as the values of all soft terms
at the scale $m_{susy}$:

\vskip .5cm

\textbf{INPUT:}
\vskip .3cm
\textbf{Point in the $(\tan \beta, m_{susy})$ plane:}
\begin{align}
\tan \beta & \ = \ 4.0\, ,  \\
m_{susy} & \ = \ 63.0 \TeV\, ,  \\
M_{GUT} & \ = \ 2.0 \times 10^{16} \GeV\, ,  \\
\mbox{sign}(\mu) & \ = \ +\, ,
\end{align}

\textbf{Soft terms at $\mathbf{M_{GUT}}$:}
\begin{align}
A_u(M_{GUT}) & \ = \ \mbox{Diag}\, (2\times 10^{-4},9\times 10^{-2},13.9) \TeV\, ,  \\
A_d(M_{GUT}) \, = \, A_e(M_{GUT}) & \ = \ \mbox{Diag}\, (-0.4,41.2,121.7) \TeV\, ,  \\
M_{1/2} & \ = \ 34.4 \TeV\, ,  \\
m_{H_u}(M_{GUT}) & \ = \ 106.2 \TeV\, ,  \\
m_{H_d}(M_{GUT}) & \ = \ 531.2 \TeV\, ,  \\
m_{10_i} & \ = \ \mbox{Diag}\, (79.9,79.9,125.7) \TeV\, ,  \\
m_{\bar 5_i} & \ = \ \mbox{Diag}\, (151.8,151.8,141.1) \TeV\, .
\end{align}

\vskip .1cm

\textbf{OUTPUT:}
\vskip .3cm
\textbf{Values of gauge and Yukawa couplings at $\mathbf{m_{susy}}$, in the supersymmetric
($m_{susy} + \epsilon$) and in the non-supersymmetric theory ($m_{susy} - \epsilon$):}
\begin{align}
g_i(m_{susy} + \epsilon) & \ = \ (0.47777, \, 0.61950, \, 0.88966)\, ,  \\
g_i(m_{susy} - \epsilon) & \ = \ (0.47777, \, 0.61900, \, 0.88742)\, ,  \\
(\lambda_{t},\lambda_{b},\lambda_{\tau})(m_{susy} + \epsilon) & \ = \ (0.74778, \, 0.06325, \, 0.04229)\, ,  \\
(h_{t},h_{b},h_{\tau})(m_{susy} - \epsilon) & \ = \ (0.72553, \, 0.01123, \, 0.01026)\, ,  \\
(\lambda_{c},\lambda_{s},\lambda_{\mu})(m_{susy} + \epsilon) & \ = \
  (2.61\times 10^{-3}, \, 3.95\times 10^{-3}, \, 2.52\times 10^{-3})\, ,  \\
(h_{c},h_{s},h_{\mu})(m_{susy} - \epsilon) & \ = \ (2.53\times 10^{-3}, \, 0.23\times 10^{-3}, \, 0.61\times 10^{-3})\, ,  \\
(\lambda_{u},\lambda_{d},\lambda_{e})(m_{susy} + \epsilon) & \ = \
  (5\times 10^{-6}, \, 19\times 10^{-6}, \, 12\times 10^{-6})\, ,  \\
(h_{u},h_{d},h_{e})(m_{susy} - \epsilon) & \ = \ (5\times 10^{-6}, \, 12\times 10^{-6}, \, 3\times 10^{-6})\, ,
\end{align}

\textbf{Soft terms at $\mathbf{m_{susy}}$:}
\begin{align}
A_u(m_{susy}) & \ = \ \mbox{Diag}\, (0,0,-25.9) \TeV\, ,  \\
A_d(m_{susy}) & \ = \ \mbox{Diag}\, (-0.9,87.6,237.2) \TeV\, ,  \\
A_e(m_{susy}) & \ = \ \mbox{Diag}\, (-0.6,55.8,163.4) \TeV\, ,  \\
M_i(m_{\tilde g}) & \ = \ \mbox{Diag}\, (16.9,29.9,62.0) \TeV\, ,  \\
m_{H_u}(m_{susy}) & \ = \ 82.1 \TeV\, ,  \\
m_{H_d}(m_{susy}) & \ = \ 483.5 \TeV\, ,  \\
m_{\tilde Q_i}(m_{susy}) & \ = \ \mbox{Diag}\, (106.2,100.0,81.4) \TeV\, ,  \\
m_{\tilde u^c_i}(m_{susy}) & \ = \ \mbox{Diag}\, (22.3,22.3,49.0) \TeV\, ,  \\
m_{\tilde e^c_i}(m_{susy}) & \ = \ \mbox{Diag}\, (138.0,132.1,120.1) \TeV\, ,  \\
m_{\tilde L_i}(m_{susy}) & \ = \ \mbox{Diag}\, (131.4,128.3,84.4) \TeV\, ,  \\
m_{\tilde d^c_i}(m_{susy}) & \ = \ \mbox{Diag}\, (172.0,164.4,76.5) \TeV\, ,  \\
\mu(m_{susy}) & \ = \ 91.6 \TeV\, ,  \\
m_A(m_{susy}) & \ = \ 507.3 \TeV\, ,
\end{align}

\textbf{Values of gauge and Yukawa couplings at $\mathbf{M_{GUT}}$:}
\begin{align}
g^{MSSM}_i(M_{GUT}) & \ = \ (0.68367, \, 0.67083, \, 0.66722)\, , \\
g_{GUT} & \ = \ 0.71339\, ,  \\
(\lambda_{t},\lambda_{b},\lambda_{\tau})(M_{GUT}) & \ = \ (0.51487, \, 0.03185, \, 0.03132)\, , \\
(\lambda_{c},\lambda_{s},\lambda_{\mu})(M_{GUT}) & \ = \
  (1.47\times 10^{-3}, \, 1.86\times 10^{-3}, \, 1.86\times 10^{-3})\, , \\
(\lambda_{u},\lambda_{d},\lambda_{e})(M_{GUT}) & \ = \ (3\times 10^{-6}, \, 9\times 10^{-6}, \, 9\times 10^{-6})\, ,
\end{align}

\textbf{GUT state masses:}
\begin{align}
m_T & \ = \ 7.1 \times 10^{16} \GeV\, ,  \\
m_{8,3} & \ = \ 5.1 \times 10^{13} \GeV\, ,  \\
m_V & \ = \ 8.8 \times 10^{16} \GeV\, ,
\end{align}

\begin{figure}[t]
\centering
\includegraphics[width=0.45\textwidth]{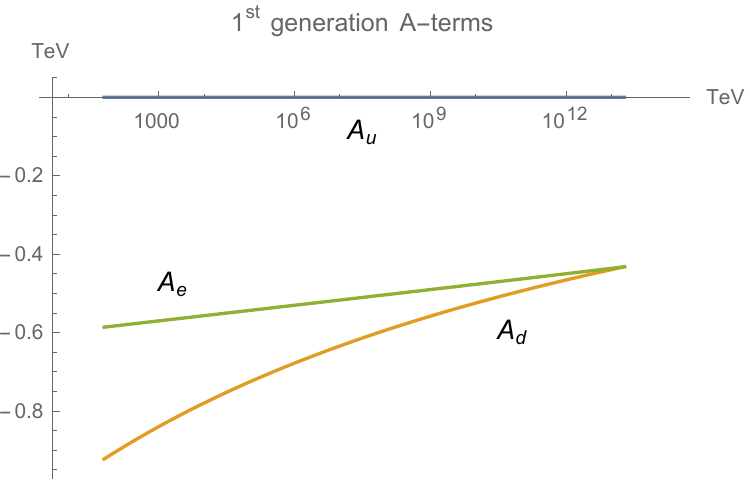}
\qquad
\includegraphics[width=0.45\textwidth]{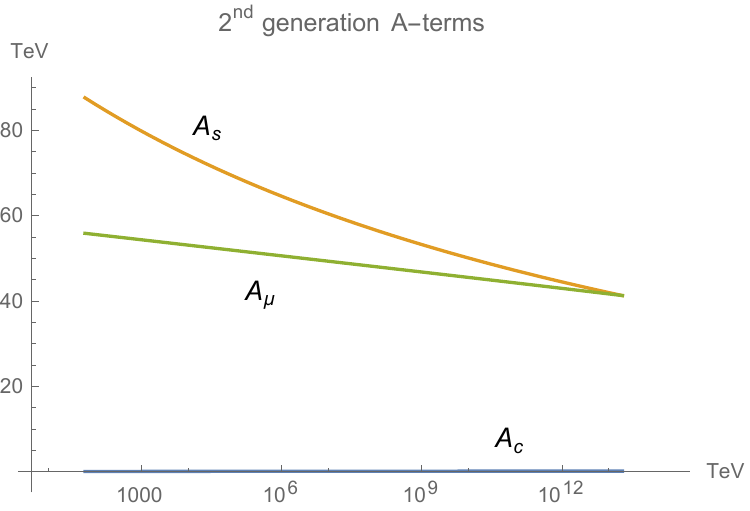}
\vskip .5cm
\includegraphics[width=0.45\textwidth]{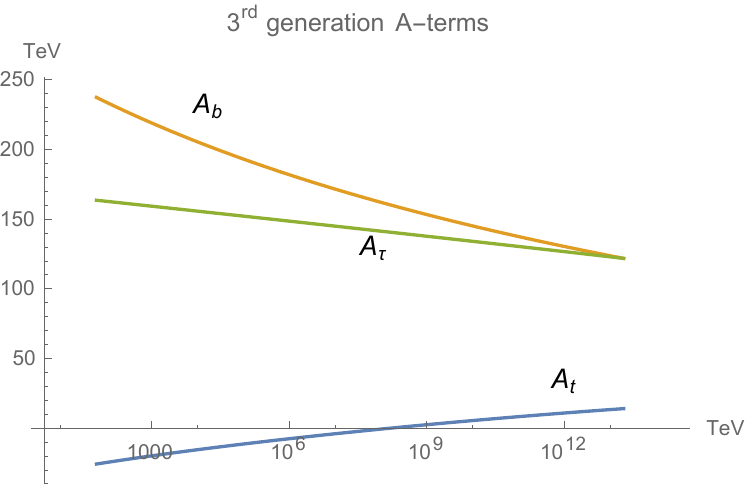}
\qquad
\includegraphics[width=0.45\textwidth]{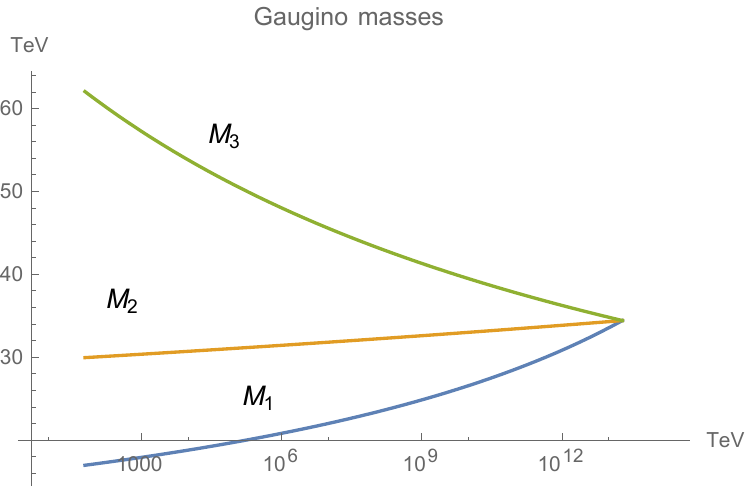}
\vskip .3cm
\caption{Running A-terms and gaugino masses for the parameter space point
described in the text (magenta dot in Fig.~\ref{fig:parameter_space_full}), obtained
using the 1-loop MSSM RGEs. Top and bottom left figures: $1^{st}$, $2^{nd}$ and $3^{rd}$
generation A-terms. The \textit{blue}, \textit{yellow} and \textit{green lines} represent
$A_{u_i}$, $A_{d_i}$ and $A_{e_i}$, respectively. Bottom right figure: the \textit{blue},
\textit{yellow} and \textit{green lines} represent $M_1$, $M_2$ and $M_3$, respectively.}
\label{fig:A-terms}
\end{figure}

\begin{figure}[t]
\centering
\includegraphics[width=0.45\textwidth]{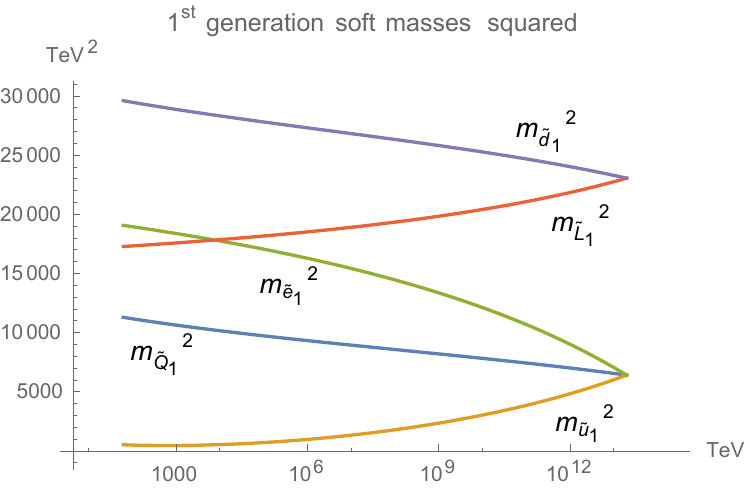}
\qquad
\includegraphics[width=0.45\textwidth]{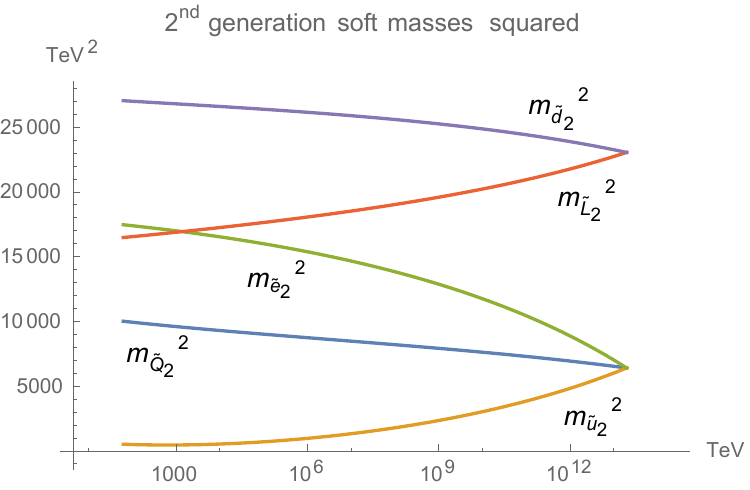}
\vskip .5cm
\includegraphics[width=0.45\textwidth]{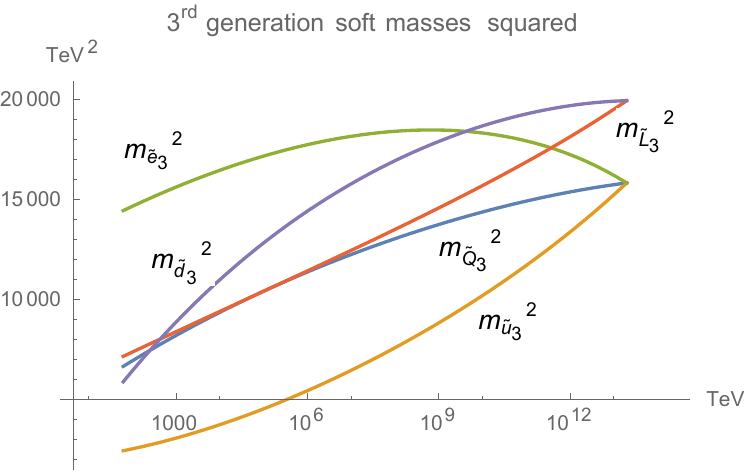}
\qquad
\includegraphics[width=0.45\textwidth]{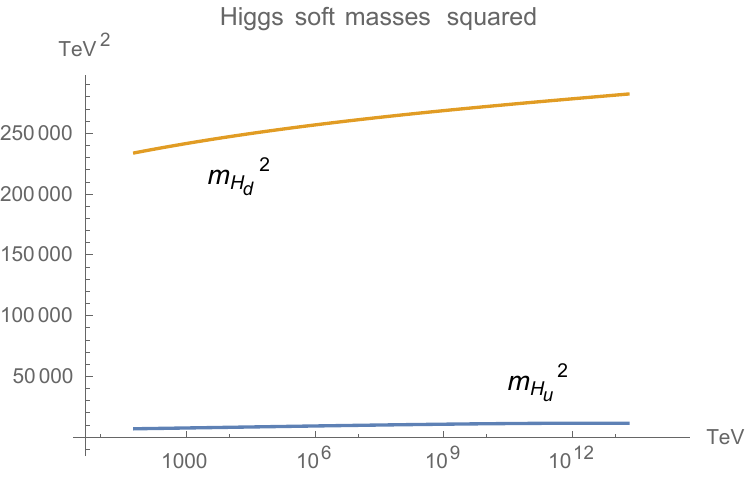}
\vskip .3cm
\caption{Running soft scalar masses for the parameter space point described in the text
(magenta dot in Fig.~\ref{fig:parameter_space_full}), obtained using the 1-loop MSSM RGEs.
Top and bottom left figures: $1^{st}$, $2^{nd}$ and $3^{rd}$ generation soft sfermion masses.
The \textit{blue}, \textit{yellow}, \textit{green}, \textit{red} and \textit{purple lines} represent
$m^2_{\tilde Q_i}$, $m^2_{\tilde u^c_i}$, $m^2_{\tilde e^c_i}$, $m^2_{\tilde L_i}$
and $m^2_{\tilde d^c_i}$, respectively. Bottom right figure: soft Higgs boson masses.
The \textit{blue line} represents $m^2_{H_u}$ and the \textit{yellow line} $m^2_{H_d}$.}
\label{fig:soft masses}
\end{figure}

\textbf{Other parameters and observables:}
\begin{align}
\alpha_{GUT}^{-1} & \ = \ 24.7\, ,  \\
\tau_p(p^+ \to K^+ \bar{\nu}) & \ = \ 3.9 \times 10^{33} \textrm{ yrs}
  \qquad  (\textrm{ for }\phi_1 = \phi_2 = \phi_3 = 0)\, ,  \\
S^{min} & \ = \ 1780\, ,
\end{align}

corresponding to a vacuum lifetime of
\beq
\tau_{vacuum}\, =\, \frac{e^{S^{min}}}{m_{susy}^4t_{universe}^3}\, \approx\, 10^{580}\, {\rm yrs}\, .
\eeq
For completeness, we also present the running of the A-terms in Fig.~\ref{fig:A-terms}
and of the soft scalar  masses in Fig.~\ref{fig:soft masses}.

Let us describe briefly the main features of this point. Most sfermions lie
between $50$ and $170 \TeV$, with however the $\tilde u_R$ and $\tilde c_R$
as light as $22.3 \TeV$. The gauginos are typically lighter than the sfermions
and the higgsinos, and the lightest supersymmetric particle (LSP) is the bino,
with mass $M_1 = 16.9 \TeV$ (the $\tilde u_R$ and the $\tilde c_R$ are the co-NLSPs,
i.e. the next-to-lightest supersymmetric particles).
The superpartners are therefore out of reach of present
and next-generation colliders, and supersymmetric contributions to flavour physics
observables are strongly suppressed, but proton decay
will be easily accessible at future large detectors. 
The Higgs sector is far into the decoupling regime,
with a standard-like lightest Higgs boson and all non-standard Higgs bosons
around $m_A \approx 500 \TeV$.
Finally, there is no suitable dark matter particle in the observable sector,
as the relic density of a $17 \TeV$ bino by far exceeds
$\Omega_{CDM} h^2 \simeq 0.12$~\cite{Ade:2015xua}.
If R-parity is conserved, a natural candidate is a gravitino in the GeV range (or lower),
to which the bino NLSP would decay without affecting Big Bang nucleosynthesis. 
The cold dark matter density would then be in the form of gravitinos 
coming from NLSP decays~\cite{Borgani:1996ag} (giving a contribution
$\Delta \Omega_{\tilde G}=\Omega_{\tilde B}m_{\tilde G}/M_1$,
from which the upper bound $m_{\tilde G} \lsim 1 \GeV$ follows)
and from thermal production during reheating~\cite{Bolz:2000fu,Pradler:2006hh}. In order
to avoid gravitino overproduction from the latter process, the reheating temperature
should lie in the TeV range, or lower. 
In the case of R-parity violation (which may be invoked to generate neutrino masses,
as an alternative to the seesaw mechanism), the only possible dark matter candidate
within this model is again the gravitino. 
Since it decays into a photon and a neutrino, acting as a source of monochromatic photons,
extragalactic gamma ray constraints put a bound on its mass. For bilinear R-parity violation, 
one obtains $m_{\tilde G}\lsim 1\;{\rm GeV}$~\cite{Takayama:2000uz}. Here, as in the R-parity 
conserving case, the maximal reheating temperature is around the TeV scale.

One may wonder whether other points exist in the parameter space
of the minimal renormalizable SU(5) model with lighter superpartners than
in the above example. While performing an extensive scan would probably
reveal the existence of such points, it appears difficult to lower significantly
the average superpartner mass scale below a few tens of $\TeV$ (some
supersymmetric particles may however be accidentally lighter). The reason
for this is the proton lifetime, which in the limit where all superpartners have
the same mass $m_{susy}$ scales as
$\tau_p \propto m_T^2 \, m_{susy}^2 \propto m^{11/3}_{susy}$
(where in the last step the approximate constraint~(\ref{eq:mT_msusy}) from gauge coupling
unification was used).
The actual proton decay constraint depends on the
individual superpartner masses\footnote{In addition, non-zero values of the high-energy
phases in the Yukawa matrices tend to increase the proton lifetime.},
and does not prevent some of them to be much
smaller than the naive lower bound on $m_{susy}$, but it seems to be difficult
to reconcile a significantly lighter superpartner spectrum with proton decay
and all other constraints. 
In fact gauge coupling unification is the main obstacle here, which makes it hard
to accommodate a low superpartner mass scale with a large colour triplet mass~\cite{Murayama:2001ur}.

What one may still try is to split the spectrum, making just a part of it light. 
This is not easy though, because of the threshold corrections to gauge couplings
at $M_{GUT}$ and to the bottom and Higgs masses at $m_{susy}$. For example,
making the higgsino light decreases the mass of the colour triplet, see Eq.~(\ref{mT}),
and therefore shortens the proton lifetime. The gluino and/or  the $\tilde b_R$ cannot
be much lighter than the $\tilde b_L$ (whose mass is of order $m_{susy}$) either,
since this would suppress the loop function in Eq.~(\ref{eq:m_di_thresholds})
and require an enormous value of $X_b$ to reproduce the observed $m_b / m_\tau$ ratio.
This large value of $X_b$ would in turn imply a large negative threshold correction
to the Higgs quartic coupling, see Eq.~(\ref{eq:Xb}), making it impossible to fit
the measured Higgs mass for moderate values of $m_{susy}$. 
All this motivates the choice of parameter intervals made in our scan, namely
$m_{\tilde Q_{1,3}} (m_{susy})$, $m_{\tilde d^c_{1,3}} (m_{susy})$ and $M_3(M_{GUT})$
in the range $[0.5 \, m_{susy},2 \, m_{susy}]$.
The only remaining possibility is to have the first two sfermion
generations lighter than the third one, an opposite situation to the one
considered e.g.  in Ref.~\cite{Murayama:2001ur}.
The investigation of this case would require the use of 2-loop RGEs for soft terms,
as a strong mass hierarchy between different sfermion generations enhances
the effect of 2-loop running and may give rise to tachyons~\cite{ArkaniHamed:1997ab}.
Such a study is beyond the scope of this paper, and we leave it for future work.


\section{Conclusions}
\label{sec:conclusions}

If one excepts the issue of neutrino masses, the minimal renormalizable supersymmetric
SU(5) model suffers from two main problems. The first one concerns the predictions
for charged fermion masses. Although the GUT-scale equality of down-type
quark and charged lepton Yukawa couplings leads to a qualitatively successful
prediction for the bottom to tau mass ratio, it quantitatively differs from the measured
value by some $20$--$30 \%$, while the discrepancy is much larger
for the first and second generations. Second, proton decay is too fast by a factor $10^4$ or so 
for typical soft terms in the TeV range, essentially because gauge coupling
unification requires a relatively light colour triplet. Each of the two problems has 
been addressed separately in the literature: large supersymmetric threshold corrections
were used to modify the fermion mass predictions
(see for example Refs.~\cite{DiazCruz:2000mn,Enkhbat:2009jt});
specific flavour structures of the soft terms~\cite{Bajc:2002bv} and a heavy superpartner
spectrum~\cite{Hisano:2013exa} were invoked to increase the proton lifetime.

In this paper, we performed a complete analysis of the minimal renormalizable
supersymmetric SU(5) model, taking into account all relevant phenomenological
and theoretical constraints: charged fermion masses, the proton lifetime,
gauge coupling unification, the Higgs mass, experimental lower bounds on superpartner
masses and flavour constraints, metastability of the vacuum and perturbativity of the model.
We showed that the model is still alive, and that the allowed region of the parameter
space spreads over a large domain with $m_{susy} \equiv  \sqrt{m_{\tilde t_1} m_{\tilde t_2}}$
ranging from around $50 \TeV$ to $10^4 \TeV$ and $\tan{\beta}$ from approximately $2$ to $5$. 
Viable points were also found outside this region of the parameter space, but they
are less frequent. Particularly interesting are the ones featuring some (relatively)
light superpartners. We studied in greater detail one such point in which the lightest
supersymmetric particle is a bino with mass $16.9 \TeV$.

A generic feature of the model is the metastability of the vacuum, which is not
a concern since its lifetime is typically much larger than the age of the universe.
A consequence of the heavy supertpartner spectrum is that the observable sector
does not contain a suitable dark matter candidate; however, a gravitino with a mass
around the GeV scale or below can play the role of cold dark matter, both in the presence
and in the absence of R-parity.

The present analysis is a good starting point for further research. 
One of the limitations is that we used the central values of the input
SM parameters; taking into account the experimental and theoretical
uncertainties will extend the viable region of the parameter space. We presented in 
a compact way all the ingredients that are needed for a more extensive scan
of the parameter space of the model. What we have shown is that the minimal renormalizable
supersymmetric SU(5) model is indeed a viable extension of the Standard Model,
in the sense that its parameter space contains points that satisfy all relevant
phenomenological and theoretical constraints. The price to pay is that the superpartner
spectrum is heavy, typically in the $100 \TeV$ region or above, although we were able
to find particular cases in which some of the supersymmetric particles can be as light
as $17$ TeV, and it is not excluded that viable points with even more split spectra
exist in some corners of the parameter space.

We did not address in this paper the issue of neutrino masses. One possibility is
to add Standard Model singlets (right-handed neutrinos) to the model and to generate
neutrino masses via the seesaw mechanism. Another possibility is to allow either
bilinear or trilinear R-parity violating terms. 
Notice that integrating out higgsinos and colour triplets with R-parity violating couplings
induces corrections to the Yukawa couplings~\cite{Bajc:2015zja}, which opens the possibility
that the SU(5) fermion mass relations are cured by the combined effect of R-parity violation
and of supersymmetric threshold corrections.

The possibilities to test experimentally the minimal renormalizable supersymmetric
SU(5) model are limited.
Barring possible isolated points of the parameter space with light remnants, 
superpartners are too heavy to be detected at the next generation of colliders or to give sizable
contributions to flavour-violating observables. Proton decay could be around the corner,
but  the allowed parameter space is too vast to give a definite prediction for the proton lifetime.
It is, however, possible to rule out the model (or at least most of its allowed parameter
space), either by discovering a light superpartner spectrum or, if the Higgs sector
is light enough to be accessible and studied at the next generation of colliders,
by measuring a relatively large value of $\tan \beta$.


\section*{Acknowledgments}

We thank Stefan Antusch, Charan Aulakh, Kaladi Babu, Abdelhak Djouadi,
Tsedenbaljir  Enkhbat, Ilia Gogoladze, Jean-Lo\"{i}c Kneur, Michal Malinsk\' y, 
Goran Senjanovi\'c, Pietro Slavich,
Constantin Sluka and Zurab Tavartkiladze for discussions.
The work of B.B. is supported by the Slovenian Research Agency.
The work of S.L. has been supported in part by the Agence Nationale de la Recherche
under contract ANR 2010 BLANC 0413 01, by the European Research Council
(ERC) Advanced Grant Higgs@LHC, and by the European Union FP7 ITN Invisibles
(Marie Curie Actions, PITN-GA-2011-289442).
The  work of T.M. is supported by the Foundation for support of science and research ``Neuron''.
T.M. also acknowledges the support from the Jo\v zef Stefan Institute in Ljubljana,
where the early part of this study was performed and wishes to express his gratitude to the IPhT
in Saclay for their warm hospitality and support during the development of this project.


\appendix

\section{\label{app:RGEs}Procedure for solving the soft term RGEs}

In this appendix, we derive the expressions for the superpartner spectrum of the
minimal renormalizable supersymmetric SU(5) model that we have used
in our analysis. These formulae account for the splitting of sfermion masses
within the same SU(5) multiplet due to renormalization group running.
They are (approximate) semi-analytic solutions to the 1-loop MSSM RGEs
for the soft terms obtained under certain assumptions (namely, the hierarchy
among Yukawa couplings allows us to neglect some terms in the RGEs
and to solve them in a sequential manner, as explained below). We have checked
that they are accurate up to the few percent level by running the full set of RGEs.
The advantage of these approximate solutions is that the low-energy soft terms
can be written as quadratic functions of the initial (GUT-scale) values of the soft
parameters, which makes it possible to perform a scan over the parameter
space of the model without having to solve the full set of RGEs for each point.
Furthermore, it is possible to scan directly over the low-energy values of the soft terms
(rather than the GUT-scale ones), since these formulae implicitly respect the SU(5) boundary conditions. 
A similar approach has already been employed in the past, e.g. 
in Ref.~\cite{Drees:1991ab} for mSUGRA and in Ref.~\cite{Rattazzi:1995gk} for SU(5).
Here we generalize this procedure to generation-dependent soft terms with SU(5) boundary conditions. 

\subsection{The procedure}
\label{subsec:procedure}

In order to perform a scan over the parameter space of the minimal renormalizable
supersymmetric SU(5) model, one has to solve a system of entangled differential equations 
describing the running of gauge, Yukawa and Higgs quartic couplings, as well as
of the soft terms, which involves
a large number of parameters (only in the supersymmetry breaking sector,
there are upon SU(5) unification 15 parameters, namely 6 A-terms, 1 gaugino mass
and 8 soft scalar masses). In addition, these parameters are subject to various
phenomenological constraints defined at different scales. This makes it hard
to solve the problem by brute force, and motivates the use of approximate
semi-analytical solutions to the soft term RGEs.

Let us first try to circumvent the issue by solving the RGEs in steps.
We first integrate numerically the system of RGEs for gauge, Yukawa and Higgs quartic couplings
at the 2-loop level. Note that these couplings depend on the soft terms only through supersymmetric
threshold corrections, so we can use this fact to determine the values of the relevant soft terms
at the matching scale $m_{susy}$. Then we can solve the 1-loop RGEs for the soft terms
-- first gaugino masses, then  A-terms and finally soft scalar masses. By working them out
in this particular order, we are able to use in each step the knowledge of the running parameters
that we have computed in the previous steps. Unfortunately this method is not efficient
since one must solve the RGEs for the soft terms every time one changes the input
(GUT-scale) values of the soft parameters. What we would like is to numerically solve
the soft term RGEs for symbolic input values. The computational procedure described
below does precisely that: when the input values of the soft parameters are changed,
only the last step (out of 3) has to be re-run.

\paragraph{Step 1}
\begin{itemize}
\item[1.]
Choose the matching scales $m_{susy}$ and $M_{GUT}$, the ratio of the MSSM Higgs vevs $\tan\beta$
and the masses of the heavy GUT states $m_T$ and $m_V$ within the perturbative regime
$m_T < m_V \ll M_{Planck}$ (note that $m_T$ and $m_V$ affect the running of the parameters
only through the high-scale threshold corrections to the Yukawa couplings, so their influence is very mild).
\item[2.]
Run the gauge, charged lepton and up-type quark Yukawa couplings as well as
the Higgs quartic coupling from their measured values
at the weak scale up to the matching scale $m_{susy}$ using the 2-loop SM RGEs.
\item[3.]
Apply the matching conditions~(\ref{eq:g1_conversion})--(\ref{eq:g3_conversion}),~(\ref{eq:Yu_conversion}) and~(\ref{eq:Ye_conversion}) to the gauge and Yukawa couplings at the scale $m_{susy}$
and use the 2-loop MSSM RGEs to run them up to the scale $M_{GUT}$.
\item[4.]
Compute the down-type quark Yukawa couplings at $M_{GUT}$ from
the values of the charged lepton Yukawa couplings using the SU(5) mass relation~(\ref{eq:Md_Me})
and the GUT threshold corrections~(\ref{eq:GUT_thresholds_Yuk}).
Run them down to the scale $m_{susy}$ using the 2-loop MSSM RGEs
(the supersymmetric threshold corrections to the leptonic Yukawa couplings are neglected
in this procedure).
\end{itemize}

\paragraph{Step 2}
\begin{itemize}
\item[5.] 
Solve the 1-loop MSSM RGEs for gaugino masses, taking into account
the unification condition~(\ref{eq:gaugino_unification}). 
\item[6.] 
Assume the hierarchy\footnote{This assumption is based on the observed hierarchy
of fermion masses, taking into account the SU(5) boundary condition $\Lambda_D = \Lambda^T_E$
and the supersymmetric threshold corrections to down-type quark masses,
and is better justified in the low to moderate $\tan \beta$ regime.
For instance, the relation $\lambda_s(M_{GUT}) = \lambda_\mu(M_{GUT})$ implies
$\lambda_s(m) > \lambda_\mu(m)$ for $m_{susy} < m < M_{GUT}$, while in the absence
of this relation and of supersymmetric threshold corrections to $m_s$ one would have
$\lambda_s(m) < \lambda_\mu(m)$ over most of the range $m_{susy} < m < M_{GUT}$.}
$\lambda_t > \lambda_b > \lambda_{\tau} > \lambda_c > \lambda_s > \lambda_{\mu} > \lambda_u > \lambda_d > \lambda_e$
and the unification of leptonic and down-type quark A-terms at the GUT scale to solve
the RGEs for A-terms in a sequential way (see details in Subsection~\ref{subsec:Aterms_approx}).
As a result, express the running A-terms as linear functions of the gaugino masses
and of their values at the scale $m_{susy}$
(which are more convenient input parameters than their values at the GUT scale,
since phenomenological constraints on A-terms apply at the scale $m_{susy}$),
with coefficients depending on the running quantities determined in Step 1.
In this way one does not need to run the A-terms every time one changes their initial values.
\item[7.] Solve the RGEs for the soft scalar masses in a similar manner.
\end{itemize}

\paragraph{Step 3}
\begin{itemize}
\item[8.] Choose the values of the input soft parameters. 
\item[9.] Make sure that no sfermion becomes tachyonic. Check that the values of the A-terms
and soft scalar masses at $m_{susy}$ reproduce the observed values of the down-type quark
masses and of the Higgs mass. Verify that all experimental lower bounds on superpartner
masses and flavour constraints are satisfied.
\item[10.] Check vacuum (meta)stability constraints and the proton lifetime. 
\end{itemize}

\subsection{Input values}

All input values of the SM parameters needed for the running were taken
either from Ref.~\cite{Buttazzo:2013uya} or from Ref.~\cite{Draper:2013oza}:
\bea
M_{Planck}&=& 2.4 \times 10^{18}\GeV\, ,  \\
v&=&174.10362 \GeV\, ,  \\ 
(m_Z,m_t,m_h)^{pole}&=&(91.1876,\, 173.10,\, 125.66)\GeV\, ,  \\
g_i(m_t) &=& (0.46167,\, 0.64822,\, 1.1666)\, ,  \\
(h_t,h_b,h_\tau)(m_t)&=&(0.93558,\, 0.0156,\, 0.0100)\, ,  \\
(m_c,m_s,m_\mu)(m_Z)&=&(619,\, 55,\, 103)\MeV\, ,  \\
(m_u,m_d,m_e)(m_Z)&=&(1.27,\, 2.9,\, 0.5)\MeV\, , \\
\lambda(m_t) &=& 0.25420\, ,
\eea
where $g_{1,2,3} (m_t)$, $h_{t,b,\tau} (m_t)$ and $\lambda (m_t)$ are evaluated at NNLO
(the NNNLO pure QCD contribution is also included in $h_t (m_t)$~\cite{Buttazzo:2013uya}),
and $h_i(m_Z) = m_i(m_Z) / v$ for the first and second generations of fermions.

\subsection{Approximate expressions for the soft terms}

In order to be able to approximately solve the 1-loop RGEs for soft terms, we shall neglect all mixings.
This means that in addition to assuming that the sfermion soft terms are aligned
with fermion masses at the GUT scale (see Subsection~\ref{subsec:SU5}),
we shall neglect the effects of the CKM matrix in the RGEs. 
This may not be fully justified, as the impact of $V_{CKM}$ on the running of the first two
generation parameters can be significant.
However, this effect is suppressed either by small Yukawa couplings or by small CKM angles
and is therefore never numerically important\footnote{Obviously, this statement does not apply
to RG-induced flavour-violating soft terms. These are not a concern, however, since
the superpartner spectrum is heavy and the first two generation squarks are almost
degenerate in mass.},
so we shall set $V_{CKM} = \mathbb{1}$ in the RGEs and omit the subleading
Yukawa contributions.

With these assumptions, we can derive approximate semi-analytic solutions
to the 1-loop RGEs for the soft terms.

\subsubsection{Gaugino masses}

The 1-loop RGEs for gaugino masses:
\begin{align}
  \frac{d}{dt}M_i & \, = \, \frac{b_i^{MSSM}}{2\pi} \, \alpha_i \, M_i \, \simeq \, \frac{d\alpha_i}{dt}\, \frac{M_i}{\alpha_i}\, ,
\end{align}
(where the last equality is only approximate because we run gauge couplings at 2 loops)
has the following simple solution respecting SU(5) symmetry:
\begin{align}
M_i (m) & \, = \, M_{1/2} \, e^{-\int_{\ln{m}}^{\ln{M_{GUT}}} dt \, \alpha_i(t) \, b_i^{MSSM} / (2\pi)}  \nonumber \\
  & \, \simeq \, M_{1/2} \, \frac{\alpha_i (m)}{\alpha_i (M_{GUT})} \, \simeq \, M_i (m_{\tilde g}) \,
  \frac{\alpha_i (m)}{\alpha_i (m_{\tilde g})} \, \simeq \, m_{\tilde g} \, \frac{\alpha_i (m)}{\alpha_3 (m_{\tilde g})} \,
  \frac{\alpha_3 (M_{GUT})}{\alpha_i (M_{GUT})}\, .
\end{align}
This solution becomes exact when the 1-loop RGEs for gauge couplings are used.

\subsubsection{A-terms}
\label{subsec:Aterms_approx}

For small to moderate $\tan \beta$, one can take advantage of the hierarchy among
Yukawa couplings (see Subsection~\ref{subsec:procedure}, Step 2) to simplify the
1-loop RGEs for A-terms and to solve them in a sequential way\footnote{In case 
the Yukawa-dependent terms in the RGEs do not follow the hierarchy
assumed for the Yukawa couplings themselves, 
one can if necessary improve the accuracy
of the solutions by iterating the procedure described below.}.
Namely, one can write the A-term RGEs as:
\begin{align}
  \frac{d}{dt}A_n & \, = \, \frac{A_n}{(4 \pi)^2} \left[ \sum_{p} J^n_p  \lambda^2_p - \sum_{i=1}^3 K^n_i g_i^2 \right]
    + \frac{\lambda_n}{(4 \pi)^2} \left[ 2 \sum_{p \neq n} J^n_p \lambda_p A_p + 2 \sum_{i=1}^3 K^n_i g_i^2 M_i \right]\, ,
\label{eq:Aterms_RGE}
\end{align}
where the indices $n$ and $p$ run over the ordered values $\{ t, b, \tau, c, s, \mu, u, d, e \}$
(such that e.g. $p < c$ means $p = t, b$ or $\tau$) and the coefficients $J^n_p$, $K^n_i$
are collected in Table~\ref{Table:aterms}.
\begin{center}
\begin{tabular}{c||*{9}{c|}l*{3}{c|}}
$p$ & $t$ & $b$ & $\tau$ & $c$ & $s$ & $\mu$ & $u$ & $d$ & $e$  \\
\hline
\hline
$J^t_p$ & $18$ & $1^*$ & & $3^*$ & & & $3^*$ & & \\
\hline
$J^b_p$ & $1$ & $18$ & $1^*$ & & $3^*$ & $1^*$ & & $3^*$ & $1^*$ \\
\hline
$J^{\tau}_p$ & & $3$ & $12$ & & $3^*$ & $1^*$ & & $3^*$ & $1^*$ \\
\hline
\hline
$J^c_p$ & $3$ & & & $18$ & $1^*$ & & $3^*$ & & \\
\hline
$J^s_p$ & & $3$ & $1$ & $1$ & $18$ & $1^*$ & & $3^*$ & $1^*$ \\
\hline
$J^{\mu}_p$ & & $3$ & $1$ & & $3$ & $12$ & & $3^*$ & $1^*$ \\
\hline
\hline
$J^u_p$ & $3$ & & & $3$ & & & $18$ & $1^*$ & \\
\hline
$J^d_p$ & & $3$ & $1$ & & $3$ & $1$ & $1$ & $18$ & $1^*$ \\
\hline
$J^e_p$ & & $3$ & $1$ & & $3$ & $1$ & & $3$ & $12$ \\
\hline
\end{tabular}
\hskip 1cm
\begin{tabular}{c||*{9}{c|}|*{3}{c|}}
$i$ & $3$ & $2$ & $1$ \\
\hline
\hline
$K^t_i$ & $\frac{16}{3}$ & $3$ & $\frac{13}{15}$ \\
\hline
$K^b_i$ & $\frac{16}{3}$ & $3$ & $\frac{7}{15}$ \\
\hline
$K^{\tau}_i$ & & $3$ & $\frac{9}{5}$ \\
\hline
\hline
$K^c_i$ & $\frac{16}{3}$ & $3$ & $\frac{13}{15}$ \\
\hline
$K^s_i$ & $\frac{16}{3}$ & $3$ & $\frac{7}{15}$ \\
\hline
$K^{\mu}_i$ & & $3$ & $\frac{9}{5}$ \\
\hline
\hline
$K^u_i$ & $\frac{16}{3}$ & $3$ & $\frac{13}{15}$ \\
\hline
$K^d_i$ & $\frac{16}{3}$ & $3$ & $\frac{7}{15}$ \\
\hline
$K^e_i$ & & $3$ & $\frac{9}{5}$ \\
\hline
\end{tabular}
\captionof{table}{\label{Table:aterms}The coefficients $J^n_p$, $K^n_i$ appearing
in the 1-loop RGEs for A-terms~(\ref{eq:Aterms_RGE}).}
\end{center}
A superscript $^*$ on a coefficient $J^n_p$ in Table~\ref{Table:aterms}  indicates
that we neglect the corresponding term $2 J^n_p \lambda_n \lambda_p A_p / (4 \pi)^2$
in the RHS of Eq.~(\ref{eq:Aterms_RGE}), consistently with the hierarchy of
Yukawa couplings\footnote{In addition, one does not need to include the terms that are suppressed by
small Yukawa couplings in the RGEs~(\ref{eq:Aterms_RGE}), 
as their effect is smaller than the precision of the 1-loop approximation. 
We nevertheless give the corresponding coefficients in Table~\ref{Table:aterms}
for completeness, allowing for the possibility of unusually large first or second
generation A-terms that would make some of these terms relevant.}.
The evolution of each A-term is then (approximately) described 
by a differential equation of the form
\begin{align}
  \frac{d}{dt} A_n(t) & \ = \ \beta^A_n(t) \, A_n(t) + \gamma^A_n(t)\, ,
\label{eq:Aterms_simplified_RGEs}
\end{align}
where the coefficients $\beta^A_n(t)$ and $\gamma^A_n(t)$ depend on the gauge
and Yukawa couplings, and in addition $\gamma^A_n(t)$ depends linearly on the gaugino
masses and on the A-terms $A_m(t)$ with $m<n$.
This makes it possible to solve the RGEs~(\ref{eq:Aterms_simplified_RGEs}) sequentially,
from $n=t$ to $n=e$. The solutions can be written as:
\beq
  A_n(t)\ =\ A_n(t_0)\; e^{\, \int_{t_0}^t dt' \beta^A_n(t')} \, +\ e^{\, \int_{t_0}^t dt' \beta^A_n(t')}
    \int_{t_0}^t dt' \, \left ( \gamma^A_n(t') \, e^{- \int_{t_0}^{t'} dt'' \beta^A_n(t'')} \right )\, ,
\label{eq:An_solution}
\eeq
where, by using the already obtained expressions for the $A_m$'s with $m < n$,
the integrals of the coefficients $\beta^A_n(t)$ and $\gamma^A_n(t)$ can be computed
numerically after having solved the MSSM RGEs for the gauge and Yukawa couplings.
As a result, one obtains the running A-terms at an arbitrary scale as linear combinations of
the 7 SU(5) soft parameters $M_{1/2}$, $A^{10}_i$, $A^{\bar 5}_i$ ($i=1,2,3$), with
numerical coefficients depending on the choice of $m_{susy}$ and $\tan\beta$
(and very mildly on $M_{GUT}$, $m_T$ and $m_V$).
In practice, it will prove convenient for the exploration of the parameter space of the model
to trade these GUT-scale parameters
for low-energy ones, so as to express the $A_n(t)$ as a function of $M_3(m_{susy})$,
$A_{d_i}(m_{susy})$ and $A_{u_i}(m_{susy})$ ($i=1,2,3$). Note that if the SU(5) boundary
conditions at $M_{GUT}$ had not been imposed, the $A_n(t)$ would depend on
12 initial parameters, namely $M_{1,2,3}$, $A_{u_i}$, $A_{d_i}$ and $A_{e_i}$.

\subsubsection{Soft scalar masses}

One can apply a similar procedure to the 1-loop RGEs for soft scalar masses.
Let us first define the following variables:
\begin{align}
  \Sigma_{u_i} & \ \equiv \ m^2_{\tilde Q_i} + m^2_{\tilde u^c_i} + m^2_{H_u}\, , 
  \label{eq:def Sigma_U} \\
  \Sigma_{d_i} & \ \equiv \ m^2_{\tilde Q_i} + m^2_{\tilde d^c_i} + m^2_{H_d}\, ,
  \label{eq:def Sigma_D} \\
  \Sigma_{e_i} & \ \equiv \ m^2_{\tilde L_i} + m^2_{\tilde e^c_i} + m^2_{H_d}\, .
  \label{eq:def Sigma_E} 
\end{align}
These combinations of masses appear on the RHS of the RGEs for soft scalar masses, 
and obey 1-loop RGEs of the form
\begin{align}
  \frac{d}{dt}\Sigma_n \, & = \, \frac{L^n_n \lambda_n^2}{(4\pi)^2}\, \Sigma_n
    + \frac{1}{(4\pi)^2} \left[ \sum_{p \neq n} L^n_p \lambda_p^2 \Sigma_p
    + \sum_p L^n_p A_p^2 - \sum_{i=1}^{3} N^n_i g_i^2 M_i^2 \right ]\, ,
\label{eq:Sigma_n_RGE}
\end{align}
where again the indices $n$ and $p$ run over the ordered values $\{ t, b, \tau, c, s, \mu, u, d, e \}$ 
and the coefficients $L^n_n$ are collected in Table~\ref{Table:Sigmas}, while $L^n_{p \neq n} = 2 J^n_p$
and $N^n_i = 4 K^n_i$.
\begin{center}
\begin{tabular}{c||*{9}{c|}}
  $n$ & $t$ & $b$ & $\tau$ & $c$ & $s$ & $\mu$ & $u$ & $d$ & $e$ \\
\hline
\hline
  $L^n_n$ & $12$ & $12$ & $8$ & $12$ & $12$ & $8$ & $12$ & $12$ & $8$ \\
\hline
\end{tabular}
\captionof{table}{\label{Table:Sigmas}The coefficients $L^n_n$ appearing in the 1-loop
RGEs for $\Sigma_n$~(\ref{eq:Sigma_n_RGE}).}
\end{center}
Neglecting\footnote{The terms suppressed by Yukawa couplings of the first and second generations
can also be dropped, as their effect is smaller than the precision of the 1-loop approximation.}
the terms $L^n_p \lambda_p^2 \Sigma_p / (4\pi)^2$ ($p \neq n$) on the RHS of
Eq.~(\ref{eq:Sigma_n_RGE}) when the coefficients $J^n_p$ are marked with a superscript $^*$
in Table~\ref{Table:aterms}, one can solve the RGEs for the $\Sigma_n$'s sequentially
from $n=t$ to $n=e$, as we did for the A-term RGEs.
Indeed, in this approximation Eq.~(\ref{eq:Sigma_n_RGE}) can be written in the form:
\begin{align}
  \frac{d}{dt} \Sigma_n(t) & \ = \ \beta^\Sigma_n(t) \, \Sigma_n(t) + \gamma^\Sigma_n(t)\, ,
\end{align}
where $\beta^\Sigma_n(t)$ is proportional to $\lambda_n^2(t)$, and $\gamma^\Sigma_n(t)$
is a linear combination of $g_i^2(t) M_i^2(t)$, $A_p^2(t)$ and $\lambda_m^2(t) \Sigma_m(t)$
with $m<n$. The solution reads:
\beq
  \Sigma_n(t)\ =\ \Sigma_n(t_0)\; e^{\, \int_{t_0}^t dt' \beta^\Sigma_n(t')}\, +\ 
    e^{\, \int_{t_0}^t dt' \beta^\Sigma_n(t')} \int_{t_0}^t dt'\, \left ( \gamma^\Sigma_n(t')\,
    e^{- \int_{t_0}^{t'} dt'' \beta^\Sigma_n(t'')} \right )\, ,
\label{eq:Sigma_n_solution}
\eeq
where the integrals involve only already known quantities, since the RGEs for the $\Sigma_m$'s
are solved in order of increasing $m$.
Using the previously obtained expressions for the running A-terms and gaugino masses,
the combinations of masses $\Sigma_n(t)$ are then expressed as quadratic
functions of the GUT-scale soft scalar masses $m_{10_i}$, $m_{\bar 5_i}$, $m_{H_u}(M_{GUT})$,
$m_{H_d}(M_{GUT})$ and of the $m_{susy}$ values of $M_3$, $A_{u,c,t}$ and $A_{d,s,b}$,
with $t$-dependent coefficients that can be computed numerically
using the known solutions to the 2-loop MSSM RGEs for gauge and Yukawa couplings.

Another combination of masses that appears in the RGEs for soft scalar masses is 
\begin{align}
  S\ & \equiv\ m^2_{H_u}-m^2_{H_d} + \Tr \left[ m^2_Q - m^2_L - 2 m^2_U + m^2_D + m^2_E \right]  \nonumber \\
  & =\ m^2_{H_u}-m^2_{H_d} + \sum_{i=1}^{3} \left[ m_{\tilde Q_i}^2 - m_{\tilde L_i}^2 - 2 m_{\tilde u_i^c}^2
    + m_{\tilde d_i^c}^2 + m_{\tilde e_i^c}^2 \right] ,
\label{eq:def S}
\end{align}
for which the 1-loop RGE takes the simple form: 
\begin{align}
  \frac{d}{dt} S &\, =\, \frac{66}{5}\, \frac{\alpha_1}{4 \pi}\, S \, = \, \frac{b_1^{MSSM}}{2 \pi}\, \alpha_1 S \,
    \simeq\, \frac{d\alpha_1}{dt}\, \frac{S}{\alpha_1}\, .
\label{eq:S_sol}
\end{align}
The last equality is only approximate because we run gauge couplings
at 2 loops. Working in the 1-loop approximation, one can solve Eq.~(\ref{eq:S_sol})
straightforwardly:
\begin{align}
  S(m)\, & = \, S(M_{GUT}) \, e^{-\int_{\ln m}^{\ln M_{GUT}} dt  \, \alpha_1(t) \, b_1^{MSSM} / (2\pi)}  \nonumber \\
  & \simeq \, S(M_{GUT}) \, \frac{\alpha_1(m)}{\alpha_{GUT}} \,
    = \left[ m^2_{H_u}(M_{GUT}) - m^2_{H_d}(M_{GUT}) \right] \frac{\alpha_1(m)}{\alpha_{GUT}}\ ,
\end{align}
where we have used the fact that $S(M_{GUT}) = m^2_{H_u}(M_{GUT}) - m^2_{H_d}(M_{GUT})$
due to the SU(5) boundary conditions on soft scalar masses.

We now have all the ingredients needed to write the solutions to the 1-loop RGEs
for soft scalar masses as:
\begin{align}
\hskip -.1cm
  m_\alpha^2(m) = m_\alpha^2(M_{GUT})+\sum_n P^\alpha_{n}I_{\Sigma_n}(m)
    +\sum_n P^\alpha_{n}I_{A_n}(m)-\sum_{i=1}^3 Q^\alpha_{i}I_{M_i}(m)+R^{\alpha}I_{S}(m) ,
\label{eq:soft_masses_solution}
\end{align}
where $\alpha = H_u, H_d, \tilde Q_i, \tilde u^c_i, \tilde e^c_i, \tilde L_i, \tilde d^c_i$,
the numerical coefficients $P^\alpha_{n}$, $Q^\alpha_{i}$ and $R^{\alpha}$ are
given in Table~\ref{Table:softmasses1},
\begin{center}
\begin{tabular}{c||*{3}{c|}|*{3}{c|}|c|}
  $\alpha$ & $P^{\alpha}_{t,c,u}$ & $P^{\alpha}_{b,s,d}$ & $P^{\alpha}_{\tau,\mu,e}$ & $Q^{\alpha}_3$
      & $Q^{\alpha}_2$ & $Q^{\alpha}_1$ & $R^{\alpha}$ \\
\hline
\hline
  $H_u$ & $6$ &  &  &  & $6$ & $\frac{6}{5}$ & $+\frac{3}{5}$ \\
\hline
  $H_d$ &  & $6$ & $2$ &  & $6$ & $\frac{6}{5}$ & $-\frac{3}{5}$ \\
\hline
\end{tabular}
\end{center}
\begin{center}	
\begin{tabular}{c||*{3}{c|}|*{3}{c|}|c|}
  $\alpha$ & $P^{\alpha}_{u_i}$ & $P^{\alpha}_{d_i}$ & $P^{\alpha}_{e_i}$ & $Q^{\alpha}_3$
      & $Q^{\alpha}_2$ & $Q^{\alpha}_1$ & $R^{\alpha}$ \\
\hline
\hline
  $\tilde Q_i$ & $2$ & $2$ & $$ & $\frac{32}{3}$ & $6$ & $\frac{2}{15}$ & $+\frac{1}{5}$ \\
\hline
  $\tilde u^c_i$ & $4$ & $$ & $$ & $\frac{32}{3}$ & $$ & $\frac{32}{15}$ & $- \frac{4}{5}$ \\
 \hline
  $\tilde e^c_i$ & $$ & $$ & $4$ & $$ & $$ & $\frac{24}{5}$ & $+\frac{6}{5}$ \\
\hline
  $\tilde L_i$ & $$ & $$ & $2$ & $$ & $6$ & $\frac{6}{5}$ & $- \frac{3}{5}$ \\
\hline
  $\tilde d^c_i$ & $$ & $4$ & $$ & $\frac{32}{3}$ & $$ & $\frac{8}{15}$ & $+\frac{2}{5}$ \\
\hline
\end{tabular}
\captionof{table}{\label{Table:softmasses1}The non-vanishing coefficients $P^\alpha_{n}$,
$Q^\alpha_{i}$ and $R^{\alpha}$ appearing in Eq.~(\ref{eq:soft_masses_solution}).}
\end{center}
and the integrals $I_{\Sigma_n}$, $I_{A_n}$, $I_{M_i}$ and $I_S$ are defined by:
\begin{align}
I_{\Sigma_n} (m)\, & \equiv\, \frac{1}{(4\pi)^2} \int_{\ln{m_{GUT}}}^{\ln{m}} dt \, \lambda_n^2(t) \Sigma_n (t)\, ,
\label{eq:I_Sigma_n}  \\
I_{A_n} (m)\, & \equiv\, \frac{1}{(4\pi)^2} \int_{\ln{m_{GUT}}}^{\ln{m}} dt \, A_n^2(t)\, ,
\label{eq:I_A_n}  \\
I_{M_i} (m)\, & \equiv\, \frac{1}{(4\pi)^2} \int_{\ln{m_{GUT}}}^{\ln{m}} dt \, g_i^2 (t) M_i^2(t)\,
  \simeq\, \frac{M_{1/2}^2}{4 b_i^{MSSM}} \left ( \frac{\alpha_i^2 (m)}{\alpha_i^2 (M_{GUT})} - 1 \right ) ,  \\
I_{S} (m)\, & \equiv\, \frac{1}{(4\pi)^2} \int_{\ln{m_{GUT}}}^{\ln{m}} dt \, g_1^2 (t) S(t)\,
  \simeq\, \frac{S(M_{GUT})}{2 b_1^{MSSM}}
  \left ( \frac{\alpha_1 (m)}{\alpha_1 (M_{GUT})} - 1 \right ) .
\end{align}
Plugging the semi-analytic expressions for $\Sigma_n(t)$ and $A_n(t)$ into Eqs.~(\ref{eq:I_Sigma_n})
and~(\ref{eq:I_A_n}), one can express the running soft scalar masses as quadratic functions of the
GUT-scale soft parameters $m_{10_i}$, $m_{\bar 5_i}$, $m_{H_u}(M_{GUT})$, $m_{H_d}(M_{GUT})$
and of the $m_{susy}$ values of $M_3$, $A_{u,c,t}$ and $A_{d,s,b}$,
with $m$-dependent coefficients that can be computed numerically.

One can further simplify Eq.~(\ref{eq:soft_masses_solution}) by writing $I_{\Sigma_n}$
as a linear combination of the differences $\Sigma_p(m) - \Sigma_p(M_{GUT})$
and of the integrals $I_{A_p}$ and $I_{M_i}$. Indeed, $\Sigma_n$ satisfies a 1-loop RGE of the form:
\begin{align}
  \frac{d}{dt} \Sigma_n(t) & \ = \ \beta^\Sigma_n(t) \, \Sigma_n(t) + \gamma^\Sigma_n(t)\, ,
\end{align}
with $\beta^\Sigma_n(t)$ equal to $12 \lambda_n^2(t) / (4\pi)^2$ for squarks and to
$8 \lambda_n^2(t) / (4\pi)^2$ for sleptons. From this and from the definition~(\ref{eq:I_Sigma_n})
of $I_{\Sigma_n}$, one immediately deduces that
\begin{align}
  I_{\Sigma_{u_i,d_i}} (m) & \ = \ \frac{\Sigma_{u_i,d_i}(m) - \Sigma_{u_i,d_i}(M_{GUT})}{12}
    - \frac{1}{12} \int_{\ln{M_{GUT}}}^{\ln{m}} dt \, \gamma^\Sigma_{u_i,d_i}(t)\, ,  \\
  I_{\Sigma_{e_i}} (m) & \ = \ \frac{\Sigma_{e_i}(m) - \Sigma_{e_i}(M_{GUT})}{8}
    - \frac{1}{8} \int_{\ln{M_{GUT}}}^{\ln{m}} dt \, \gamma^\Sigma_{e_i}(t)\, ,
\end{align}
where the integrals on the RHS can be expressed in terms of $I_{\Sigma_p}$,
$I_{A_p}$ and $I_{M_i}$, so that eventually all $I_{\Sigma_n}$'s can be written
as linear combinations of $\Sigma_p(m) - \Sigma_p(M_{GUT})$, $I_{A_p}$ and $I_{M_i}$. 
This leads to the following expressions for the running soft scalar masses:
\begin{align}
  m_\alpha^2(m)\, =\, m_\alpha^2(M_{GUT}) & + \sum_n T^\alpha_{n} \left(\Sigma_n(m) - \Sigma_n(M_{GUT}) \right)
    \nonumber \\
  & + \sum_n U^\alpha_{n}I_{A_n}(m) + \sum_{i=1}^3 V^\alpha_{i}I_{M_i}(m) + R^{\alpha}I_{S}(m)\, ,
\label{eq:soft_masses_solution2}
\end{align}
with the numerical coefficients $T^\alpha_{n}$, $U^\alpha_{n}$, $V^\alpha_{i}$
(resp. $R^\alpha$) given in Table~\ref{Table:softmasses2} (resp. Table~\ref{Table:softmasses1}).

The quantities $\Sigma_n(m)$, $I_{A_n}(m)$, $I_{M_i}(m)$ and $I_S(m)$ on the RHS of
Eq.~(\ref{eq:soft_masses_solution2}) -- hence the running soft scalar masses $m^2_\alpha(m)$ --
are quadratic functions of chosen initial parameters, with coefficients depending on the scale $m$,
$\tan\beta$ and $m_{susy}$ (and very weakly on $M_{GUT}$, $m_T$ and $m_V$).
In the above derivation, the initial parameters were taken to be the GUT-scale soft masses
$m_{10_i}$, $m_{\bar 5_i}$, $m_{H_u}(M_{GUT})$, $m_{H_d}(M_{GUT})$
and the $m_{susy}$ values of $M_3$, $A_{u,c,t}$ and $A_{d,s,b}$.
Alternatively, one can trade the 8 GUT-scale masses $m_{10_i}$, $m_{\bar 5_i}$,
$m_{H_u}(M_{GUT})$ and $m_{H_d}(M_{GUT})$ for 8 low-energy soft scalar masses
by inverting 8 of the 17 equations~(\ref{eq:soft_masses_solution2}), so as to express
all running soft parameters (soft scalar masses, gaugino masses and $A$-terms)
as functions of the $m_{susy}$ value of 15 of them.

\begin{center}	
\begin{tabular}{c||*{9}{c|}|*{3}{c|}}
$n, i$ & $t$ & $b$ & $\tau$ & $c$ & $s$ & $\mu$ & $u$ & $d$ & $e$ & $3$ & $2$ & $1$ \\
\hline
\hline
  $T^{H_u}_n$ & $\frac{1}{8}$ & & & $\frac{1}{4}$ & & & $\frac{1}{2}$ & & & & & \\
  $U^{H_u}_n$ & & $-\frac{1}{4}$ & & $-\frac{3}{4}$ & $-\frac{1}{2}$ & & $-\frac{9}{4}$ & $-1$  & & & & \\
  $V^{H_u}_i$ & & & & & & & & & & $\frac{56}{3}$ & $\frac{9}{2}$ & $\frac{11}{6}$ \\
\hline
  $T^{H_d}_n$ & $\frac{19}{1024}$ & $\frac{27}{512}$ & $\frac{9}{256}$ & $\frac{1}{128}$ & $\frac{9}{64}$
      & $\frac{3}{32}$ & $-\frac{1}{16}$ & $\frac{3}{8}$ & $\frac{1}{4}$ & & & \\
  $U^{H_d}_n$ & &$-\frac{19}{512}$ & $-\frac{27}{256}$ & $-\frac{57}{512}$ & $-\frac{139}{256}$
      & $-\frac{117}{256}$ & $-\frac{81}{512}$ & $-\frac{463}{256}$ & $-\frac{357}{256}$ & & & \\
  $V^{H_d}_i$ & & & & & & & & & & $\frac{545}{48}$ & $\frac{1263}{256}$ & $\frac{9461}{3840}$ \\
\hline
\hline
  $T^{\tilde Q_3}_n$ & $\frac{5}{36}$ & $\frac{1}{6}$ & & & & & & & & & & \\
  $U^{\tilde Q_3}_n$ & & $-\frac{5}{18}$ & $-\frac{1}{3}$ & $-\frac{5}{6}$ & $-1$ & $-\frac{1}{3}$
      & $-\frac{5}{6}$ & $-1$ & $-\frac{1}{3}$ & & & \\
  $V^{\tilde Q_3}_i$ & & & & & & & & & & $-\frac{112}{27}$ & $-\frac{7}{3}$ & $\frac{89}{135}$ \\
\hline
  $T^{\tilde u^c_3}_n$ & $\frac{1}{3}$ & & & & & & & & & & & \\
  $U^{\tilde u^c_3}_n$ & & $-\frac{2}{3}$ & & $-2$ & & & $-2$ & & & & & \\
  $V^{\tilde u^c_3}_i$ & & & & & & & & & & $-\frac{32}{9}$ & $4$ & $-\frac{44}{45}$ \\
\hline
  $T^{\tilde e^c_3}_n$ & $\frac{1}{24}$ & $-\frac{1}{4}$ & $\frac{1}{2}$ & & & & & & & & & \\
  $U^{\tilde e^c_3}_n$ & & $-\frac{1}{12}$ & $\frac{1}{2}$ & $-\frac{1}{4}$ & $-\frac{3}{2}$ & $-\frac{1}{2}$
      & $-\frac{1}{4}$ & $-\frac{3}{2}$ & $-\frac{1}{2}$ & & & \\
  $V^{\tilde e^c_3}_i$ & & & & & & & & & & $-\frac{40}{9}$ & $\frac{7}{2}$ & $-\frac{137}{90}$ \\
\hline
  $T^{\tilde L_3}_n$ & $\frac{1}{48}$ & $-\frac{1}{8}$ & $\frac{1}{4}$ & & & & & & & & & \\
  $U^{\tilde L_3}_n$ & & $-\frac{1}{24}$ & $\frac{1}{4}$ & $-\frac{1}{8}$ & $-\frac{3}{4}$ & $-\frac{1}{4}$
      & $-\frac{1}{8}$ & $-\frac{3}{4}$ & $-\frac{1}{4}$ & & & \\
  $V^{\tilde L_3}_i$ & & & & & & & & & & $-\frac{20}{9}$ & $-\frac{17}{4}$ & $\frac{79}{180}$ \\
\hline
  $T^{\tilde d^c_3}_n$ & $-\frac{1}{18}$ & $\frac{1}{3}$ & & & & & & & & & & \\
  $U^{\tilde d^c_3}_n$ & & $\frac{1}{9}$ & $-\frac{2}{3}$ & $\frac{1}{3}$ & $-2$ & $-\frac{2}{3}$
      & $\frac{1}{3}$ & $-2$ & $-\frac{2}{3}$ & & & \\
  $V^{\tilde d^c_3}_i$ & & & & & & & & & & $-\frac{128}{27}$ & $\frac{10}{3}$ & $-\frac{14}{135}$ \\
\hline
\hline
  $T^{\tilde Q_2}_n$ & $-\frac{17}{288}$ & $-\frac{1}{16}$ & $-\frac{1}{24}$ & $\frac{5}{36}$
      & $\frac{1}{6}$ & & & & & & & \\
  $U^{\tilde Q_2}_n$ & & $\frac{17}{144}$ & $\frac{1}{8}$ & $\frac{17}{48}$ & $\frac{25}{72}$
      & $-\frac{1}{8}$ & $-\frac{23}{48}$ & $-\frac{3}{8}$ & $-\frac{1}{8}$ & & & \\
  $V^{\tilde Q_2}_i$ & & & & & & & & & & $-\frac{182}{27}$ & $-\frac{103}{24}$ & $\frac{41}{1080}$ \\
\hline
  $T^{\tilde u^c_2}_n$ & $-\frac{1}{6}$ & & & $\frac{1}{3}$ & & & & & & & & \\
  $U^{\tilde u^c_2}_n$ & & $\frac{1}{3}$ & & $1$ & $-\frac{2}{3}$ & & $-1$ & & & & & \\
  $V^{\tilde u^c_2}_i$ & & & & & & & & & & $-\frac{64}{9}$ & $2$ & $-\frac{14}{9}$ \\
\hline
  $T^{\tilde e^c_2}_n$ & $-\frac{1}{192}$ & $-\frac{3}{32}$ & $-\frac{1}{16}$ & $\frac{1}{24}$
      & $-\frac{1}{4}$ & $\frac{1}{2}$ & & & & & & \\
  $U^{\tilde e^c_2}_n$ & & $\frac{1}{96}$ & $\frac{3}{16}$ & $\frac{1}{32}$ & $\frac{41}{48}$
      & $\frac{13}{16}$ & $-\frac{7}{32}$ & $-\frac{9}{16}$ & $-\frac{3}{16}$ & & & \\
  $V^{\tilde e^c_2}_i$ & & & & & & & & & & $-\frac{59}{9}$ & $\frac{25}{16}$ & $-\frac{1559}{720}$ \\
\hline
  $T^{\tilde L_2}_n$ & $-\frac{1}{384}$ & $-\frac{3}{64}$ & $-\frac{1}{32}$ & $\frac{1}{48}$
      & $-\frac{1}{8}$ & $\frac{1}{4}$ & & & & & & \\
  $U^{\tilde L_2}_n$ & & $\frac{1}{192}$ & $\frac{3}{32}$ & $\frac{1}{64}$ & $\frac{41}{96}$
      & $\frac{13}{32}$ & $-\frac{7}{64}$ & $-\frac{9}{32}$ & $-\frac{3}{32}$ & & & \\
  $V^{\tilde L_2}_i$ & & & & & & & & & & $-\frac{59}{18}$ & $-\frac{167}{32}$ & $\frac{169}{1440}$ \\
\hline
  $T^{\tilde d^c_2}_n$ & $\frac{7}{144}$ & $-\frac{1}{8}$ & $-\frac{1}{12}$ & $-\frac{1}{18}$
      & $\frac{1}{3}$ & & & & & & & \\
  $U^{\tilde d^c_2}_n$ & & $-\frac{7}{72}$ & $\frac{1}{4}$ & $-\frac{7}{24}$ & $\frac{49}{36}$
      & $-\frac{1}{4}$ & $\frac{1}{24}$ & $-\frac{3}{4}$ & $-\frac{1}{4}$ & & & \\
  $V^{\tilde d^c_2}_i$ & & & & & & & & & & $-\frac{172}{27}$ & $\frac{17}{12}$ & $-\frac{83}{108}$ \\
\hline
\end{tabular}
\begin{tabular}{c||*{9}{c|}|*{3}{c|}}
$n, i$ & $t$ & $b$ & $\tau$ & $c$ & $s$ & $\mu$ & $u$ & $d$ & $e$ & $3$ & $2$ & $1$ \\
\hline
\hline
  $T^{\tilde Q_1}_n$ & $-\frac{83}{2304}$ & $-\frac{3}{128}$ & $-\frac{1}{64}$ & $-\frac{17}{288}$
      & $-\frac{1}{16}$ & $-\frac{1}{24}$ & $\frac{5}{36}$ & $\frac{1}{6}$ & & & & \\
  $U^{\tilde Q_1}_n$ & & $\frac{83}{1152}$ & $\frac{3}{64}$ & $\frac{83}{384}$ & $\frac{203}{576}$
      & $\frac{13}{64}$ & $\frac{73}{128}$ & $\frac{335}{576}$ & $-\frac{3}{64}$ & & & \\
  $V^{\tilde Q_1}_i$ & & & & & & & & & & $-\frac{865}{108}$ & $-\frac{997}{192}$ & $-\frac{2101}{8640}$ \\
\hline
  $T^{\tilde u^c_1}_n$ & $-\frac{1}{12}$ & & & $-\frac{1}{6}$ & & & $\frac{1}{3}$ & & & & & \\
  $U^{\tilde u^c_1}_n$ & & $\frac{1}{6}$ & & $\frac{1}{2}$ & $\frac{1}{3}$ & & $\frac{3}{2}$
      & $-\frac{2}{3}$ & & & & \\
  $V^{\tilde u^c_1}_i$ & & & & & & & & & & $-\frac{80}{9}$ & $1$ & $-\frac{83}{45}$ \\
\hline
  $T^{\tilde e^c_1}_n$ & $-\frac{19}{1536}$ & $-\frac{9}{256}$ & $-\frac{3}{128}$ & $-\frac{1}{192}$
      & $-\frac{3}{32}$ & $-\frac{1}{16}$ & $\frac{1}{24}$ & $-\frac{1}{4}$ & $\frac{1}{2}$ & & & \\
  $U^{\tilde e^c_1}_n$ & & $\frac{19}{768}$ & $\frac{9}{128}$ & $\frac{19}{256}$ & $\frac{139}{384}$
      & $\frac{39}{128}$ & $\frac{27}{256}$ & $\frac{463}{384}$ & $\frac{119}{128}$ & & & \\
  $V^{\tilde e^c_1}_i$ & & & & & & & & & & $-\frac{545}{72}$ & $\frac{91}{128}$ & $-\frac{14069}{5760}$ \\
\hline
  $T^{\tilde L_1}_n$ & $-\frac{19}{3072}$ & $-\frac{9}{512}$ & $-\frac{3}{256}$ & $-\frac{1}{384}$
      & $-\frac{3}{64}$ & $-\frac{1}{32}$ & $\frac{1}{48}$ & $-\frac{1}{8}$ & $\frac{1}{4}$ & & & \\
  $U^{\tilde L_1}_n$ & & $\frac{19}{1536}$ & $\frac{9}{256}$ & $\frac{19}{512}$ & $\frac{139}{768}$
      & $\frac{39}{256}$ & $\frac{27}{512}$ & $\frac{463}{768}$ & $\frac{119}{256}$ & & & \\
  $V^{\tilde L_1}_i$ & & & & & & & & & & $-\frac{545}{144}$ & $-\frac{1445}{256}$ & $-\frac{49}{2304}$ \\
\hline
  $T^{\tilde d^c_1}_n$ & $\frac{13}{1152}$ & $-\frac{3}{64}$ & $-\frac{1}{32}$ & $\frac{7}{144}$
      & $-\frac{1}{8}$ & $-\frac{1}{12}$ & $-\frac{1}{18}$ & $\frac{1}{3}$ & & & & \\
  $U^{\tilde d^c_1}_n$ & & $-\frac{13}{576}$ & $\frac{3}{32}$ & $-\frac{13}{192}$ & $\frac{107}{288}$
      & $\frac{13}{32}$ & $-\frac{23}{64}$ & $\frac{527}{288}$ & $-\frac{3}{32}$ & & & \\
  $V^{\tilde d^c_1}_i$ & & & & & & & & & & $-\frac{385}{54}$ & $\frac{59}{96}$ & $-\frac{4501}{4320}$ \\
\hline
\end{tabular}
\captionof{table}{\label{Table:softmasses2}The non-vanishing coefficients $T^\alpha_{n}$,
$U^\alpha_n$ and $V^{\alpha}_i$ appearing in Eq.~(\ref{eq:soft_masses_solution2}).}
\end{center}
%

\section{\label{app:pdk}Proton lifetime computation}

In this appendix, we outline the computation of the higgsino-mediated (D=5)
proton decay rate~\cite{ENR82,ACN85,HMY93,Nath06} in the minimal
renormalizable supersymmetric SU(5) model. Integrating out the heavy colour
triplet components of the $5_H$ and $\bar 5_H$ superfields generates
the following D=5 operators:
\beq
  \left. \frac 1 2\, \frac{\kappa_{ijkl}}{m_T}\, Q_i Q_j Q_k L_l\ \right|_{\theta^2}\,
    +\ \left. \frac 1 2\, \frac{\kappa'_{ijkl}}{m_T}\,
    u^c_i u^c_j d^c_k e^c_l\ \right|_{\theta^2}\, +\ \mbox{h.c.}\, ,
\label{eq:D=5_ops}
\eeq
where the parameters $\kappa_{ijkl}$ and $\kappa'_{ijkl}$ can be expressed
in terms of the $SU(5)$ Yukawa couplings $\Lambda^{10}_{ij}$ and $\Lambda^{\bar 5}_{ij}$
defined in Eq.~(\ref{eq:W_Y}):
\beq
  \kappa_{ijkl}\ =\ \Lambda^{10}_{ij} \Lambda^{\bar 5}_{lk}\, , \qquad
  \kappa'_{ijkl}\ =\ \Lambda^{10}_{il} \Lambda^{\bar 5}_{kj} - \Lambda^{10}_{jl} \Lambda^{\bar 5}_{ki}\, ,
\eeq
and the contraction of gauge indices is understood.
Since colour invariance implies $\kappa_{iiil} = 0$ and $\kappa'_{iikl} = 0$,
the dominant proton decay modes arising from the above operators
involve a kaon, and in practice $p \rightarrow K^+ \bar \nu$ dominates.
The corresponding amplitude is obtained by ``dressing'' the D=5 operators
of Eq.~(\ref{eq:D=5_ops}) with gaugino/higgsino loops.
Over the region of the parameter space considered in this paper,
the dominant contribution comes from the wino dressing of the $QQQL$
operator\footnote{The gluino dressing of the $QQQL$ operator can be neglected
as the mass difference between the first two generations of squarks is small,
and the charged higgsino dressing of the $u^c u^c d^c e^c$ operator
is significant only for large values of $\tan \beta$. Although we consider
large A-terms, the left-right sfermion mixing remains small
($A v / m^2_{\rm sfermions} \ll 1$), hence the wino dressing of the
$u^c u^c d^c e^c$ operator is not relevant either.},
and the decay rate reads:
\beq
  \Gamma (p \rightarrow K^+ \bar \nu)\
    =\ \frac{m_p}{32 \pi} \left( 1 - \frac{m^2_{K^+}}{m^2_p} \right)^2\,
    \sum_l\, \left| W^{112}_0\, C_{112l} + W^{121}_0\,  C_{121l} \right|^2\, ,
\label{eq:pdecay_D=5}
\eeq
where $C_{ijkl} (\mu)$ is the Wilson coefficient of the 4-fermion operator $u_i d_j d_k \nu_l$
and the hadronic parameter $W^{ijk}_0$ is defined by
$\langle K^+ | (u_{Li} d_{Lj}) d_{Lk} | p \rangle \simeq W^{ijk}_0 P_L u_p$,
where $u_p$ is the proton spinor. The lattice computation of  Ref.~\cite{Aoki13}
gives $W^{112}_0 (\mu = 2 \GeV) = (0.111 \pm 0.027) \GeV^2$
and $W^{121}_0 (\mu = 2 \GeV) = (0.036 \pm 0.014) \GeV^2$,
with statistical and systematic errors added in quadrature.
In order to minimize the dependence of the proton decay rate on
the renormalization scale $\mu$, the $C_{ijkl}$'s are evaluated at the scale
$\mu = 2 \GeV$ rather than at $\mu = m_p$.

The Wilson coefficients $C_{ijkl}$ are computed at the matching scale $m_{susy}$
at which supersymmetric partners are integrated out
by ``dressing'' the $QQQL$ operators with loops containing a wino:
\beq
  C_{ijkl} (m_{susy})\ =\ \sum_{m,n,p}\, V_{mj} V_{nk}\, \frac{\tilde \kappa_{pmnl}}{m_T}
    \left( \sum_q V^*_{iq} V_{pq}\, I (\tilde W, {\tilde d}_{Lq},\tilde u_{Lm})
    + \delta_{pi}\, I (\tilde W, \tilde u_{Ln}, \tilde e_{Ll}) \right) ,
\label{eq:Cijkl}
\eeq
in which $V$ is the CKM matrix, the parameters
$\tilde \kappa_{pmnl} \equiv \kappa_{pmnl} - \kappa_{nmpl}$
are expressed in the super-CKM basis in which the up quarks are mass eigenstates,
and the possible misalignment between the fermion and the sfermion mass eigenstate
bases is neglected.
In Eq.~(\ref{eq:Cijkl}), the CKM matrix entries and the $\tilde \kappa_{pmnl}$'s
are renormalized at the scale $m_{susy}$. The loop function $I(\tilde W, a, b)$ is given by:
\beq
  I(\tilde W, a, b)\, =\, \frac{\alpha_2}{4\pi}\, m_{\tilde w}\, I_3\! \left( m^2_a, m^2_b, m^2_{\tilde w} \right) ,
\eeq
where the function $I_3(x,y,z)$ has been defined in Eq.~(\ref{eq:I3}).
For equal sfermion masses ($m_a = m_b \equiv m_{susy}$), $I(\tilde W, a, b)$ reads:
\beq
  I(\tilde W, a, b)\, =\, \frac{\alpha_2}{4\pi}\ \frac{1}{m_{susy}}\, g \left( \frac{m_{susy}^2}{m^2_{\tilde w}} \right) ,
  \qquad g(x)\, =\, \frac{\sqrt{x}\, (x - 1 - \ln x)}{(x-1)^2}\ .
\eeq
The parameters $\tilde \kappa_{ijkl}$ depend on the details of the Yukawa sector
of the Grand Unified Theory. In the minimal supersymmetric $SU(5)$ model,
they are given by (neglecting the running between $M_{GUT}$ and the triplet mass scale):
\beq
  \tilde \kappa_{ijkl} (m_T)\, =\, \lambda_{u_i} e^{i \phi_i} \delta_{ij} V^*_{kl} \lambda_{e_l}
    - \lambda_{u_j} e^{i \phi_j} \delta_{jk} V^*_{il} \lambda_{e_l}\, ,
\label{eq:kappa_tilde}
\eeq
where the Yukawa couplings and the CKM matrix are evaluated at the scale
$m_T$, and the $\phi_i$ are high-energy phases satisfying the constraint
$\phi_1 + \phi_2 + \phi_3 = 0$. These parameters must then be evolved
down to the scale $m_{susy}$ by solving the appropriate RGEs.
Neglecting the Yukawa couplings, the running simply amounts to an overall rescaling: 
\begin{equation}
  \tilde \kappa_{ijkl} (m_{susy})\, =\, A_{SD}\, \tilde \kappa_{ijkl} (m_T)\, ,
\end{equation}
where in the minimal renormalizable supersymmetric $SU(5)$ model:
\beq
  A_{SD}\, =\, \left( \frac{\alpha_1 (m_{susy})}{\alpha_1 (m_T)} \right)^{\! - \frac{1}{33}}\!
     \left( \frac{\alpha_2 (m_{susy})}{\alpha_2 (m_3)} \right)^{\! -3}\!
     \left( \frac{\alpha_2 (m_3)}{\alpha_2 (m_T)} \right)^{\! -1}\!
     \left( \frac{\alpha_3 (m_{susy})}{\alpha_3 (m_T)} \right)^{\! +\frac{4}{3}}\!
     \left( \frac{m_T}{m_8} \right)^{\! \frac{2 \alpha_3(m_T)}{\pi}}\!\!\! .
\label{eq:A_SD_SU5}
\eeq
Finally, the Wilson coefficients $C_{112l} (m_{susy})$ and $C_{121l} (m_{susy})$
must be renormalized down to the scale $\mu = 2 \GeV$ before being inserted into
Eq.~(\ref{eq:pdecay_D=5}):
\bea
  C_{112l} (\mu = 2 \GeV) & = & A^{112}_{LD}\ C_{112l} (m_{susy})\, ,  \\
  C_{121l} (\mu = 2 \GeV) & = & A^{121}_{LD}\ C_{121l} (m_{susy})\, ,
\eea
where $A^{112}_{LD}$ and $A^{121}_{LD}$ are renormalization factors given by
(using the formulae of Ref.~\cite{Abbott80}):
\bea
  & A^{112}_{LD}\, =\, \frac{1}{3}
  \left[\, 2 + \left( \frac{\alpha_2 (m_Z)}{\alpha_2 (m_{susy})} \right)^{-36/19}\, \right]\,
   A_{LD}\, , &  \\
  & A^{121}_{LD}\, =\, \frac{1}{3}
  \left[\, 4 - \left( \frac{\alpha_2 (m_Z)}{\alpha_2 (m_{susy})} \right)^{-36/19}\, \right]\,
   A_{LD}\, , &  \\
  & A_{LD}\, =\, \left( \frac{\alpha_3 (\mu = 2 \GeV)}{\alpha_3 (m_b)} \right)^{6/25}\!
   \left( \frac{\alpha_3 (m_b)}{\alpha_3 (m_t)} \right)^{6/23}\!
   \left( \frac{\alpha_3 (m_t)}{\alpha_3 (m_{susy})} \right)^{6/21}\!
   \left( \frac{\alpha_2 (m_Z)}{\alpha_2 (m_{susy})} \right)^{45/19}\!
     \left( \frac{\alpha_1 (m_Z)}{\alpha_1 (m_{susy})} \right)^{-1/41}\!\! . \ \ \ \ &
\label{eq:A_LD}
\eea
The fact that $A^{112}_{LD} \neq A^{121}_{LD}$ is due to
the RG-induced mixing between 4-fermion operators with different flavour
indices; when this (numerically small) mixing is neglected,
$A^{112}_{LD} = A^{121}_{LD} = A_{LD}$.

\section{\label{app:UFB}UFB constraints}

In this appendix, we derive the constraints on the model parameters associated with the absence
of directions along which the scalar potential is unbounded from below. In doing this we will follow 
Ref.~\cite{Casas:1995pd}. 

\subsection{UFB-1}

The absence of this direction (where $H_u=H_d$ to cancel the D-terms
and all other vevs are vanishing) requires
\beq
m_{H_u}^2+m_{H_d}^2+2|\mu|^2\, \geq\, 2B\, ,
\eeq
which can be rewritten (for finite $m_Z$) as
\beq
\label{UFB1}
m_{H_d}^2-m_{H_u}^2\, \geq\, m_Z^2\frac{\tan^2{\beta}-1}{\tan^2{\beta}+1}\, ,
\eeq
where 
\begin{align}
\label{mufinitemz}
|\mu|^2\,&=\, \frac{m_{H_d}^2-m_{H_u}^2\tan^2{\beta}}{\tan^2{\beta}-1}-\frac{m_Z^2}{2}\ ,  \\
\label{Bfinitemz}
B\,&=\, \frac{(m_{H_d}^2-m_{H_u}^2)\tan{\beta}}{\tan^2{\beta}-1}-\frac{m_Z^2\tan{\beta}}{\tan^2{\beta}+1}\ .
\end{align}
Eq.~(\ref{UFB1}) is equivalent to 
\beq
B\, \geq\, 0\, ,
\eeq
which is automatically satisfied by assumption. As a cross-check, one can notice that
neglecting $m_Z$ (since we typically consider situations where $m_{H_{d,u}}^2\gg m_Z^2$),
the constraint~(\ref{UFB1}) reduces to:
\beq
m_{H_d}^2\, \geq\,  m_{H_u}^2 ,
\eeq
which is again automatic since $|\mu|^2 \geq 0$.

\subsection{UFB-2}

We now allow nonzero vevs for $H_u$, $H_d$ and $\tilde\nu$ only:
\begin{align}
V\, =\, \left(m_{H_d}^2+|\mu|^2\right)\! H_d^2\, + \left(m_{H_u}^2+|\mu|^2\right)\! H_u^2\,
  &-\, 2B H_dH_u\, +\, m_{\tilde L}^2\tilde\nu^2  \nonumber \\
&+\, \frac{g'^2+g_2^2}{8} \left(H_u^2-H_d^2-\tilde\nu^2\right)^2 .
\label{UFB2}
\end{align}
In the limit $m_Z\to0$, one can recast Eq.~(\ref{UFB2}) into a sum of three positive terms:
\beq
\label{VUFB2}
V\, =\, \frac{\left(m_{H_d}^2-m_{H_u}^2\right)\left(H_d\tan{\beta}-H_u\right)^2}{\tan^2{\beta}-1}\,
+\, m_{\tilde L}^2\tilde\nu^2\, +\, \frac{g'^2+g_2^2}{8} \left(H_u^2-H_d^2-\tilde\nu^2\right)^2 ,
\eeq
hence there cannot be any minimum lower than the electroweak vacuum $V=0$.
For finite $m_Z$ only the first term in Eq.~(\ref{VUFB2}) is modified, but this does not affect
the minimization of $V$ with respect to $\tilde\nu$:
\beq
\tilde\nu\, \left[\, \tilde\nu^2-\left(H_u^2-H_d^2-\frac{4m_{\tilde L}^2}{g'^2+g_2^2}\right)\, \right]\, =\, 0\, .
\label{eq:min_nu}
\eeq
The solution $\tilde\nu=0$ brings us back to the UFB-1 case, so we are only interested
in the non-trivial $\tilde\nu \neq 0$ solution, which is however consistent only for $H_u^2>H_d^2$.
One can check both analytically (using for example the command Reduce in Mathematica)
and numerically that in this case there is an UFB direction only if 
\beq
\label{m2Hvsm2Z}
m_{H_d}^2-m_{H_u}^2<m_Z^2
\eeq
(plus some extra constraints), or
\begin{align}
\label{UFB2a}
m_{H_d}^2-m_{H_u}^2\, &\geq\, m_Z^2\, ,  \\
\label{UFB2b}
m_{\tilde L}^2\, & \leq\, \frac{m_Z^2}{2} \ \frac{\tan^2{\beta}-1}{\tan^2{\beta}+1}
\end{align}
(plus some extra constraints). Since none of these conditions is satisfied in our analysis of the minimal 
renormalizable supersymmetric SU(5) model, we can conclude that there are no UFB-2 directions.

Naively one could think that this conclusion may change if instead of $\tilde\nu \neq 0$
one would allow a nonzero $\tilde e=\tilde e^c$, because it contributes with the opposite
sign to the D-term. This is however not the case: while the non-trivial solution $\tilde e=\tilde e^c \neq 0$
is now consistent for $H^2_u<H^2_d$,  there is also a sign change in Eq.~(\ref{eq:min_nu}).
Hence also in this case  there are no dangerous UFB-2 direction.

\subsection{UFB-3}

One can avoid the UFB-3 direction (with nonzero vevs for $H_u$, $\tilde\nu$ and $\tilde d=\tilde d^c$)
if the conditions 
\begin{align}
m_{H_u}^2 +m_{\tilde L}^2& \, > \, 0
\end{align}
are satisfied for any of the three slepton generations.

\section{\label{app:CCB}Decays into CCB vacua}

In this appendix, we describe a method to estimate the lifetime
of the electroweak vacuum in the presence of charge and colour breaking (CCB)
vacua induced by large A-terms from any sector and generation. To this end,
we will approximate the minimal value of the bounce action in
Euclidean space~\cite{Coleman:1977py,Callan:1977pt,Coleman:1977th}:
\beq
S\, =\, 2\pi^2\int_0^\infty dr r^3 \left({\cal L}_{kin}+V\right)(\phi_i(r))
\eeq
for the solution of the equations of motion
\beq
\phi_i''(r)+\frac{3}{r}\phi_i'(r)\, =\, \frac{\partial V}{\partial\phi_i}\ ,
\label{eq:eom_phi}
\eeq
subject to the boundary conditions (describing fields sitting in the false vacuum at $r=\infty$
and tending to a lower minimum at $r=0$):
\beq
\phi_i(\infty)=\phi_i^{\infty}\, , \qquad \phi_i'(0)=0\, .
\eeq
The general problem can be solved only numerically~\cite{Claudson:1983et,Kusenko:1995jv,
Kusenko:1996jn,Cline:1999wi,Konstandin:2006nd,Park:2010rh,Wainwright:2011kj}. 
We will simplify it by minimizing the bounce action along a constant direction
in field space~\cite{Dasgupta:1996pz,Dasgupta:1996qu,Borzumati:1999sp},
which effectively reduces the problem to a single differential equation. This approximation
provides an upper bound on the minimal value of the correct multi-field bounce 
action~\cite{Coleman:1987rm,Dasgupta:1996qu}. This means that the points of the parameter
space ruled out in this way are indeed characterized by a too short vacuum lifetime,
but the method retains points that after a more careful and complete analysis
would be rejected.

What we have to do is therefore to minimize an Euclidean action of the form:
\beq
\label{S}
S\, =\, 2\pi^2\int_0^\infty dr\;r^3\;\left(\frac{z}{2}\phi'^2+M_2^2\phi^2-M_3\phi^3+\lambda\phi^4\right) ,
\eeq
where the field $\phi(r)$ parametrizes a direction in field space, and check whether its minimal 
value is large enough for the lifetime of the electroweak vacuum to be larger than the age 
of the universe, namely
\beq
S\, >\, 400\, .
\label{eq:metastability}
\eeq
This is done by solving the equation of motion~(\ref{eq:eom_phi}) and plugging the solution
back into Eq.~(\ref{S}).  A semi-analytical approximate solution for the action can be found
in Ref.~\cite{Sarid:1998sn} (see also Ref.~\cite{Dine:1992wr}):
\beq
S\, =\, \frac{z^2M_2^2}{M_3^2}\hat S(\kappa)\, ,
\label{eq:S_solution}
\eeq
where 
\beq
\label{Shat}
\hat S(0)\, \approx\, 45.4\, ,
\eeq
and 
\beq
\kappa\, =\, \lambda\frac{M_2^2}{M_3^2}\, .
\eeq
$\kappa<0$ indicates that the potential is unbounded from below, while $\kappa\geq1/4$ 
ensures that the minimum at $\phi=0$ is absolutely stable. Hence we only have to check
that the metastability constraint~(\ref{eq:metastability}) is satisfied for $0\leq\kappa<1/4$.
In this range of $\kappa$ values, one has:
\beq
\hat S(\kappa)\, \geq\, \hat S(0)\, \approx\, 45.4\, .
\eeq

\subsection{An example of CCB constraint}

As a concrete example, let us first consider the vacuum decay to CCB minima induced
by a large A-term in the down squark sector~\cite{Casas:1995pd}; then we will generalize
the discussion to other A-terms. The relevant Lagrangian is: 
\begin{align}
{\cal L}_{kin}\, &=\, |h_u'(r)|^2+|h_d'(r)|^2+|\tilde d'(r)|^2+|(\tilde d^c)'(r)|^2+|\tilde e'(r)|^2+|(\tilde e^c)'(r)|^2\, ,
\nonumber\\
V\, &= \, m_{H_u}^2|h_u|^2+m_{H_d}^2|h_d|^2+m_{\tilde Q}^2|\tilde d|^2+m_{\tilde d}^2|\tilde d^c|^2+
m_{\tilde L}^2|\tilde e|^2+m_{\tilde e}^2|\tilde e^c|^2 \nonumber\\
& \quad +\left(A_d\tilde d^c\tilde dh_d-\lambda_d\mu^*\tilde d^c\tilde dh_u^*
-Bh_uh_d + \mbox{h.c.} \right)  \nonumber \\
& \quad +|\mu|^2\left(|h_u|^2+|h_d|^2\right)+|\lambda_d|^2\left(|\tilde d|^2|\tilde d^c|^2+|\tilde d^c|^2|h_d|^2
+|h_d|^2|\tilde d|^2\right)\nonumber\\
& \quad +\frac{3g_1^2/5}{8}\left(-|h_d|^2+|h_u|^2+\frac{1}{3}|\tilde d|^2
+\frac{2}{3}|\tilde d^c|^2-|\tilde e|^2+2|\tilde e^c|^2\right)^2 \nonumber\\
& \quad +\frac{g_2^2}{8}\left(|h_d|^2-|h_u|^2-|\tilde d|^2-|\tilde e|^2\right)^2
+\frac{g_3^2}{6}\left(|\tilde d|^2-|\tilde d^c|^2\right)^2 ,
\end{align}
where $h_{u,d}$, $\tilde d\, (\tilde d^c)$ and $\tilde e\ (\tilde e^c)$ stand for the neutral
components of the MSSM Higgs doublets, the scalar partners of the left(right)-handed
down-type quarks and of the left(right)-handed charged leptons, respectively.
Generation indices are omitted, and all mixing angles are neglected.
Finally, the parameters $\mu$ and $B$ are determined by electroweak symmetry breaking,
i.e. in the limit $m_Z\to0$:
\begin{align}
\label{mu}
|\mu|^2\, &=\, \frac{m_{H_d}^2-m_{H_u}^2\tan^2{\beta}}{\tan^2{\beta}-1}\ ,  \\
\label{B}
B\, &=\, \frac{(m_{H_d}^2-m_{H_u}^2)\tan{\beta}}{\tan^2{\beta}-1}\ .
\end{align}
From now on, all quantities are assumed to be real. This is equivalent to assuming that 
the relative phase between $\lambda_d$ and $\mu$ equals (modulo $\pi$) the relative
phase between $A_d$ and $B$ (a condition which may be realized in some scenarios
of supersymmetry breaking), because in this special case all phases can be absorbed
in the fields involved.

For all CCB constraints except the one associated with $A_t$ (which will be discussed separately) 
and the one associated with $A_b$ in the large $\tan{\beta}$ regime (which will not be
considered because, in the minimal renormalizable supersymmetric SU(5) model, large values
of $\tan \beta$ are excluded by a combination of constraints, see Section~\ref{sec:results}) 
one has $\lambda_{u,d,e}^2\ll g_{1,2,3}^2$, therefore one can impose the constraint
of vanishing SU(3), SU(2) and U(1) $D$-terms~\cite{Casas:1995pd}. 

As explained at the beginning of the appendix, we will search for solutions to the field
equations of motion of the form:
\beq
(h_u, h_d,\tilde d,\tilde d^c, \tilde e,\tilde e^c)(r)\, =\, (H_u, H_d, d, d, e, e)\;\phi(r)\, ,
\eeq
with $H_{u}$, $H_{d}$, $d$, $e$ constant, where we have in addition
imposed the SU(3) D-term constraint
(which also implies that $d$ and $d^c$ point in the same SU(3) direction)
\beq
d^c=d
\eeq
(the alternative sign choice $d^c=-d$ would just flip the sign of $M_3$, see Eq.~(\ref{eq:M3})
below, without affecting the action~(\ref{eq:S_solution}), which depends on $M_3^2$), as well as 
the compatibility of the U(1) and SU(2) D-term constraints:
\beq
e^c=e\, .
\eeq
Finally, the vanishing of the SU(2) D-term:
\beq
\label{Dtermcase12}
-d^2-e^2+H_d^2-H_u^2\, =\, 0\, ,
\eeq
will be considered as a constraint in the following.

With these assumptions, the coefficients of the bounce action~(\ref{S}) can be expressed as:
\begin{align}
\label{M22}
M_2^2\, &=\, \left(m_{\tilde Q}^2+ m_{\tilde d}^2\right)d^2
+\left(m_{\tilde L}^2+ m_{\tilde e}^2\right)e^2  \non \\
& \quad + \left(m_{H_u}^2+|\mu|^2\right)H_u^2+\left(m_{H_d}^2+|\mu|^2\right)H_d^2-2BH_u H_d\, ,  \\
M_3\, &=\, -2d^2(A_dH_d-\lambda_d\mu H_u)\, ,
\label{eq:M3}  \\
z\, &=\, 2 \left(2d^2+2e^2+H_u^2+H_d^2\right) ,  \\
\lambda\, &=\, \lambda_d^2\, d^2(d^2+2H_d^2)\, .
\end {align}
Notice that we chose $B>0$ (see Eq.~(\ref{B}); this amounts to fix the sign of $H_u H_d$). 
Without loss of generality, we can also choose
\beq
\lambda_d\, >\, 0\, ,  \qquad  H_u\, >\, 0\, ,
\eeq
while the signs of $A_d$, $\mu$ and $H_d$ can in principle be arbitrary. For a given point
in parameter space (for which the signs of $A_d$ and $\mu$ are fixed), one needs to
consider both signs of $H_d$.

\subsection{General CCB constraints}

Let us now generalize the previous discussion to all CCB constraints (except for the one
associated with $A_t$). The coefficients of the bounce action~(\ref{S}) can be written
compactly as:
\begin{align}
M_2^2(\psi,\bar\psi,H,\bar H)\, &=\, M_\psi^2\psi^2+M_{\bar\psi}^2\bar\psi^2  \nonumber \\
& \quad +\left(m_{H}^2+|\mu|^2\right)H^2+\left(m_{\bar H}^2+|\mu|^2\right)\bar H^2-2BH \bar H\, ,  \\
M_3(\psi,H,\bar H)\, &=\, -2\psi^2(A_\psi H-\lambda_\psi \mu \bar H)\, ,  \\
z(\psi,\bar\psi,H,\bar H, c_z)\, &=\, 2\left(2\psi^2+ c_z\bar\psi^2+H^2+\bar H^2\right) ,  \\
\lambda(\psi,H)\, &=\, \lambda_\psi^2\psi^2(\psi^2+2H^2)\, ,
\end {align}
and the D-term constraint is given by:
\beq
\label{Dterm}
0\, =\, -\psi^2+(3-2c)\bar \psi^2+ H^2-\bar H^2\, ,
\eeq
where the 6 different possibilities are summarized in Table~\ref{cases}
(the example considered previously was case 2).
\begin{center}
$
\begin{array}{c||c|c|c|c||c|c|c|c|c|c|}
& \psi & \bar \psi & H & \bar H & M_\psi^2 & M_{\bar \psi}^2 & m_H^2 & m_{\bar H}^2 & c_z & c\\ 
\hline
\hline
1 &d\;(=d^c) & \nu & H_d & H_u & m_{\tilde Q}^2+m_{\tilde d}^2 & m_{\tilde L}^2 & m_{H_d}^2 & m_{H_u}^2 & 1 & 1\\ 
2 & d\;(=d^c) & e\;(=e^c) & H_d & H_u &  m_{\tilde Q}^2+m_{\tilde d}^2 & m_{\tilde L}^2 +m_{\tilde e}^2 & m_{H_d}^2 & m_{H_u}^2 & 2 & 2\\ 
3 & u\;(=u^c) & \nu & H_u & H_d & m_{\tilde Q}^2+m_{\tilde u}^2 & m_{\tilde L}^2 & m_{H_u}^2 & m_{H_d}^2 & 1 & 2\\ 
4 & u\;(=u^c) & e\;(=e^c) & H_u & H_d & m_{\tilde Q}^2+m_{\tilde u}^2 & m_{\tilde L}^2 +m_{\tilde e}^2 & m_{H_u}^2 & m_{H_d}^2 & 2 & 1\\ 
5 & e\;(=e^c) & u\;(=u^c) & H_d & H_u &  m_{\tilde L}^2+m_{\tilde e}^2 & m_{\tilde Q}^2 +m_{\tilde u}^2 & m_{H_d}^2 & m_{H_u}^2 & 2 & 1\\ 
6 & e\;(=e^c) & d\;(=d^c) & H_d & H_u & m_{\tilde L}^2+m_{\tilde e}^2 & m_{\tilde Q}^2 +m_{\tilde d}^2 & m_{H_d}^2 & m_{H_u}^2 & 2 & 2\\ 
\hline
\end{array}
$
\captionof{table}{\label{cases}The six different cases of CCB constraints considered.
The D-term constraints are given in parenthesis. Whenever $d$ and $d^c$ ($u$ and $u^c$)
are involved, they point in the same SU(3)$_C$ direction.}
\end{center}

For each case, there are several subcases to be checked according to which
generation indices are carried by the $\psi$ and $\bar \psi$ fields. For a given
generation $\psi_i$ (the 2 $\psi$ fields belong to the same generation),
one must consider all generation choices for the $\bar \psi$
fields that do not give a substantial contribution to the F-terms due to the size
of the corresponding Yukawa couplings
(otherwise there would be no CCB minimum lower than the electroweak vacuum). 
Namely,
\begin{itemize}
\item 
when a single $\bar \psi$ field is involved  (Cases 1 and 3), all generation indices
should be considered;
\item
when two $\bar \psi$ fields are involved (Cases 2, 4, 5 and 6), all combinations
of two different generation indices should be considered;
\item
the cases of two $\bar \psi$ fields with the same generation index should also be
considered when the Yukawa coupling of the corresponding lepton is much smaller
than $\lambda_\psi$ (i.e. one needs to consider only generations of $\bar \psi$'s
whose fermionic partners are lighter than the fermionic partner of $\psi$).
\end{itemize}
In practice, it is enough to check generations of the $\bar\psi$ fields leading to the
smallest value of $M_{\bar\psi}^2$, since the bounce action increases with
$M_{\bar\psi}^2$. There are thus 16 possible cases: 3 generations for case 1 with 
$\psi=(d, s, b)$, 3 generations for case 2 with $\psi=(d, s, b)$, 2 generations for case 3 with $\psi=(u, c)$, 
2 generations for case 4 with $\psi=(u, c)$, 3 generations for case 5 with $\psi=(e, \mu, \tau)$ and 
3 generations for case 6 with $\psi=(e, \mu, \tau)$.

Finally, starting from the semi-analytic approximation~(\ref{eq:S_solution}), it is easy to show that:
\begin{align}
S\, &=\, 4 \hat S(\kappa) \left[\frac{\lambda}{\lambda_\psi^2}+2\psi^2\left((3+ c_z-2 c)\bar\psi^2+H^2\right)
  + \left((3+ c_z-2 c)\bar\psi^2+2H^2\right)^2\, \right]\frac{M_2^2}{M_3^2}  \nonumber \\
&\geq\, 4 \hat S(\kappa) \frac{\kappa}{\lambda_\psi^2}\ ,
\end{align}
which is very large unless $\kappa$ is very small. Thus only very small values of $\kappa$
may lead to a short vacuum lifetime,
and we can approximate $\hat S(\kappa) \simeq \hat S(0) \approx 45.4$. 
To a good approximation, the quantity that we have to minimize is therefore
\beq
\label{Sapprox}
{\cal S}\, \equiv\, \frac{S}{\hat S(0)}\, =\, \frac{z^2M_2^2}{M_3^2}(\psi,\bar\psi,H,\bar H, c, c_z)\, ,
\eeq
subject to the D-term constraint~(\ref{Dterm}), and we must check whether 
\beq
{\cal S}^{min}\, \gtrsim\, \frac{400}{45.4}\, \approx\, 9\, ,
\label{Smin}
\eeq
hereafter referred to as the metastability constraint. This condition must be satisfied
for all six cases in Table~\ref{cases} to be able to conclude (within our approximate
method) that the lifetime of the electroweak vacuum is longer than the age of the universe.

Notice that one needs to minimize $\cal S$ with respect to two variables only. Indeed,
one of the variables $\psi,\bar\psi,H,\bar H$ is fixed by the D-term constraint~(\ref{Dterm}),
and the quantity~(\ref{Sapprox}) only depends on ratios of fields. 
The minimization is performed numerically. A point of the parameter space is admitted
if the bounce action satisfies Eq.~(\ref{Smin}) in all cases listed in Table~\ref{cases}.

The D-term constraint~(\ref{Dterm}) is most easily taken into account in the
minimization of $\cal S$ by choosing a suitable parametrization of its general solution,
for instance:

\subsubsection{Cases 1, 4, 5 ($c=1$)}

For $c=1$, the most general solution of Eq.~(\ref{Dterm}) can be parametrized as:
\begin{align}
H\, &=\, s_H\bar H\cosh{\theta}\cos{\alpha}\, ,  \\
\bar\psi\, &=\, \bar H\cosh{\theta}\sin{\alpha}\, ,  \\
\psi\, &=\, \bar H\sinh{\theta}\, ,
\end{align}
with $s_H=\pm1$, $\theta\in[0,\infty[$ and $\alpha\in[0,\pi/2]$.
The actual value of $\bar H\ne0$ is irrelevant, as (\ref{Sapprox}) only depends on
ratios of fields.

\subsubsection{Cases 2, 3, 6 ($c=2$)}

For $c=2$, the most general solution of Eq.~(\ref{Dterm}) can be parametrized as:
\begin{align}
H\, &=\, s_H \bar H\cosh{\theta}\, ,  \\
\bar\psi\, &=\, \bar H\sinh{\theta}\sin{\alpha}\, ,  \\
\psi\, &=\, \bar H\sinh{\theta}\cos{\alpha}\, ,
\end{align}
with again $s_H=\pm1$, $\theta\in[0,\infty[$ and $\alpha\in[0,\pi/2]$. As in the previous case, 
the actual value of $\bar H\ne0$ is irrelevant.

\subsection{The special case of the stop}

In this case too we will make suitable approximations for estimating the minimal value
of the bounce action (for more details on the subject, see 
Refs.~\cite{Kusenko:1996jn,Camargo-Molina:2013sta,Blinov:2013uda,Chowdhury:2013dka,Blinov:2013fta}).
As explained in Ref.~\cite{Casas:1995pd}, one cannot assume exact cancellation of the D-terms here,
but we will nevertheless stick to the relatively good approximation of vanishing SU(3) D-terms,
namely $\tilde t=\tilde t^c$. Keeping the contributions of all other fields, one can write:
\begin{align}
{\cal L}_{kin}\, &=\, |h_u'(r)|^2+|h_d'(r)|^2+2|\tilde t'(r)|^2\, ,  \nonumber \\
V\, &=\, m_{H_u}^2|h_u|^2+m_{H_d}^2|h_d|^2+\left(m_{\tilde Q}^2+m_{\tilde t}^2\right)|\tilde t|^2  \nonumber \\
&\quad + \left( A_t \tilde t^2 h_u - \lambda_t \mu^* \tilde t^2 h_d^*
- B h_u h_d + \mbox{h.c.} \right)  \nonumber \\
&\quad +\, |\mu|^2\left(|h_u|^2+|h_d|^2\right)+|\lambda_t|^2\left(|\tilde t|^4+2|\tilde t|^2|h_u|^2\right)  \nonumber \\
&\quad +\, \frac{g'^2+g_2^2}{8}\left(|h_d|^2-|h_u|^2+|\tilde t|^2\right)^2\, .
\end{align}
Assuming again all quantities to be real, we take the same ansatz as for the other CCB constraints:
\beq
(h_u, h_d,\tilde t)(r)\, =\, (H_u, H_d, t)\;\phi(r)\, ,
\eeq
with $H_u$, $H_d$ and $t$ constant. From this we obtain:
\begin{align}
M_2^2\, &=\, \left(m_{\tilde Q}^2+ m_{\tilde t}^2\right)t^2
+\left(m_{H_u}^2+|\mu|^2\right)H_u^2+\left(m_{H_d}^2+|\mu|^2\right)H_d^2-2BH_u H_d\, ,  \\
M_3\, &=\, -2t^2(A_tH_u-\lambda_t\mu H_d)\, ,  \\
z\, &=\, 2\left(2t^2+H_u^2+H_d^2\right) ,  \\
\lambda\, &=\, \lambda_t^2t^2(t^2+2H_u^2)+\frac{g'^2+g_2^2}{8}\left(H_u^2-H_d^2-t^2\right)^2 .
\end{align}

Due to the large Yukawa coupling of the top quark, we also have to check
the lifetime of the electroweak vacuum for sizable values of $\kappa$, as opposed
to the previous cases. Hence we can no longer approximate
$\hat S(\kappa)$ by $\hat S(0)$, and we must minimize the following
(normalized) bounce action:
\beq
{\cal S}\, =\, \left(1+\frac{f(\kappa)-f(0)}{\hat S(0)}\right)\frac{z^2M_2^2}{M_3^2}\, ,
\eeq
with~\cite{Sarid:1998sn}
\beq
f(\kappa)\, =\, \frac{\pi^2/6}{(1-4\kappa)^3}+\frac{16.5}{(1-4\kappa)^2}+\frac{28}{1-4\kappa}\, ,
\eeq
and $\hat S(0) \approx 45.4$. The minimization goes again over two variables,
for example the ratios $H_u/t$ and $H_d/t$, and is done numerically.
It is important to keep in mind that $\cal S$ should
be minimized only over the range $0\leq \kappa <1/4$, for which the potential is metastable.
For $\kappa<0$ the potential is unbounded from below and for $M_2^2<0$ it is unstable,
while for $\kappa\geq 1/4$ the electroweak vacuum is the global minimum of the model.

\subsection{How to improve the estimate}

We can improve the estimates described above in several ways:
\begin{itemize}
\item
on top of the 6 cases described in Table~\ref{cases}, one could consider the case
where both $\phi$ and $\bar\phi$ are leptons. One would then need to perform a
minimization with respect to three fields, since there is no constraint due to SU(3) D-terms;
\item 
one could consider more general radius-dependent directions in field space, which
would require the use of numerical algorithms as in Ref.~\cite{Wainwright:2011kj};
\item 
we have calculated transitions between vacua of the tree-level scalar potential
at the fixed scale $m_{susy}$. To improve our estimates higher order corrections
and/or RG-improved potentials should be considered. Such corrections can be
relevant in some cases~\cite{Casas:1995pd}.
\end{itemize}
All these generalizations are however quite involved and are beyond the scope of this paper.


\end{document}